\documentclass[prd,twocolumn,nofootinbib,reprint,preprintnumbers,showkeys]{revtex4-1}
\usepackage{color}
\usepackage{amsmath,amsfonts,bm}
\usepackage{graphicx, epstopdf,scrextend}
\usepackage{mathrsfs,mathtools}

\usepackage{pgfplots,pgfplotstable}
\usepgfplotslibrary{fillbetween}
\usepgfplotslibrary{groupplots}
\usetikzlibrary{intersections}
\pgfplotsset{compat=newest}

\pgfplotsset{every axis/.style={
    width=8.6cm,
    height=8.6cm,
    grid=both,
    scaled ticks=false,
    yticklabel style={/pgf/number format/.cd, fixed,precision=5}
  }
}

\newcommand{\errorband}[6][]{
  \pgfplotstableread{#3}\datatable
  \addplot[name path=pluserror,draw=none,no markers,forget plot]  table [x={#4},y expr=\thisrow{#5}+\thisrow{#6}]{\datatable};
  \addplot[name path=minuserror,draw=none,no markers,forget plot] table [x={#4},y expr=\thisrow{#5}-\thisrow{#6}]{\datatable};
  \addplot[forget plot,#2] fill between[on layer={},of=pluserror and minuserror];
  \addplot [#1,no markers] table [x={#4},y={#5}]{\datatable};
}

\newcommand{\ratioplot}[8][]{
  \pgfplotstableread{#3}\datatable
    
  \addplot[name path=pluserror,draw=none,no markers,forget plot]  table
  [ x={#4},
    y expr=\thisrow{#5}/\thisrow{#7} 
    + abs(\thisrow{#6}/\thisrow{#7})
    + abs(\thisrow{#5}*\thisrow{#8}/\thisrow{#7}/\thisrow{#7})
  ]{\datatable};
    
  \addplot[name path=minuserror,draw=none,no markers,forget plot]  table
  [ x={#4},
    y expr=\thisrow{#5}/\thisrow{#7}
    - abs(\thisrow{#6}/\thisrow{#7})
    - abs(\thisrow{#5}*\thisrow{#8}/\thisrow{#7}/\thisrow{#7})
  ]{\datatable};
    
  \addplot[forget plot,#2] fill between[on layer={},of=pluserror and minuserror];
  \addplot [#1,no markers] table [x={#4},y expr=\thisrow{#5}/\thisrow{#7}]{\datatable};
}

\definecolor{darkgreen}{rgb}{0,0.4,0} 
\definecolor{darkblue}{rgb}{0,0,0.6} 
\usepackage[colorlinks=true,linkcolor=darkblue,citecolor=darkgreen]{hyperref}

\newcommand{\as}{\alpha_{\mathrm{s}}}

\newcommand{\LA}{\mathrm{A}}

\newcommand{\LE}{\mathrm{E}}
\newcommand{\LF}{\mathrm{F}}

\newcommand{\LL}{\mathrm{L}}
\newcommand{\LR}{\mathrm{R}}
\newcommand{\LS}{\mathrm{S}}
\newcommand{\LT}{\mathrm{T}}
\newcommand{\LU}{\mathrm{U}}
\newcommand{\La}{\mathrm{a}}
\newcommand{\Lb}{\mathrm{b}}
\newcommand{\Lc}{\mathrm{c}}
\newcommand{\Lf}{\mathrm{f}}

\newcommand{\Lg}{\mathrm{g}}
\newcommand{\Lp}{\mathrm{p}}
\newcommand{\Ls}{\mathrm{s}}
\newcommand{\Lu}{\mathrm{u}}

\newcommand{\LCP}{{\textsc{lc}\raisebox{+0.5 pt}{$\scriptscriptstyle{+}$}}}

\newcommand{\cH}{\mathcal{H}}

\newcommand{\cN}{\mathcal{N}}
\newcommand{\cO}{\mathcal{O}}
\newcommand{\cP}{\mathcal{P}}

\newcommand{\cU}{\mathcal{U}}
\newcommand{\cV}{\mathcal{V}}

\newcommand{\cX}{\mathcal{X}}

\newcommand{\GeV}{\ \mathrm{GeV}}
\newcommand{\TeV}{\ \mathrm{TeV}}

\definecolor{red}{rgb}{1,0,0}
\newcommand{\NOTE}[1]{\textcolor{red}{ \bf[NOTE: #1]}}

\def\mi{{\mathrm i}}

\def\ket#1{\big|{#1}\big\rangle}
\def\bra#1{\big\langle{#1}\big|}
\def\brax#1{\big\langle{#1}}   
\def\<>#1{\big\langle{#1}\big\rangle}
\def\[]#1{\big[{#1}\big]}

\def\sket#1{\big|{#1}\big)}
\def\sbra#1{\big({#1}\big|}
\def\sbrax#1{\big({#1}}        

\newbox\charbox
\newbox\slabox
\def\s#1{{      
        \setbox\charbox=\hbox{$#1$}
        \setbox\slabox=\hbox{$/$}
        \dimen\charbox=\ht\slabox
        \advance\dimen\charbox by -\dp\slabox
        \advance\dimen\charbox by -\ht\charbox
        \advance\dimen\charbox by \dp\charbox
        \divide\dimen\charbox by 2
        \raise-\dimen\charbox\hbox to \wd\charbox{\hss/\hss}
        \llap{$#1$}
}}

\newif\ifusefigs
\usefigstrue

\begin{document}

\title{Parton showers with more exact color evolution}

\author{Zolt\'an Nagy}

\affiliation{
  DESY,
  Notkestrasse 85,
  22607 Hamburg, Germany
}

\email{Zoltan.Nagy@desy.de}

\author{Davison E. Soper}

\affiliation{
Institute of Theoretical Science,
University of Oregon,
Eugene, OR  97403-5203, USA
}

\email{soper@uoregon.edu}

\begin{abstract}
Parton shower event generators typically approximate evolution of QCD color so that only contributions that are leading in the limit of an infinite number of colors are retained. Our parton shower generator, \textsc{Deductor}, has used an ``LC+'' approximation that is better, but still quite limited. In this paper, we introduce a new scheme for color in which the approximations can be systematically improved. That is, one can choose the theoretical accuracy level, but the accuracy level that is practical is limited by the computer resources available.

\end{abstract}

\preprint{DESY 19-013}
\date{7 March 2019}

\maketitle

\section{\label{sec:intro}Introduction}

Parton shower event generators for hadron collisions, such as \textsc{Herwig}  \cite{Herwig}, \textsc{Pythia} \cite{Pythia}, and \textsc{Sherpa} \cite{Sherpa},  are essential for the analysis of experiments at the Large Hadron Collider. They treat QCD color in the leading color (LC) approximation, that is at leading order in an expansion in powers of $1/N_\Lc^2$, where $N_\Lc = 3$ is the number of colors. Previous versions of our parton shower generator, \textsc{Deductor} \cite{NSI, NSII, NSspin, NScolor, Deductor, ShowerTime, PartonDistFctns, ColorEffects, NSThreshold, NSThresholdII}, use what we call the LC+ approximation \cite{NScolor}, which includes some terms suppressed by powers of $1/N_\Lc^2$.

This paper concerns improvements in the color treatment in \textsc{Deductor} that could be applicable to other parton shower generators. The treatment of parton spin is also important, but we simply ignore spin here.\footnote{In the implementation of the parton shower in \textsc{Deductor}, we average over spins at each stage. This is evidently an approximation. We analyzed what to do with spin in Ref.~\cite{NSspin}, but this strategy for spin is not implemented in \textsc{Deductor}.} It is also of interest to consider how color works in a parton shower at an arbitrary order of perturbation theory for the splitting functions that generate the shower \cite{NSAllOrder}. However, we work only to first order in $\as$ for the splitting functions here.

One can define an evolution equation for a parton shower with leading order splitting functions such that the evolution is exact in color. This is relatively straightforward \cite{NSI}, as we explain briefly below. Once we understand the exact evolution equation, we are faced with trying to implement it as a computer program. This is, so far as we know, impossible with any available computer. However, one should not be discouraged. What we really need is not a numerical answer that is exact with respect to color for the calculation of a cross section $\sigma$ of interest, but rather an approximate answer that can be systematically improved.

What we seek is an algorithm for a parton shower that depends on parameters that control the level of approximation with respect to color. We can begin with the cross section calculated at the lowest level of approximation, call it $\sigma(0)$. Then we can change parameters in the calculation so that we get successively better approximations, $\sigma(1), \sigma(2), \dots$. We will find that the successive calculations use more and more computer resources, so that there will be a practical limit to how exact we can be.  Note that $|\sigma(n) - \sigma(n-1)|$ serves as an error estimate for $\sigma(n)$. If $|\sigma(n) - \sigma(n-1)|$ is not as small as we would like by the time that we run out of computer resources, we will have to admit the limitations of the calculation. However, even in that case, we obtain a calculation with an error estimate. With just $\sigma(0)$, we have a less precise calculation with no error estimate.

There has been work on extending the accuracy of parton shower algorithms beyond the leading order in an expansion in powers of $1/N_\Lc^2$ \cite{NScolor, ColorEffects, PlatzerSjodahl, Seymour2018, Isaacson:2018zdi, PlatzerSjodahlThoren}. To date, however, what we see as the main obstacle to a systematically improvable treatment of color has not been overcome. This obstacle has been nicely stated in Ref.~\cite{PlatzerSjodahlThoren}: ``To fully include all subleading $N_\Lc$ terms in the soft and collinear limits, virtual color rearranging terms associated with the same singularity structure should also be kept. To accomplish this, a full resummation of virtual exchanges is needed. Unfortunately, within the current event generator structure these contributions cannot be included, and we postpone their inclusion for future work.'' The aim of the present paper is to provide an event generator structure that does include these contributions in a systematically improvable fashion.

Before we turn to approximate versions of color in a parton shower, we briefly describe a shower evolution equation that is exact in color, as in Ref.~\cite{NSI}. We view shower evolution as an application of the renormalization group. At any stage, we include hard interactions and remain inclusive over softer interactions. Then the picture changes when we change the definition of where ``hard'' interactions end. We define $t = \log(\mu_0^2/\mu^2)$, where $\mu^2$ is a measure of hardness\footnote{As reviewed in Ref.~\cite{NSThresholdII}, the default hardness measure in \textsc{Deductor} is a variable $\Lambda^2$ that is proportional to the virtuality of the splitting. See Eq.~(\ref{eq:showertime}). However, an alternative in \textsc{Deductor} is an appropriately defined transverse momentum squared $k_\LT^2$ in the splitting.} in a splitting and $\mu_0^2$ is a reference value of $\mu^2$. The shower develops with increasing $t$ as the interactions that are included become softer and softer.

At any stage of the shower we have $m$ final state partons plus two initial state partons, ``a'' and ``b,'' with momenta and flavors $\{p,f\}_m = \{p_\La,f_\La, p_\Lb,f_\Lb, p_1,f_1, \dots, p_m,f_m\}$. These partons carry QCD color, so we can describe them using a quantum amplitude $\ket{M(\{p,f\}_m)}$, which is a vector in the color space of $m$ final state partons plus two initial state partons. In order to keep the notation simple, we ignore spin in this paper, but if we were to include spin, then $\ket{M(\{p,f\}_m)}$ would be a vector in the color$\times$spin space.

Shower evolution is probabilistic since, at any stage of the shower, we are integrating over probabilities for potential splittings that are too soft to be visible at that stage. For this reason, we use the language of quantum statistical mechanics to describe shower evolution. With this language, we do not use pure color amplitudes but rather a color density operator,
\begin{equation}
\rho(\{p,f\}_m) =
\sum_n \ket{M_n(\{p,f\}_m)}\bra{M_n(\{p,f\}_m)}
\;.
\end{equation}
We use the color density operator as the basis for parton shower evolution. We note that it is sometimes also used explicitly for analytical summations of large logarithms \cite{CaronHuot, Becher:2016mmh, Balsiger:2018ezi, Balsiger:2019tne}.

We can expand $\rho(\{p,f\}_m)$ in color basis states,
\begin{equation}
\label{eq:rhomatrix}
\rho(\{p,f\}_m) = \sum_{\{c,c'\}_m}
\rho(\{p,f,c,c'\}_m)
\ket{\{c\}_m}\bra{\{c'\}_m}
\;.
\end{equation}
\textsc{Deductor} uses the trace basis (which might better be called the color string basis) for color basis states $\{c\}_m$ \cite{NSI}. The set of all functions $\rho(\{p,f,c,c'\}_m)$ constitutes a vector space, which we call the statistical space.\footnote{To be more precise, $\rho$ is a function of $m$ and $\{p,f,c,c'\}_m$. Additionally, $\rho$ depends parameters $x$ derived from prior states, $\{p,f,c,c'\}_{m'}$ with $m' < m$. For example, $\rho$ depends on the vector $Q_0$ equal to the total momentum of the final state partons at the start of the shower, which is used to define the shower evolution variable. Normally we suppress the dependence on other parameters $x$ in statistical states and in functions.} We represent the function $\rho$ as a ket vector, $\sket{\rho}$. The rounded end of the ket is meant to distinguish a vector in the statistical space from a vector in the quantum color space.

Note that $\sket{\rho}$ represents the whole function $\rho$, which gives the distribution of simulated events as a function of the number of partons and their momenta, flavors, and colors. A parton shower event generator generates particular events by Monte Carlo sampling from this distribution. In each event, there is, for instance, a certain choice for the number of partons and their momenta. 

We find it useful to use a notation in which linear operators act on vectors $\sket{\rho}$ in the statistical space. Thus we might write
\begin{equation}
\label{eq:A3A2A1rho}
\sket{\rho'} = A_3 A_2 A_1 \sket{\rho}
\;.
\end{equation}
Mart\'inez, De Angelis, Forshaw, Pl\"atzer, and Seymour \cite{Seymour2018} have recently analyzed the influence of color on parton evolution via soft gluon emission using the color density operator, but with a different notation from ours. If the operators $A_i$ in Eq.~(\ref{eq:A3A2A1rho}) have a suitable product form, these authors would write $\rho$ and $\rho'$ as density operators as in Eq.~(\ref{eq:rhomatrix}) and write
\begin{equation}
\label{eq:A3A2A1rhoA1A2A3}
\rho' = A^{(L)}_3 A^{(L)}_2 A^{(L)}_1 \rho\, A^{(R)}_1 A^{(R)}_2 A^{(R)}_3
\;.
\end{equation}
There is no physics difference between these two notations. We will use the notation of Eq.~(\ref{eq:A3A2A1rho}).

\section{Evolution equation exact in color}

We can now discuss the evolution equation that forms the basis for the approximations used in \textsc{Deductor} \cite{NSI, NSII, NScolor}.  The statistical state changes as the hardness resolution varies, so that it is a function $\sket{\rho(t)}$ of $t$. We can write
\begin{equation}
\sket{\rho(t)} = \cU(t,t_0) \sket{\rho(t_0)}
\end{equation}
for any $t$ and $t_0$. There is also a threshold factor $\cU_\cV$ that appears at the start of the shower \cite{NSAllOrder, NSThresholdII}. See Eq.~(\ref{eq:sigmaU8}). We return to this factor later. 

The shower evolution equation operator $\cU(t,t_0)$ with full color obeys the evolution equation,
\begin{equation}
\label{eq:evolutionU}
\frac{d}{dt}\,\cU(t,t_0)
= [\cH_I(t) - \cV(t)]\,\cU(t,t_0)
\;.
\end{equation}
The operator $\cH_I(t)$ creates a parton splitting, increasing the number of partons by one. The new parton carries color and the colors of the old partons change. The operator $\cV(t)$ leaves the number of partons unchanged. It does, however, change the color state of the partons. It carries the color structure of virtual graphs.

These two operators are related by an identity,
\begin{equation}
\label{eq:1HV}
\sbra{1}\cH_I(t) = \sbra{1}\cV(t)
\;.
\end{equation}
Here, for any parton state $\sket{\rho}$, to calculate $\sbrax{1}\sket{\rho}$, we integrate over all of the parton momenta in $\sket{\rho}$, sum over flavors, and take the trace over colors. This gives us the total probability associated with $\sket{\rho}$. Because of Eq.~(\ref{eq:1HV}), together with the initial condition $\cU(t_0,t_0) = 1$, we have
\begin{equation}
\label{eq:unitarity}
\sbra{1} \cU(t,t_0)\sket{\rho} = \sbrax{1}\sket{\rho}
\end{equation}
for any statistical state $\sket{\rho}$. This says that probability is conserved in the evolution of the state.

The operator $\cV(t)$ contains a term that we can call $\cV_{\mi\pi}(t)$ that we calculate from the imaginary part of virtual graphs. This operator has the form (assuming massless partons) given in Eq.~(10.14) of Ref.~\cite{NScolor},
\begin{equation}
\label{eq:Vipi}
\cV_{\mi\pi}(t) = - 4\mi \pi \frac{\as}{2\pi}
\big([(T_\La\cdot T_\Lb)\otimes 1]
- [1\otimes (T_\La\cdot T_\Lb)]\big)
\;.
\end{equation}
Here $T_\La$ represents the insertion of a color matrix $T^c$ on incoming parton line ``a'', $T_\Lb$ represents the insertion of a color matrix $T^c$ on incoming parton line ``b'', and the dot in $(T_\La\cdot T_\Lb)$ represents a summation over the octet color index $c$. In $[(T_\La\cdot T_\Lb)\otimes 1]$ the color matrices act on the ket state, while in $[1\otimes (T_\La\cdot T_\Lb)]$, they act on the bra state. When we take the color trace, we get $\sbra{1} \cV_{\mi\pi}(t) = 0$. 

The operator $\cV(t)$ also contains terms with real coefficients that reflect the color structure of the real parts of virtual graphs,
\begin{equation}
\label{eq:Vreal}
\cV_{lk}(t) \propto
\big([(T_l\cdot T_k)\otimes 1]
+ [1\otimes (T_l\cdot T_k)]\big)
\;.
\end{equation}
Here $[(T_l\cdot T_k)\otimes 1]$ inserts color matrices on the color lines of partons with indices $l$ and $k$ in the ket state, while $[1 \otimes (T_l\cdot T_k)]$ inserts color matrices on the color lines of partons with indices $l$ and $k$ in the bra state. Here $l$ or $k$ or both can be the indices ``a'' and ``b'' of the initial state partons and $k$ can equal $l$.

\section{Evolution in the LC+ approximation}

We can use the LC+ approximation described in Ref.~\cite{NScolor}:
\begin{equation}
\label{eq:evolutionLCplusbis}
\frac{d}{dt}\,\cU^\LCP(t,t_0)
= [\cH^\LCP(t) - \cV^\LCP(t)]\,
\cU^\LCP(t,t_0)
\;.
\end{equation}
Ref.~\cite{NScolor} does not give an LC+ approximation for $\cV_{\mi\pi}(t)$.\footnote{The suggestion in Ref.~\cite{NScolor} that the first term in Eq.~(10.8) of that paper might be included in the LC+ approximation does not work because Eq.~(\ref{eq:1His1V}) below would fail.} In this paper, we simply take\footnote{One can, at least in principle, include $\cV_{\mi\pi}(t)$ in the LC+ approximation. We leave the exploration of this possibility to future work.}
\begin{equation}
\cV^\LCP_{\mi\pi}(t) = 0
\;.
\end{equation}
The LC+ operators $\cH^\LCP(t)$ and $\cV^\LCP(t)$ obey
\begin{equation}
\label{eq:1His1V}
\sbra{1}\cH^\LCP(t)
= \sbra{1}\cV^\LCP(t)
\;.
\end{equation}
This gives us
\begin{equation}
\sbra{1} \cU^\LCP(t,t_0) = \sbra{1}
\;,
\end{equation}
so that probability is conserved in LC+ evolution.

The differential equation (\ref{eq:evolutionLCplusbis}) can be solved in the form
\begin{equation}
  \label{eq:evolutionsolutionLCplus}
  \begin{split}
    \cU^\LCP{}&(t,t_0) =
    \cN^\LCP(t,t_0)
    \\
    &+
    \int_{t_0}^t\!d\tau\ 
    \cU^\LCP(t,\tau)\,
    \cH^\LCP(\tau)\,
    \cN^\LCP(\tau,t_0) 
    \;.
  \end{split}
\end{equation}
Here $\cN^\LCP(t_{2},t_1)$ is the no-splitting operator,
\begin{equation}
\label{eq:NLCplus}
\cN^\LCP(t_2,t_1) = \exp\left[
-\int_{t_1}^{t_2} d\tau\ \cV^\LCP(\tau)\right]
\;.
\end{equation}
It is well to recall here an essential point: the operator $\cV^\LCP(\tau)$ is diagonal in the color basis that we use, the trace basis, so that it is practical to calculate its exponential.

\section{Evolution beyond the LC+ approximation}
\label{sec:generalevolution}

Now, what if we want shower evolution with full color? Then we need
\begin{equation}
  \label{eq:evolutionU2}
  \begin{split}
    \frac{d}{dt}\,\cU(t,t_0)
    = [{}&\cH^\LCP(t) - \cV^\LCP(t)
    \\
    &
    + \Delta \cH(t) - \Delta\cV(t)]\,\cU(t,t_0)\;,
  \end{split}
\end{equation}
where
\begin{equation}
\begin{split}
\label{eq:DeltaHDeltaV}
\Delta \cH(t) ={}& \cH_I(t) - \cH^\LCP(t)
\;,
\\
\Delta \cV(t) ={}& \cV(t) - \cV^\LCP(t)
\;.
\end{split}
\end{equation}
Note that since we have set $\cV^\LCP_{\mi\pi}(t) = 0$, $\cV_{\mi\pi}(t)$ is included in $\Delta \cV(t)$:
\begin{equation}
\label{eq:DeltaVreDeltaVipi}
\Delta \cV(t) = \Delta \cV_\textrm{Re}(t) + \cV_{\mi\pi}(t)
\;.
\end{equation}

This differential equation can be solved in the form
\begin{equation}
  \label{eq:evolutionU4}
  \begin{split}
    \cU{}&(t,t_0) = 
    \cN(t,t_0)
    \\
    &+
    \int_{t_0}^t\!d\tau\ 
    \cU(t,\tau)\,
    \big[\cH^\LCP(\tau) + \Delta \cH(\tau)\big]\,\cN(\tau,t_0)\;.
  \end{split}
\end{equation}
Here $\cN(t_{2},t_1)$ is the no-splitting operator,
\begin{equation}
\cN(t,t_0) = \mathbb{T} \exp\!\left[
-\int_{t_0}^{t} d\tau\ \left[
\cV^\LCP(\tau) + \Delta\cV(\tau)
\right]\right]
\;,
\end{equation}
where $\mathbb{T}$ denotes time ordering for the non-commuting operators in the exponent, with latest times to the left.

Unfortunately, we cannot include $\Delta\cV(\tau)$ here in exponentiated form because of its complicated color structure. Thus we write the evolution equation for $\cN(t,t_0)$:
\begin{equation}
\label{eq:evolutionN}
\frac{d}{dt}\,\cN(t,t_0)
= -[\cV^\LCP(t) +  \Delta\cV(t)]\,
\cN(t,t_0)
\;.
\end{equation}
We can solve this in the form
\begin{equation}
  \label{eq:evolutionN2}
  \begin{split}
    \cN(t,t_0) = {}&
    \cN^\LCP(t,t_0)
    \\
    &
    - \int_{t_0}^t\!d\tau\ 
    \cN(t,\tau)\,
    \Delta \cV(\tau)\,
    \cN^\LCP(\tau,t_0) 
    \;.
  \end{split}
\end{equation}
This can be solved iteratively. The first three terms are
\begin{equation}
  \begin{split}
    \label{eq:Nexpansion}
    \cN{}&(t,t_0) = \cN^\LCP(t,t_0)
    \\ & 
    - \int_{t_0}^t\!d\tau\ 
    \cN^\LCP(t,\tau)\,
    \Delta \cV(\tau)\,
    \cN^\LCP(\tau,t_0)
    \\ & 
    + \int_{t_0}^t\!d\tau_2\ \int_{t_0}^{\tau_2}\!d\tau_1\ 
    \cN^\LCP(t,\tau_2)\,
    \Delta \cV(\tau_2)\,
    \\
    &\qquad\times
    \cN^\LCP(\tau_2,\tau_1)\,
    \Delta \cV(\tau_1)\,
    \cN^\LCP(\tau_1,t_0)
    \\&
    + \cdots 
    \;.
  \end{split}
\end{equation}
It is convenient to write the solution as
\begin{equation}
\label{eq:Xdef}
\cN(t,t_0) = \cX(t,t_0)\, \cN^\LCP(t,t_0)
\;,
\end{equation}
where
\begin{equation}
  \begin{split}
    \label{eq:NDeltaexpansion}
    \cX{}&(t,t_0) = 1
    \\ & 
    - \int_{t_0}^t\!d\tau\ 
    \cN^\LCP(t,\tau)\,
    \Delta \cV(\tau)\,
    \cN^\LCP(t,\tau)^{-1}
    \\ & 
    + \int_{t_0}^t\!d\tau_2\ \int_{t_0}^{\tau_2}\!d\tau_1\ 
    \cN^\LCP(t,\tau_2)\,
    \Delta \cV(\tau_2)\,
    \\
    &\qquad\times
    \cN^\LCP(\tau_2,\tau_1)\,
    \Delta \cV(\tau_1)\,
    \cN^\LCP(t,\tau_1)^{-1}
    \\&
    + \cdots 
    \;.
  \end{split}
\end{equation}
When we use $\cX(t,t_0)$, we understand that it is expanded to whatever order in $\Delta \cV(\tau)$ that we need.

Using the operator $\cX$, we write the evolution equation Eq.~(\ref{eq:evolutionU4}) in the form
\begin{equation}
  \label{eq:evolutionU5}
  \begin{split}
    \cU{}&(t,t_0) = 
    \cX(t,t_0)\, \cN^\LCP(t,t_0)
    \\&
    + \int_{t_0}^t\!d\tau\ 
    \cU(t,\tau)\,
    \big[\cH^\LCP(\tau) + \Delta \cH(\tau)\big]\,
    \\
    &\qquad\times\cX(\tau,t_0)\, \cN^\LCP(\tau,t_0) 
    \;.
  \end{split}
\end{equation}
This generates a shower, at least in principle. At each step in the shower the splitting operator is
\begin{equation}
\label{eq:netsplitting}
\cO(\tau) = 
\big[\cH^\LCP(\tau) + \Delta \cH(\tau)\big]\,
\cX(\tau,t_0)
\;.
\end{equation}
Notice that $\cX(\tau,t_0)$ is an operator on the color space, but does not create any new partons. Once the color is transformed by $\cX(\tau,t_0)$, the operator $\big[\cH^\LCP(\tau) + \Delta \cH(\tau)\big]$ creates a new parton and further modifies the color. The net operator in Eq.~(\ref{eq:netsplitting}) is then a splitting operator in the sense that it creates a new parton. It is also a non-trivial operator on the color state.

Our goal now is to treat the operator $\Delta \cH(\tau)$ and the operator $\Delta \cV(\tau)$ in $\cX$ perturbatively. To avoid confusion, we note that expanding in powers of $\Delta \cH(\tau)$ and $\Delta \cV(\tau)$ is not equivalent to expanding in powers of $1/N_\Lc^2$.

In order to make the evolution equation~(\ref{eq:evolutionU5}) practical for a computer program, we will need to rearrange it. To do that, we first review the singularities that control the evolution and analyze the different roles of soft and collinear singularities.

\section{Splitting variables and singularities}
\label{sec:splittingvariables}

The splitting operator $\cH_I(t)$ is singular in the limit of very large shower times $t$, corresponding to very small splitting virtualities. In order to study this limit, it is convenient to define a dimensionless virtuality variable $y$. For a final state splitting in which a massless parton with momentum $p_l$ splits into massless daughter partons with momenta $\hat p_l$ and $\hat p_{m+1}$, we define (Ref.~\cite{NSI}, Eq.~(4.19))
\begin{equation}
\label{eq:ydefFS}
y = \frac{(\hat p_l + \hat p_{m+1})^2}{2 p_l \cdot Q}
\;,
\end{equation}
where $Q$ is the total momentum of the final state partons before the splitting. 
For an initial state splitting in which a massless initial state parton with momentum $p_l$ ($l =$ a or b) splits in backward evolution into a new massless initial state parton with momentum $\hat p_l$ and a massless final state parton with momentum $\hat p_{m+1}$, we define (Ref.~\cite{PartonDistFctns}, Eq.~(4.1))
\begin{equation}
\label{eq:ydefIS}
y = -\frac{(\hat p_l - \hat p_{m+1})^2}{2 p_l \cdot Q}
\;.
\end{equation}

In addition to $y$, the splitting functions in $\cH_I(t)$ depend on a momentum fraction $z$ (Ref.~\cite{NSThresholdII}, Eqs.~(7) and (9)). Various definitions of $z$ are possible. For a final state splitting, \textsc{Deductor} uses
\begin{equation}
\label{eq:zdefFS}
\frac{\hat p_{m+1}\cdot \tilde n_l}{\hat p_{l}\cdot \tilde n_l}
= \frac{1-z}{z}
\;,
\end{equation}
where the auxiliary lightlike vector $\tilde n_l$ is defined using the total momentum $Q$ of the final state partons:
\begin{equation}
\label{tildenldef}
\tilde n_l = \frac{2 p_l\cdot Q}{Q^2}\, Q - p_l
\;.
\end{equation}
For the splitting of initial state parton ``a'' from hadron A, $z$ is the ratio of momentum fractions $\eta_\La$ before and $\hat \eta_\La$ after the splitting: 
\begin{equation}
\label{eq:zdefIS}
z = \frac{\eta_\La}{\hat \eta_\La}  
= \frac{p_\La\cdot p_\Lb}{\hat p_\La\cdot p_\Lb}
\;.
\end{equation}
Here $p_\Lb$ is the momentum of the initial state parton from hadron B. There is a third splitting variable, the azimuthal angle $\phi$ of $\hat p_{m+1}$ about the direction of the mother parton, $p_l$ in the rest frame of $Q$.

We will denote the splitting variables $\{y,z,\phi\}$ collectively by $\zeta_\Lp$ and the flavor choice in the splitting by $\zeta_\Lf$. We denote $\{\zeta_\Lp, \zeta_\Lf\}$ by $\zeta$, as described in more detail in Appendix \ref{sec:HandVresult}.

The default choice of the shower ordering variable in \textsc{Deductor} is $\Lambda^2$, defined by 
\begin{equation}
\begin{split}
\label{eq:Lambdadef}
\Lambda^2 ={}& \frac{(\hat p_l + \hat p_{m+1})^2}{2 p_l\cdot Q_0}\ Q_0^2
\hskip 1 cm {\rm final\ state},
\\
\Lambda^2 ={}&  \frac{|(\hat p_l - \hat p_{m+1})^2|}{2 p_l \cdot Q_0}\ Q_0^2
\hskip  1 cm {\rm initial\ state},
\end{split}
\end{equation}
where $Q_0$ is the total momentum of the final state particles at the start of the shower. (See Ref.~\cite{NSThresholdII}, Eq.~(5)). Then the shower time $t$ is the logarithmic variable (Ref.~\cite{NSThresholdII}, Eq.~(A.6)),
\begin{equation}
\label{eq:showertime}
t = \log(Q_0^2/\Lambda^2) =
\log\left(\frac{1}{y}\,
\frac{p_l\cdot Q_0}{p_l\cdot Q}\right)
\;,
\end{equation}
so that $y\to 0$ implies $t \to \infty$.

\section{Soft versus collinear contributions.}
\label{sec:softcollinear}

The splitting functions in $\cH_I(t)$ are singular in the limit $y \to 0$. There are two kinds of singularities. There are collinear singularities, corresponding to $y \to 0$ at fixed $z$. There are also fixed angle, soft singularities, corresponding to  emission of a soft gluon carrying momentum fraction $(1-z)$:  $y \to 0$ and $z \to 1$ with fixed $y/(1-z)$. There are also collinear$\times$soft singularities, in which $y \to 0$, $z \to 1$, and $y/(1-z) \to 0$. We recall \cite{NScolor} that the LC+ approximation is exact for collinear splittings and for collinear$\times$soft splittings. It it approximate only for fixed angle soft splittings.

In this section we explore the contribution of the collinear and collinear$\times$soft regions compared to the contribution of the fixed angle soft contribution to the operators $\Delta \cH(\tau)$ and $\cX(\tau,t_0)$. We do this by decomposing the operators involved into parts that get contributions only from the fixed angle soft region and parts that get contributions from everywhere else. The \textsc{Deductor} code is not organized using this decomposition, but nevertheless the analysis of this section can help us to understand the behavior of the approximations that we use to go beyond the LC+ approximation.

We can divide $\cH_I(t)$ into two contributions \cite{NSI},
\begin{equation}
\cH_I(t) = \cH_{\textrm{coll.}}(t) + \cH_{\textrm{soft}}(t)
\;.
\end{equation}
The contribution $\cH_{\textrm{coll.}}(t)$ contains the collinear and also the collinear$\times$soft singularities, while $\cH_{\textrm{soft}}(t)$ contains only the fixed angle, soft singularities. We review the definitions for this decomposition in Appendix.~\ref{sec:softsplitting}. Here we state only a few important properties of the two contributions.

\begin{figure}
\begin{center}
\includegraphics[width = 8.6 cm]{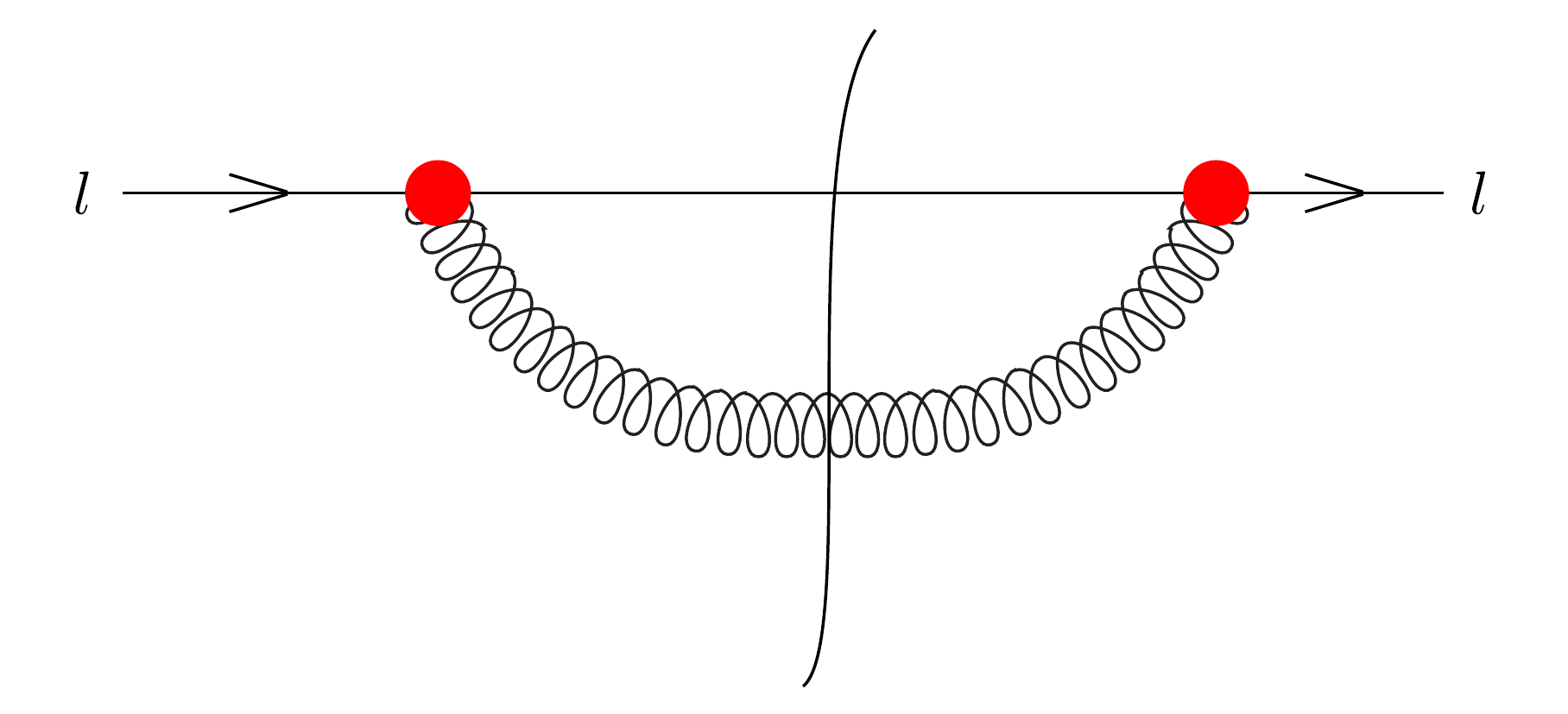}
\end{center}
\caption{
Collinear real emission from parton $l$. In this cut diagram, the vertical line represents the final state.
}
\label{fig:directsplitting}
\end{figure}

The contribution $\cH_{\textrm{coll.}}(t)$ comes from the square of a Feynman graph in a physical gauge in which the new parton $m+1$ is emitted from a given parton ({\it e.g.} $l$ or $\La$), as illustrated in Fig.~\ref{fig:directsplitting}. This gives a function whose $y \to 0$ limit is a DGLAP splitting function $P(z)$, including its $1/(1-z)$ singularity. 

The contribution $\cH_{\textrm{soft}}(t)$ comes from interference of two emissions: a soft gluon emitted from one parton in the ket state and the same soft gluon emitted from a different parton in the bra state, as illustrated in Fig.~\ref{fig:interferencesplitting}. As we will see in Appendix \ref{sec:softsplitting}, in a physical gauge, this contribution has at most integrable singularities in the limit in which the emitted gluon becomes collinear to either of the two emitting partons \cite{NSII}.

In Eq.~(\ref{eq:DeltaHDeltaV}), we defined another decomposition of $\cH_I(t)$,
\begin{equation}
\cH_I(t) = \cH^\LCP(t) 
+ \Delta\cH(t)
\;.
\end{equation}
According to the definition of the LC+ approximation in Ref.~\cite{NScolor}, the LC+ approximation is exact for splittings corresponding to the square of a Feynman graph in a physical gauge. We can divide $\cH_I^{\textrm{coll.}}(t)$ into its LC+ approximation and a remainder, as in Appendix \ref{sec:softsplitting},  
\begin{equation}
\cH_{\textrm{coll.}}(t)
=
\cH_{\textrm{coll.}}^\LCP(t) + \Delta\cH_{\textrm{coll.}}(t)
\;.
\end{equation}
We have
\begin{equation}
\Delta\cH_{\textrm{coll.}}(t) = 0
\;.
\end{equation}
That is, $\Delta\cH(t)$ has only a fixed angle, soft contribution, but has no collinear contribution. On the other hand, the LC+ approximation is not exact for $\cH_{\textrm{soft}}(t)$, so that in the decomposition
\begin{equation}
\cH_{\textrm{soft}}(t) = \cH_{\textrm{soft}}^\LCP(t) 
+ \Delta\cH_{\textrm{soft}}(t)
\;,
\end{equation}
both contributions are non-zero. 

\begin{figure}
\begin{center}
\includegraphics[width = 8.6 cm]{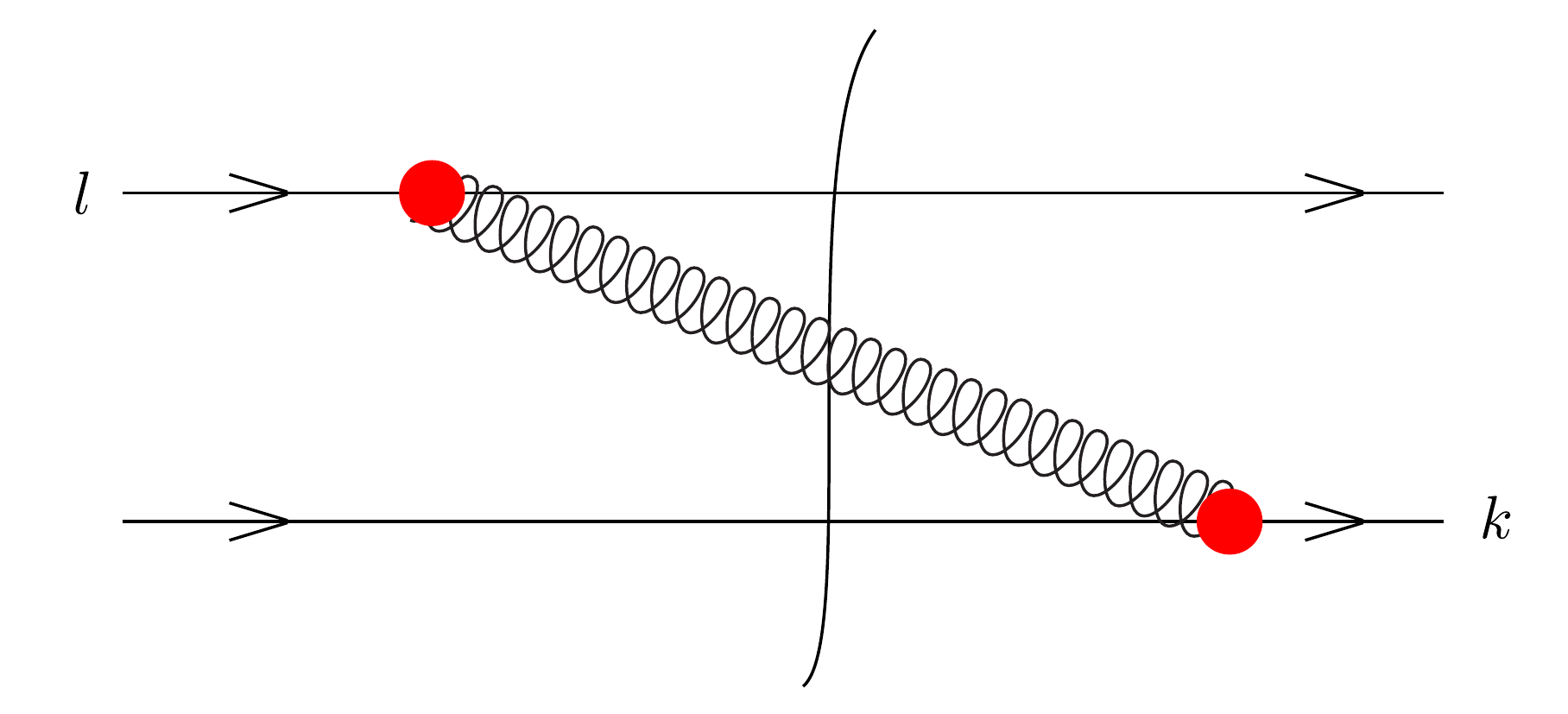}
\end{center}
\caption{
Soft real emission from parton $l$ interfering with emission from parton $k$.
}
\label{fig:interferencesplitting}
\end{figure}

The Sudakov exponent $\cV(t)$ can be decomposed in an analogous manner,
\begin{equation}
\cV(t) = \cV_{\textrm{coll.}}(t) + \cV_{\textrm{soft}}(t) + \cV_{\mi\pi}(t)
\;.
\end{equation}
Here $\cV_{\mi\pi}(t)$ is the contribution from the imaginary part of virtual graphs and is given in Eq.~(\ref{eq:Vipi}). The splitting functions $\cV_{\textrm{coll.}}(t) $ and $\cV_{\textrm{soft}}(t)$ represent splittings that did not happen. They are the same as those in $\cH_I(t)$ except for their color structure and except that we now integrate over $z$ and $\phi$. Thus if we write 
\begin{equation}
\begin{split}
\cV_{\textrm{coll.}}(t) ={}& \cV_{\textrm{coll.}}^\LCP(t) + \Delta\cV_{\textrm{coll.}}(t)
\;,
\\
\cV_{\textrm{soft}}(t) ={}& \cV_{\textrm{soft}}^\LCP(t) 
+ \Delta\cV_{\textrm{soft}}(t)
\;,
\end{split}
\end{equation}
we have
\begin{equation}
\label{eq:DeltaVcoll}
\Delta\cV_{\textrm{coll.}}(t) = 0
\;.
\end{equation}
In this paper, we do not introduce an LC+ approximation for $\cV_{\mi\pi}(t)$, so
\begin{equation}
\cV_{\mi\pi}(t) = \Delta\cV_{\mi\pi}(t)
\;,
\end{equation}
as in Eq.~(\ref{eq:DeltaVreDeltaVipi}).

\begin{figure}
\begin{center}
\includegraphics[width = 8.6 cm]{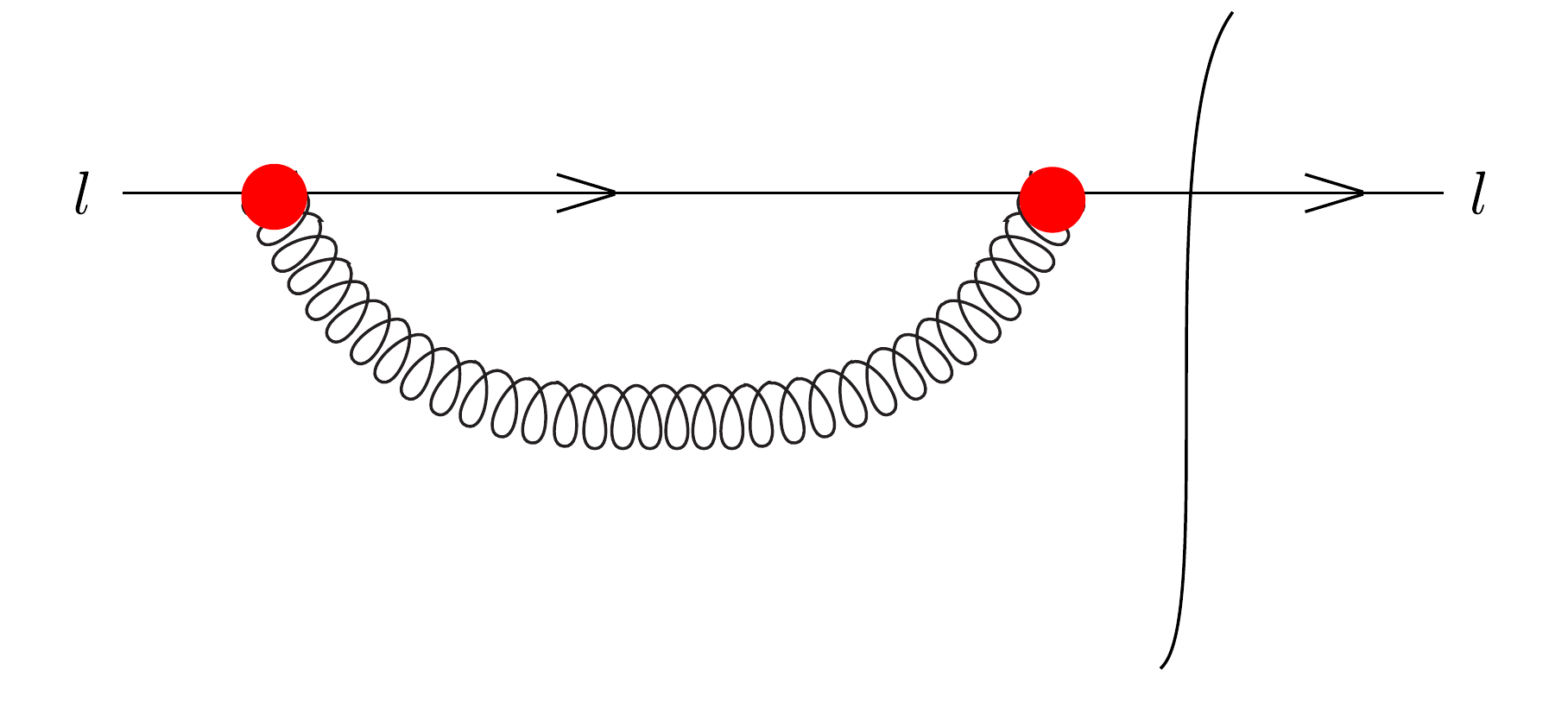}
\end{center}
\caption{
Collinear virtual emission from parton $l$.
}
\label{fig:Vll}
\end{figure}

\begin{figure}
\begin{center}
\includegraphics[width = 8.6 cm]{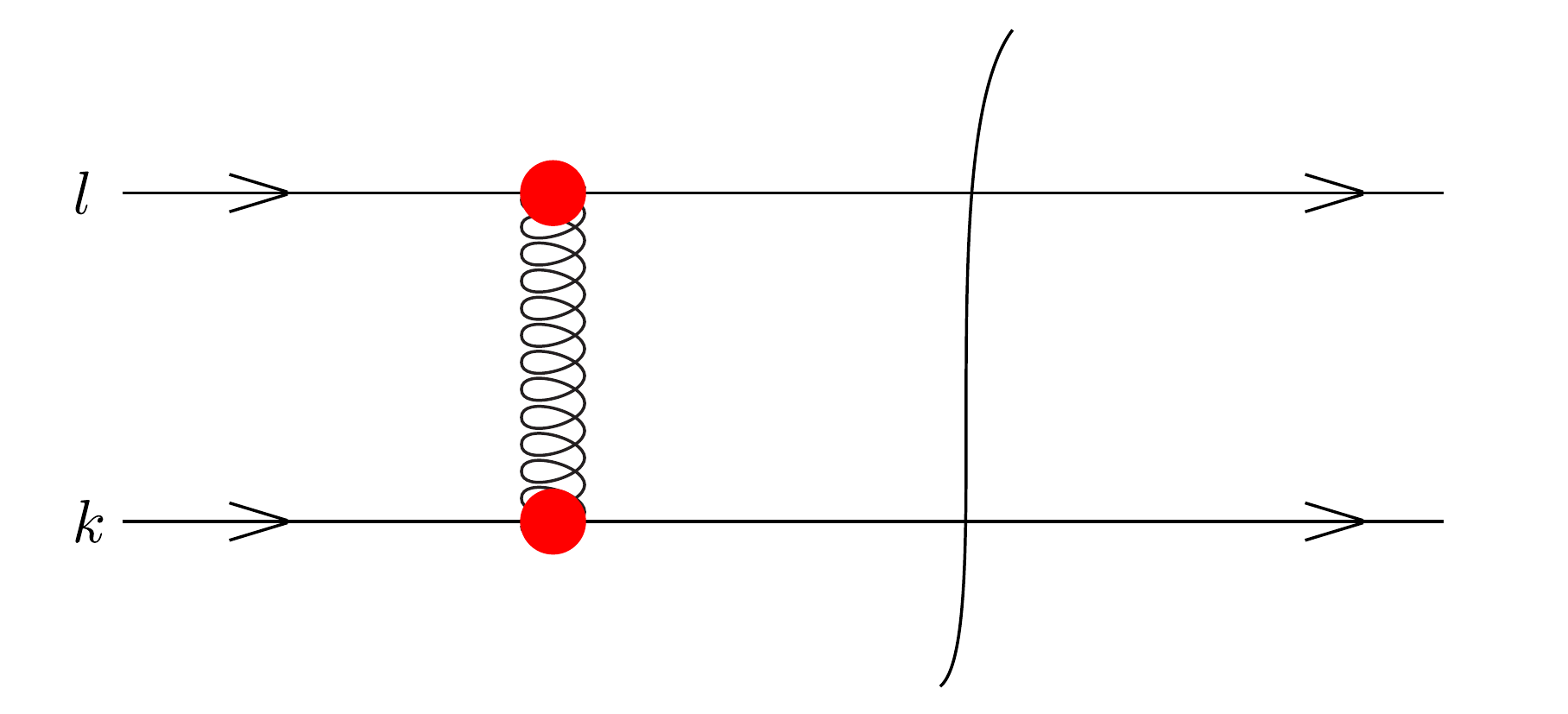}
\end{center}
\caption{
Soft virtual exchange between partons $l$ and $k$.
}
\label{fig:Vlk}
\end{figure}

The corresponding color structures \cite{NScolor} are illustrated in Fig.~\ref{fig:Vll} for $\cV_{\textrm{coll.}}(t)$ and Fig.~\ref{fig:Vlk} for $\cV_{\textrm{soft}}(t)$. The color structure of $\cV_{\textrm{soft}}(t)$ is non-trivial, so that we cannot exponentiate $\cV_{\textrm{soft}}(t)$ in any simple way. Fortunately, the color structure of $\cV_{\textrm{coll.}}(t)$ is very simple: for gluon emission from a quark, it is proportional to a unit operator on the color space times a factor $C_\LF$, for gluon emission from a gluon, it is proportional to $C_\LA$, and for a $\Lg \to q + \bar q$ splitting, it is proportional to $T_\LR = 1/2$ . 

The simplicity of the color structure of $\cV_{\textrm{coll.}}(t)$ has an important consequence: even if the color state to which we apply $\cV_{\textrm{coll.}}(t)$ changes, applying $\cV_{\textrm{coll.}}(t)$ to this state still returns the state times $C_\LF$, $C_\LA$, or $T_\LR$ depending on the flavor of the splitting parton. This means that $\cV_{\textrm{coll.}}^\LCP(t)$ commutes with $\cV_{\textrm{soft}}^\LCP(t)$ and the full $\Delta\cV(t)$. Because $\cV_{\textrm{coll.}}^\LCP(t)$ commutes with $\cV_{\textrm{soft}}^\LCP(t)$, $\cN^\LCP(t_2,t_1)$ in  Eq.~(\ref{eq:NLCplus}) takes the form
\begin{equation}
  \label{eq:NLCplusmod}
  \begin{split}
    \cN^\LCP(t_2,t_1) = {}&\exp\!\left[-\int_{t_1}^{t_2} d\tau\ \cV_{\textrm{coll.}}^\LCP(\tau)\right]
    \\
    &\times\exp\!\left[-\int_{t_1}^{t_2} d\tau\ \cV_{\textrm{soft}}^\LCP(\tau)\right]
    \;.
  \end{split}
\end{equation}
Additionally, in Eq.~(\ref{eq:NDeltaexpansion}) for $\cX(t,t_0)$, all of the factors of exponentials of $\cV_{\textrm{coll.}}^\LCP(\tau)$ commute with the other factors in $\cX(t,t_0)$, giving
\begin{equation}
  \begin{split}
    \cN_\textrm{coll.}^\LCP(t,\tau_n)\,&
    \cN_\textrm{coll.}^\LCP(\tau_n,\tau_{n-1})\cdots
    \\
    &\cdots\cN_\textrm{coll.}^\LCP(\tau_2,\tau_1)\,
    \cN_\textrm{coll.}^\LCP(t,\tau_1)^{-1}
    = 1
    \;.
  \end{split}
\end{equation}
That is, only the soft contribution to $\cV^\LCP(\tau)$ contributes to $\cX(t,t_0)$. Furthermore, according to Eq.~(\ref{eq:DeltaVcoll}), only $\cV_{\mi\pi}(\tau)$ and the soft contribution to $\cV(\tau)$ contributes to $\Delta\cV(\tau)$. Thus everywhere in Eq.~(\ref{eq:NDeltaexpansion}) for $\cX(t,t_0)$, we can replace $\cV(\tau) \to \cV_\textrm{soft}(\tau) + \cV_{\mi\pi}(\tau)$, dropping $\cV_\textrm{coll.}(\tau)$. Thus the operator $\cV_\textrm{coll.}(\tau)$ does not appear at all in $\cX(t,t_0)$. That is,
\begin{align}
\label{eq:NDeltaexpansionmod}
    \cX{}&(t,t_0) = 1
    \notag\\ & 
    - \int_{t_0}^t\!d\tau\ 
    \cN^\LCP_{\textrm{soft}}(t,\tau)\,
    [\Delta \cV_{\textrm{soft}}(\tau) + \cV_{\mi\pi}(\tau)]\,
    \cN^\LCP_{\textrm{soft}}(t,\tau)^{-1}
    \notag\\ & 
    + \int_{t_0}^t\!d\tau_2\ \int_{t_0}^{\tau_2}\!d\tau_1\ 
    \cN^\LCP_{\textrm{soft}}(t,\tau_2)\,
    [\Delta \cV_{\textrm{soft}}(\tau_2) + \cV_{\mi\pi}(\tau_2)]\,
    \notag\\&\quad\times
    \cN^\LCP_{\textrm{soft}}(\tau_2,\tau_1)
    [\Delta \cV_{\textrm{soft}}(\tau_1) + \cV_{\mi\pi}(\tau_1)]\,
    \cN^\LCP_{\textrm{soft}}(t,\tau_1)^{-1}
    \notag\\&
    + \cdots 
    \;.
\end{align}
where
\begin{equation}
\label{eq:NLCplussoft}
\cN^\LCP_{\textrm{soft}}(t_2,t_1) = 
\exp\!\left[
-\int_{t_1}^{t_2} d\tau\ \cV_{\textrm{soft}}^\LCP(\tau)\right]
\;.
\end{equation}
We note that in calculating $\cX(t,t_0)$ according to Eq.~(\ref{eq:NDeltaexpansionmod}), we exponentiate $\cV_\textrm{soft}^\LCP(\tau)$ and expand perturbatively in powers of $\Delta\cV_\textrm{soft}(\tau)$ and $\cV_{\mi\pi}(\tau)$. We also note that the operator $\cV_\textrm{coll.}(\tau)$ carries both collinear and collinear$\times$soft singularities and thus can contribute double large logarithms for each power of $\as$ for some observables. Since this operator does not appear at all in $\cX(t,t_0)$, the factor $\cX(t,t_0)$ in Eq.~(\ref{eq:evolutionU5}) gives only single logarithms for each power of $\as$.
 
\section{Color states}
\label{sec:colorstates}

In the previous versions of \textsc{Deductor}, we always evolved a single basis state and after every step of the shower we always expanded the color state on basis states. The corresponding sums over basis states were then performed by Monte Carlo summation: picking one term at random. This is useful in the implementation since it keeps the code rather simple. However when we consider the effect of the fixed angle soft radiation, we can have serious numerical problems because performing all sums over color basis states by Monte Carlo summation leads to greater fluctuations than one wants. Instead, we can always select a unique basis state in the momentum and flavor space but in the color space we can use a linear combination of the color basis states. For this purpose, we define a statistical state with definite momentum and flavor choice but a more general color state as 
\begin{equation}
\label{eq:phi-state}
\sket{\{p, f, \psi\}_m} = \sket{\{p,f\}_m}\otimes \sum_{\{c',c\}_m} \psi(\{c',c\}_m)\sket{\{c',c\}_m}
\;.
\end{equation}
The state is labeled by the function $\psi$ giving the coefficients of the expansion of the color state in color basis states.


\section{The LC+ no splitting operator}
\label{sec:LCplusN}

From Eq.~(\ref{eq:evolutionU5}), we see that the no-splitting operator in the LC+ approximation plays an important role
in shower evolution. The LC+ approximation is defined in such a way that every basis state $\sket{\{p,f,c',c\}_m}$ is an
eigenstate of $\cN^\LCP(t_2, t_1)$ \cite{NScolor},  
\begin{equation}
  \label{eq:NLCPonbasis}
  \begin{split}
    \cN^\LCP{}&(t_2, t_1) \sket{\{p,f,c',c\}_m}
    \\
    &= \Delta(t_2, t_1, \{p,f,c',c\}_m) \sket{\{p,f,c',c\}_m}
    \;.
  \end{split}
\end{equation}
When the LC+ no-splitting operator acts on a generic state we have 
\begin{align}
  \label{eq:LCplusgeneric}
    &\cN^\LCP(t_2, t_1)\sket{\{p, f, \psi\}_m} 
    = \sket{\{p,f\}_m}
   \\
    &\quad\otimes \sum_{\{c',c\}_m} 
    \Delta(t_2, t_1, \{p,f,c',c\}_m)\,
    \psi(\{c',c\}_m)\sket{\{c',c\}_m}
    .
     \notag
\end{align}
The result is a linear combination of the basis vectors, as in $\sket{\{p, f, \psi\}_m} $, but the terms are weighted by the  corresponding Sudakov factor. In this section, we rewrite the LC+ no-splitting operator as the product of an average Sudakov factor $\Delta(t_2, t_1, \{p,f,\psi\}_m)$ for the generic color state $\psi$ and a weight factor given by an operator $\Phi$. We can use the average Sudakov factor to select splitting variables.

The Sudakov factor $\Delta(t_2, t_1, \{p,f,c',c\}_m)$ is an exponential,\footnote{The function $\lambda(\{p,f,c',c\}_m, \tau)$ here was called $\lambda^\LCP(\{p,f,c\}_m, \tau) + \lambda^\LCP(\{p,f,c'\}_m, \tau)$ in Eq.~(7.4) of Ref.~\cite{NScolor}.}
\begin{equation}
  \label{eq:Deltadef}
  \begin{split}
    \Delta(t_2,{}& t_1, \{p,f,c',c\}_m)
    \\
    &= 
    \exp\!\left(
      -\int_{t_1}^{t_2}d\tau\,\lambda(\{p,f,c',c\}_m, \tau)
    \right)
    \;.
  \end{split}
\end{equation}
The Sudakov exponent has the form\footnote{The function $\lambda_{lk}(\{p,f\}_m, \zeta)\,\chi(k,l,\{c',c\}_m)$ was called $\lambda(\{p,f,c',c\}_m, l,k,\{\hat p,\hat f\}_{m+1},))$  in Eq.~(7.6) of Ref.~\cite{NScolor}, although the $c'$ argument seems to be missing in Ref.~\cite{NScolor}.}
\begin{equation}
  \label{eq:lambdatotaldef}
  \begin{split}
    \lambda({}&\{p,f,c',c\}_m,t)
    \\
    &= 
    \sum_l\int d\zeta\,
    \delta\big(t - T_l(\zeta_\Lp,\{p\}_{m})\big)
    \\
    &\qquad\quad\times
    \sum_{k} \lambda_{lk}(\{p,f\}_m, \zeta)\,\chi(k,l,\{c',c\}_m)
    \;.
  \end{split}
\end{equation}
Recall from Sec.~\ref{sec:splittingvariables} that $\zeta = \{\zeta_\Lp, \zeta_\Lf\}$ is a shorthand notation for the splitting variables in momentum and flavor. Then $d\zeta$ stands for integrating over the momentum splitting variables and summing over the new flavors as in Eqs.~(\ref{eq:dzetaF}) and (\ref{eq:dzetaI}) below. The function $T$ specifies the shower time according to eq.~(\ref{eq:showertime}). The function $\lambda_{lk}(\{p,f\}_m, \zeta)$ is a rather complicated non-negative color-independent function given in Eq.~(\ref{eq:partiallambda}). The function $\chi(k,l,\{c',c\}_m)$ is a simple momentum-independent function
\begin{equation}
\label{eq:chidef1}
\chi(k,l,\{c',c\}_m) = \chi(k,l,\{c\}_{m}) + \chi(k,l,\{c'\}_{m})
\;,
\end{equation}
where 
\begin{equation}
  \label{eq:chidef2}
  \begin{split}
    \chi(k,{}&l,\{c\}_{m})
    \\
    &= 
    \begin{cases}
      1\qquad &\text{if $k=l$}\\
      1\qquad &\text{if $k$ is color connected to $l$ in $\{c\}_{m}$} \\
      0\qquad &\text{otherwise}
    \end{cases}
    \;.
  \end{split}
\end{equation}

We want to define an averaged Sudakov exponent that is simple in color and has all the soft-collinear and collinear
singularities. This can be done many ways but in this paper we define this via an averaged characteristic function, $\xi(k,l,\{\psi\}_m)$. In general it can depend on the $\{\psi\}_m$ state. In order to recover the exact collinear and soft-collinear singularities in the averaged Sudakov exponent we should have  
\begin{equation}
  \begin{split}
    \sum_{k\neq l} \xi(k,l, \{\psi\}_m) = {}& \sum_{k\neq l} \chi(k,l, \{c',c\}_m)
    \\
    ={}& 2+2\theta(f_l = {\rm g})\;.
  \end{split}
\end{equation}
We also require
\begin{equation}
\xi(l,l, \{\psi\}_m) 
= 2\;.
\end{equation}
We give two possible definitions of the averaged characteristic function $\xi(k,l,\{\psi\}_m)$ below. Using whatever definition is chosen, we can define an averaged Sudakov exponent 
\begin{equation}
  \label{eq:lambdapsidef}
  \begin{split}
    \lambda(\{{}&p,f,\psi\}_m, t)
    \\
    &= 
    \sum_l\int d\zeta\,
    \delta\big(t - T_l(\zeta_\Lp,\{p\}_{m})\big)
    \\
    &\qquad\quad\times
    \sum_{k} \lambda_{lk}(\{p,f\}_m, \zeta)\,\xi(k,l,\{\psi\}_m)
  \end{split}
\end{equation}
and the corresponding Sudakov factor
\begin{equation}
  \begin{split}
    \Delta(t_2,{}& t_1, \{p,f,\psi\}_m)
    \\
    &= 
    \exp\!\left(
      -\int_{t_1}^{t_2}d\tau\,\lambda(\{p,f,\psi\}_m, \tau)
    \right)
    \;.
  \end{split}
\end{equation}

Now the no-splitting operator is the product of the new Sudakov factor $\Delta(t_2, t_1, \{p,f,\psi\}_m)$ and an operator $\Phi$ that supplies a correction factor:
\begin{equation}
  \label{eq:NLCPmod}
  \begin{split}
    &\cN^\LCP(t_2, t_1)\sket{\{p, f, \psi\}_m}
    \\
    &\quad
    = \Delta(t_2, t_1, \{p,f,\psi\}_m)\,\Phi(t_2,t_1; \psi)\sket{\{p,f, \psi\}_m}
    \;,
  \end{split}
\end{equation}
where the operator $\Phi(t_2,t_1; \psi)$ is defined as 
\begin{equation}
  \label{eq:phidef}
  \begin{split}
    \Phi(t_2{}&,t_1; \psi)  \sket{\{p,f, c',c\}_m}
    \\&=  
    \frac{\Delta(t_2, t_1, \{p,f,c',c\}_m)}
    {\Delta(t_2, t_1, \{p,f, \psi\}_m)}
    \sket{\{p,f, c',c\}_m}
    \;.
  \end{split}
\end{equation}
This definition gives us
\begin{align}
\label{eq:Phiexponential}
   &\Phi(t_2,t_1; \psi)  \sket{\{p,f, \psi\}_m}
   \notag
   \\&  =
   \sum_{\{c',c\}_m}  \psi(\{c',c\}_m)\sket{\{p,f,c',c\}_m}
   \\&\times
   \exp\!\left(
    \! - \!
    \int_{t_1}^{t_2}\! d\tau
    \left[\lambda(\{p,f,c',c\}_m, \tau)
          -\lambda(\{p,f, \psi\}_m, \tau)\right]
    \right)
    .
    \notag
\end{align} 
Our expectation is that the difference between $\lambda(\{p,f,c',c\}_m, \tau)$ and $\lambda(\{p,f,\psi\}_m, \tau)$ will be reasonably small, so that the weight factor created by $\Phi(t_2,t_1; \psi)$ will be close to 1. The reason for this expectation is that splitting functions in $\lambda(\{p,f,c',c\}_m,\tau)$ are independent of the color state $\{c',c\}_m$ in the limit of collinear and collinear$\times$soft splittings. This means that the difference between $\lambda(\{p,f,c',c\}_m, \tau)$ and $\lambda(\{p,f,\psi\}_m, \tau)$ in Eq.~(\ref{eq:Phiexponential}) is sensitive only to  fixed-angle soft splittings, which we expect are not numerically very important.

In designing \textsc{Deductor} 3.0.0, we considered and implemented two versions of the averaged characteristic function. Define a function $p(\psi, \{c',c\}_m)$ that assigns probabilities to the basis states $|\{c',c\}_m)$. The probabilities are positive and normalized to
\begin{equation}
  \sum_{\{c',c\}_m} p(\psi, \{c',c\}_m) = 1\;.
\end{equation}
The choice of the probabilities is largely arbitrary and the physical quantities are independent of them. One choice tries to emphasize the importance of the basis state,
\begin{equation}
  p(\psi, \{c',c\}_m) = \frac{|\psi(\{c',c\}_m)|\, N_c^{-I_0(\{c',c\}_m)}}
  {\displaystyle \sum_{\{\tilde c', \tilde c\}_m} |\psi(\{c',c\}_m)|\, N_c^{-I_0(\{\tilde c', \tilde c\}_m)}}
 \;.
\end{equation}
Here $I_0(\{c',c\}_m)$ is defined using the color overlap of color basis states using the $U(N_\Lc)$ group instead of $SU(N_\Lc)$:
\begin{equation}
\label{eq:I0def}
\brax{\{\tilde c'\}_m}\ket{\{\tilde c\}_m}_{U(N_\Lc)}
= \frac{\textrm{const.}}{N_\Lc^{I_0}}
\left\{1 + \cO\!\left(\frac{1}{N_\Lc}\right)\right\}
\;.
\end{equation}
This choice of the probability function emphasizes the color basis states in $\psi$ that don't have big $I_0$.

Now, with these probabilities, the first averaged characteristic function is defined as
\begin{equation}
  \xi_1(k,l,\{\psi\}_m) = \sum_{\{c',c\}_m} p(\psi, \{c',c\}_m) \chi(k,l,\{c',c\}_m)\;.
\end{equation}
Note that when $k=l$ then $\xi_1(k,l,\{\psi\}_m) = 2$.

The other choice is much simpler. It doesn't depend on the color state at all but depends only on the flavor and number of the partons. We define
\begin{equation}
  \xi_2(k,l,\{\psi\}_m) = 2\theta(k=l) + 2\theta(k\neq l) \frac{1+\theta(f_l = \mathrm{g})}{m+1}\;\;.
\end{equation}

Both choices of the characteristic function have some advantages and disadvantages. We will discuss them in the next section.

 
\section{Evolution}
\label{sec:fancycolorevolution}

We are now in a position to rewrite the evolution equation Eq.~(\ref{eq:evolutionU5}) in a computationally useful
form. We start by simply applying Eq.~(\ref{eq:evolutionU5}) to a state $\sket{\{p,f,\psi\}_m}$:
\begin{equation}
  \begin{split}
    \label{eq:evolutionpsi1}
    \cU{}&(t,t_0)\sket{\{p,f,\psi\}_m} =
    \cX(t,t_0)\, \cN^\LCP(t,t_0)\sket{\{p,f,\psi\}_m}
    \\&
    + \int_{t_0}^t\!d\tau\ 
    \cU(t,\tau)\,
    \cH_I(\tau)
    \cX(\tau,t_0)\, \cN^\LCP(\tau,t_0) 
    \sket{\{p,f,\psi\}_m}
    \;.
  \end{split}
\end{equation}

Next, we use Eq.~(\ref{eq:NLCPmod}) for $\cN^\LCP$ and, inside the integral, we multiply and divide by the normalization
factor $\lambda(\{p,f,\psi\}_m, \tau)$. This gives us
\begin{widetext}
\begin{equation}
  \label{eq:evolutionpsi}
  \begin{split}
    \cU(t, t_0)\sket{\{p,f,\psi\}_m}
    ={}&
    \Delta(t, t_0, \{p,f,\psi\}_m)\, 
    \cX(t, t_0)\,\Phi(t,t_0;\psi)\sket{\{p,f, \psi\}_m}
    \\ & 
    + \int_{t_0}^{t}d\tau\, 
    \lambda(\{p,f,\psi\}_m, \tau)\, \Delta(\tau, t_0, \{p,f,\psi\}_m)
    \\
    &\qquad\quad\times
    \cU(t,\tau)
    \frac{\cH_I(\tau)}{\lambda(\{p,f,\psi\}_m, \tau)}\,
    \cX(\tau, t_0)\,\Phi(\tau,t_0;\psi)\sket{\{p,f, \psi\}_m}
    \;.
  \end{split}
\end{equation}
In order to use this, we need to define the color coefficients of $\cX\Phi\sket{\psi}$:
\begin{equation}
  \begin{split}
    \cX{}&(\tau, t_0)\,\Phi(\tau,t_0;\psi)
    \sket{\{p,f,\psi\}_m}
    = \sum_{\{c',c\}_m}
    X(\tau,t_0,\{p,f,c',c\}_m,\psi)
    \sket{\{p,f,c',c\}_m}
    \;.
  \end{split}
\end{equation}

Then
\begin{equation}
  \begin{split}
    \label{eq:evolutionpsi2}
    \cU(t, t_0)\sket{\{p,f,\psi\}_m}
    = {}& \Delta(t, t_0, \{p,f,\psi\}_m)
    \sum_{\{c',c\}_m}
    \sket{\{p,f,c',c\}_m}
    X(t,t_0,\{p,f,c',c\}_m,\psi)
    \\ & 
    + \int_{t_0}^{t}d\tau\, 
    \lambda(\{p,f\}_m, \tau)\, 
    \Delta(\tau, t_0, \{p,f,\psi\}_m)  
    \\&\qquad\times
    \cU(t,\tau)\sum_{\{c',c\}_m}
    \frac{\cH_I(\tau)\sket{\{p,f,c',c\}_m}}
    {\lambda(\{p,f,\psi\}_m, \tau)}\,
    X(\tau,t_0,\{p,f,c',c\}_m,\psi)
    \;.
  \end{split}
\end{equation}

Now we examine what the splitting operator $\cH_I(\tau) = \cH^\LCP(\tau) + \Delta \cH(\tau)$ does when applied to an arbitrary basis state, 
  \begin{equation}
    \begin{split}
      \cH_I(t)\sket{\{p,f,c',c\}_{m}}
      = {}&
      \sum_{l} 
      \int\!d\zeta\
      \delta(t - T_l(\zeta_\Lp,\{p\}_{m}))\,
      \sum_{\{\hat c',\hat c\}_{m+1}} \sket{\{\hat p,\hat f,\hat c',\hat c\}_{m+1}}
      \\&\qquad\times
      \Big[
      \lambda_{ll}(\{p,f\}_{m},\zeta)\, 
      G(l,l,\zeta_\Lf, \{\hat c',\hat c\}_{m+1},\{c',c\}_m)
      \\&  
      \qquad\qquad+ 
      \sum_{k'\neq l}\lambda_{lk'}(\{p,f\}_{m},\zeta)\, 
      G(k',l,\zeta_\Lf, \{\hat c',\hat c\}_{m+1},\{c',c\}_m)
      \Big]
      \;.
    \end{split}
  \end{equation}
Here $\zeta = \{\zeta_\Lp,\zeta_\Lf\}$ stands for the splitting variables and the function $T_l(\zeta_\Lp,\{p\}_{m})$ defines the shower time $t$ as described in Sec.~\ref{sec:splittingvariables}. Then the function $\lambda_{lk}(\{p,f\}_{m},\zeta)$ is a rather complicated function that defines the splitting functions in $\cH_I$. See Eqs.~(\ref{eq:partiallambda}) and (\ref{eq:HIdef4}). The first term under the square bracket represents the contributions of pure collinear radiation while the second term represents the contributions of the collinear$\times$soft and pure soft emissions.  The function $G$, defined in Eq.~(\ref{eq:Gdef}), contains the color structure. It contains two parts, corresponding to the division of $\cH_I(t)$ into $\cH^\LCP(\tau)$ and $\Delta \cH(\tau)$:
\begin{equation}
\begin{split}
G(k,l,\zeta_\Lf, \{\hat c',\hat c\}_{m+1},\{c',c\}_m)
={}&
G^\LCP(k,l,\zeta_\Lf, \{\hat c',\hat c\}_{m+1},\{c',c\}_m)
+ \Delta G(k,l,\zeta_\Lf, \{\hat c',\hat c\}_{m+1},\{c',c\}_m)
\;.
\end{split}
\end{equation}
When $k = l$, only the LC+ term is present:
\begin{equation}
\begin{split}
G(l,l,\zeta_\Lf, \{\hat c',\hat c\}_{m+1},\{c',c\}_m)
={}&
G^\LCP(l,l,\zeta_\Lf, \{\hat c',\hat c\}_{m+1},\{c',c\}_m)
\;.
\end{split}
\end{equation}

Now we factor  $\sum_{k \ne l}\lambda_{lk}(\{p,f\}_{m},\zeta)\, \xi(k,l,\{\psi\}_{m})$
from the second term. Then we have 
\begin{equation}
  \begin{split}
    \label{eq:HILCPdef3}
    \cH_I(t)\big|\{p,f,c',c\}_{m}\big)
    ={}&
    \sum_{l} 
    \int\!d\zeta\
    \delta(t - T_l(\zeta_\Lp,\{p\}_{m}))\,
    \sket{\{\hat p,\hat f\}_{m+1}}\,\otimes
    \sum_{\{\hat c',\hat c\}_{m+1}} \sket{\{\hat c',\hat c\}_{m+1}}
    \\&\qquad\times
    \sum_{k}\lambda_{lk}(\{p,f\}_{m},\zeta)\, \xi(k,l,\{\psi\}_{m})\,
    \\&\qquad\times
    \Bigg[
    \theta(k=l)\frac{G(l,l,\zeta_\Lf, \{\hat c',\hat c\}_{m+1},\{c',c\}_m)}
    {\xi(l,l,\{\psi\}_{m})}
    \\&\qquad\qquad 
    + 
    \theta(k\ne l)\,
    \frac{ \sum_{k'\neq l}\lambda_{lk'}(\{p,f\}_{m},\zeta)\,  G(k',l,\zeta_\Lf, \{\hat c',\hat c\}_{m+1},\{c',c\}_m)}
    { \sum_{\tilde k \neq l}\lambda_{l\tilde{k}}(\{p,f\}_{m},\zeta)\,\xi(\tilde{k},l,\{\psi\}_{m})}
    \Bigg]\;.
  \end{split}
\end{equation}
Inserting this into Eq.~\eqref{eq:evolutionpsi2} and using $\xi(l,l,\{\psi\}_m) = 2$, we obtain
\begin{equation}
  \begin{split}
    \label{eq:evolutionpsi3}
    \cU(t, t_0)\sket{\{p,f,\psi\}_m}
    = {}& \Delta(t, t_0, \{p,f,\psi\}_m) 
    \sum_{\{c',c\}_m}
    \sket{\{p,f,c',c\}_m}\,
    X(t,t_0,\{p,f,c',c\}_m,\psi)
    \\ & 
    + \int_{t_0}^{t}d\tau\, 
    \lambda(\{p,f\}_m, \tau)\, 
    \Delta(\tau, t_0, \{p,f,\psi\}_m)
    \\&\qquad\times 
    \sum_{l,k} 
    \int\!d\zeta\
    \delta(\tau - T_l(\zeta_\Lp,\{p\}_{m}))\,
    \frac{\lambda_{lk}(\{p,f\}_{m},\zeta)\,\xi(k,l,\{\psi\}_m)}{\lambda(\{p,f,\psi\}_m, \tau)}
    \\&\qquad\times
    \sum_{\{\hat c',\hat c\}_{m+1}} 
    \cU(t,\tau)\sket{\{\hat p,\hat f,\hat c',\hat c\}_{m+1}}
    \sum_{\{c',c\}_m} X(\tau,t_0,\{p,f,c',c\}_m,\psi)
    \\
    &\qquad\times
    \Bigg[\frac{1}{2}\,
    \theta(k = l)\,G(l,l,\zeta_\Lf, \{\hat c',\hat c\}_{m+1},\{c',c\}_m)
    \\&  \qquad\qquad
    + 
    \theta(k \ne l)\,
    \frac{ \sum_{k'\neq l}\lambda_{lk'}(\{p,f\}_{m},\zeta)\, G(k',l,\zeta_\Lf, \{\hat c',\hat c\}_{m+1},\{c',c\}_m)}
    {\sum_{\tilde k \neq l}\lambda_{l\tilde{k}}(\{p,f\}_{m},\zeta)\,\xi(\tilde{k},l,\{\psi\}_{m})}
    \Bigg]
    \;.
  \end{split}
\end{equation}
\end{widetext}

The color matrix $X$ transforms the original color state $\psi$ for $m$ partons to a linear combination of color basis states $\{c',c\}_m$ under the color transformation provided by the operators $\cX$ and $\Phi$. Then the expression under the square brackets is the color matrix that transforms the color state $\{c',c\}_m$ to $\{\hat c', \hat c\}_{m+1}$. This matrix depends on the hard state $\{p,f\}_m$, the splitting variables $\zeta$, and the parton labels $l,k$.

We now have a form that is useful for calculations. The factors $\lambda \Delta\,d\tau$ in the integral over $\tau$ say that we should pick a next shower time $\tau$ with a probability 
\begin{equation}
\begin{split}
\label{eq:dPpsi}
dP_\tau ={}& \lambda(\{p,f\}_m, \tau)\, \Delta(\tau, t_0, \{p,f,\psi\}_m)\, d\tau
\\ ={}& - \frac{d}{d\tau}\Delta(\tau, t_0, \{p,f,\psi\}_m)\ d\tau
\\ ={}& - d \Delta(\tau, t_0, \{p,f,\psi\}_m)
\end{split}
\end{equation}
determined by the total color state $\psi$. This is different from the probability that we would have with a single
color basis state $\{c',c\}_m$. However, we account for the probabilities for each basis state in the factor
$\Delta(\tau, t_0, \{p,f,c',c\}_m)/\Delta(\tau, t_0, \{p,f,\psi\}_m)$ in $\Phi$, Eq.~(\ref{eq:phidef}), and in the
perturbative factor $\cX$. Additionally, in Eq.~(\ref{eq:evolutionpsi}) we have multiplied by $\lambda(\{p,f, \psi\}_m,
\tau)$, but we have divided by this same factor. These factors remain in Eq.~(\ref{eq:evolutionpsi3}).

Having fixed $\tau$, we also need to pick $l$, $k$, and the splitting variables $\zeta$. We pick $l$, $k$, and $\zeta$ with probability
\begin{equation}
  \begin{split}
    dP_{l,k,\zeta} ={}& d\zeta\,
    \delta(t - T(\zeta,\{p\}_{m}))\,
    \\
    &\times
    \frac{\lambda_{lk}(\{p,f\}_{m},\zeta)\,\xi(k,l,\{\psi\}_m)}
    {\lambda(\{p,f,\psi\}_{m},t)}\,
    \;.
  \end{split}
\end{equation}
According to Eq.~(\ref{eq:lambdapsidef}), this gives us a properly normalized probability, $\sum_{lk}\int dP_{l,k,\zeta} = 1$.

We are left with a sum over colors $\{\hat c',\hat c\}_{m+1}$ for the $m+1$ partons after the splitting. To perform this sum, we must chart a sensible course \cite{Homer}. At one extreme is Scylla: we could perform this sum Monte Carlo style, picking one color state $\{\hat c',\hat c\}_{m+1}$ at random and accumulating a weight factor given by the remaining factors in Eq.~(\ref{eq:evolutionpsi3}). This leads to large fluctuations that render the calculation useless. At the other extreme waits   Charybdis: we could accumulate all of the color states $\{\hat c',\hat c\}_{m+1}$ into a new combined color state $\psi$. This leads to color states $\psi$ that contain so many basis states $\{\hat c',\hat c\}_{m+1}$ that the calculation crashes. We choose a middle course. The algorithm can then be adjusted to optimize performance. Its details are not important, at least conceptually.

The algorithm described above is implemented in \textsc{Deductor} with both $\xi_1(k,l,\{\psi\}_m)$ and $\xi_2(k,l,\{\psi\}_m)$ averaged characteristic functions.


\section{Approximations}
\label{sec:approximations}

We now need some approximations in order to have an algorithm for color evolution that can be executed on a finite size computer in a finite amount of time. The key requirement is that one should be able to control the level of approximation, so that one can obtain more nearly exact results if one has greater computer resources available.

To start, we put a limit on the color suppression index $I$ in the shower evolution operator $\cU(t,t_0)$. As defined in Ref.~\cite{NScolor}, the color suppression index is obtained from two factors. The first is the number $p_\LE$ of explicit powers of $1/N_\Lc$ that arise from choosing the color suppressed term in the Fierz identity for $g \to q + \bar q$ splittings that have occurred in the shower history so far. The second is the number of powers of $1/N_\Lc$ in the overlap of the bra and ket color states using $\LU(N_\Lc)$ color:
\begin{equation}
\label{eq:Idef}
\frac{1}{N_\Lc^{p_\LE}}\brax{\{\tilde c'\}_m}\ket{\{\tilde c\}_m}_{U(N_\Lc)}
= \frac{\textrm{const.}}{N_\Lc^{I}}
\left\{1 + \cO\!\left(\frac{1}{N_\Lc}\right)\right\}
\;.
\end{equation}
The cross section at the end of the shower will be suppressed by a color factor that is at least as small as $1/N_\Lc^I$ \cite{NScolor}, so there is little point in keeping contributions with large $I$. Thus we use the full $\LS\LU(N_\Lc)$ evolution for color states for which $I - I_\textrm{hard} \le I_\textrm{max}$, where $I_\textrm{hard}$ is the color suppression index of the hard scattering state at the start of the shower and $I_\textrm{max}$ is a parameter that we choose.

A sensible choice for $I_\textrm{max}$ is $I_\textrm{max} = 4$. We can improve the approximation that results from limiting $I$ by increasing $I_\textrm{max}$, but there is a cost. Increasing $I_\textrm{max}$ increases statistical fluctuations in the results after a fixed amount of computer time, so that we need more computer time to achieve the same statistical accuracy.

If the shower operator reaches $I - I_\textrm{hard} \ge I_\textrm{max}$, it switches to an approximate shower based on the color group $\LU(N_\Lc)$ instead of $\LS\LU(N_\Lc)$. We also omit any further contributions from $\Delta \cH$ and $\Delta \cV$. Thus contributions proportional to $1/N_\Lc^I$ are calculated only approximately.

Next, we put a limit on the number of times, $N_\Delta^\textrm{thr.}$, that the operator $\cV_{\mi \pi}$ defined in Eqs.~(\ref{eq:DeltaHDeltaV}) and (\ref{eq:DeltaVreDeltaVipi}) appears in the threshold operator $\cU_\cV$. This operator gives results as an expansion in powers of $\cV_{\mi \pi}$ and we retain only those terms with no more than $N_\Delta^\textrm{thr.}$ factors of $\cV_{\mi \pi}$.

Then, we put limits on the number of times that the operators $\Delta \cH$, $\Delta \cV_\textrm{Re}$ and $\cV_{\mi\pi}$ appear in the shower evolution operator $\cU(t_2,t_1)$. As described in Sec.~\ref{sec:fancycolorevolution}, $\cU(t_2,t_1)$ produces terms proportional to $[\Delta \cH]^A [\Delta \cV_\textrm{Re}]^B [\cV_{\mi \pi}]^C$. We choose parameters $N_\Delta$, $N_\textrm{Re}$ and $N_{\mi\pi}$. We retain only terms with $A+B \le N_\textrm{Re}$ and $C \le N_{\mi\pi}$ and $A + B + C \le N_\Delta$.

Increasing $N_\textrm{Re}$, $N_{\mi\pi}$, $N_\Delta$, or $N_\Delta^\textrm{thr.}$ gives more accurate results in the limit of long computer running times and large computer memory. However, it could use more memory than is available and it increases statistical fluctuations in the results after a fixed amount of computer time, so that we need more computer time to achieve the same statistical accuracy.

We can prescribe different accuracy parameters $N_\textrm{Re}$, $N_{\mi \pi}$, $N_\Delta$ and $I_\textrm{max}$ to successive shower time intervals. One would do this if one suspected that, for the observable being measured, the first splitting steps of the shower are more important than later steps, with softer splittings. In the simplest application, we assign $N_\textrm{Re}$, $N_{\mi \pi}$, $N_\Delta$, and $I_\textrm{max}$ to the interval $t^{(0)} < t < t^{(1)}$. Here $t^{(0)}$ is the starting time of the shower as determined by the hard interaction that initiates the shower and $t^{(1)} > t^{(0)}$, perhaps $t^{(1)} = t^{(0)} + 5$. Then we can use an LC+ shower ($N_\Delta = 0$) with the same $I_\textrm{max}$ until the shower ends. The shower ends because the splitting kernel has a cutoff built into it that stops splittings at a lower cutoff for the transverse momentum in a splitting. We choose this cutoff to be $k_\LT^{\rm min} = 5 \GeV$.

In a more elaborate calculation, one might have shower time intervals $t^{(0)} < t < t^{(1)}$, $t^{(1)} < t < t^{(2)}$,\dots. Then we would specify $N_\textrm{Re}$, $N_{\mi \pi}$, $N_\Delta$, and $I_\textrm{max}$ for each interval.

The \textsc{Deductor} 3.0.0 user interface also allows one to specify accuracy parameters for successive shower intervals determined by a fixed number of splitting steps.

\section{Probability conservation}
\label{sec:unitarity}

Version 3.0.0 of \textsc{Deductor} implements the algorithm outlined above. It is capable of producing cross sections, as we will see below in Sec.~\ref{sec:jets}. However, it is not so easy to check whether it is producing correct cross sections including non-leading color effects, as intended. 

In this section, we present a test of the inner workings of the program by testing the probability conservation property (\ref{eq:unitarity}),
\begin{equation}
\label{eq:unitaritybis}
\sbra{1} \cU(t,t_0)\sket{\rho(t_0)} = \sbrax{1}\sket{\rho(t_0)}
\;.
\end{equation}
Here $\sket{\rho(t_0)}$ is an arbitrary statistical state at shower time $t_0$. Then  $\sbrax{1}\sket{\rho(t_0)}$ is the total cross section measured for that state, obtained by integrating the differential cross section over all parton variables and taking the trace of the color density matrix. After generating a shower with $\cU(t,t_0)$, we have a much more complicated statistical state with typically many more partons. We then measure the total cross section associated with this state. The total cross section should be the same.

Note that it is a simple consequence of the shower evolution equation (\ref{eq:evolutionU2}) that probability conservation, Eq.~(\ref{eq:unitaritybis}), holds in the LC+ approximation and also with the inclusion of contributions from the operators $\Delta\cH$, $\Delta\cV_\textrm{Re}$, and $\cV_{\mi \pi}$. The evolution equation (\ref{eq:evolutionU2}) leads us to the evolution equation (\ref{eq:evolutionU5}). In this form of the evolution equation, it is no longer self-evident that probability conservation works, since $\Delta\cH$ and $\Delta\cV$ are treated very differently now. With further manipulations, we arrive at the algorithm implemented in \textsc{Deductor}, Eq.~(\ref{eq:evolutionpsi3}). Now, probability conservation must still work but it is not a property that one would guess from examining Eq.~(\ref{eq:evolutionpsi3}) without knowledge of its derivation. Thus it can be a powerful test of Eq.~(\ref{eq:evolutionpsi3}) and its implementation in code to see whether probability conservation (\ref{eq:unitaritybis}) works within the statistical accuracy of the calculation. 

We will check whether 
\begin{equation}
\label{eq:unitaritytest}
\frac{\sbra{1} \cU(t,t_0)\sket{\rho(t_0)}}{\sbrax{1}\sket{\rho(t_0)}}
=1
\end{equation}
for each value of $t_0$. We will expand the left-hand side of (\ref{eq:unitaritytest}) in powers of $\Delta \cH$, $\Delta \cV_\textrm{Re}$, and $\cV_{\mi \pi}$ and examine the terms proportional to $[\Delta \cH]^A [\Delta \cV_\textrm{Re}]^B [\cV_{\mi \pi}]^C$. The term with $A = B = C = 0$ gives us the LC+ approximation to $\cU(t,t_0)$, which obeys Eq.~(\ref{eq:unitaritytest}). Thus the term with $A = B = C = 0$ gives us the 1 on the right hand side of Eq.~(\ref{eq:unitaritytest}). Consequently, the other times must combine to give zero. Furthermore, if we were to replace $\Delta \cH - \Delta \cV_\textrm{Re}$ by $\lambda_\textrm{Re}(\Delta \cH - \Delta \cV_\textrm{Re})$ and $\cV_{\mi \pi}$ by $\lambda_{\mi \pi}\cV_{\mi \pi}$, the relations  $\sbra{1} (\Delta \cH - \Delta \cV_\textrm{Re}) = 0$ and $\sbra{1} \cV_{\mi \pi} = 0$ tell us that Eq.~(\ref{eq:unitaritytest}) holds order by order in $\lambda_\textrm{Re}$ and $\lambda_{\mi \pi}$. Thus all of the contributions with fixed values of $A+B$ and $C$, other than $A = B = C = 0$, must sum to zero.

In this test, the shower is limited to the shower time interval $t_0 < t < t_0 + 5$.  We choose the maximum color suppression index to be $I_\textrm{max} = 4$. We also limit the number of $\Delta \cH$ and $\Delta \cV$ operators as specified below.

In our test, $\sket{\rho(t_0)}$ is the state produced by a $2 \to 2$ hard scattering. The two final state partons have absolute value of transverse momenta $p_{1,\LT} = p_{2,\LT} = p_\LT$ and c.m. energy squared $(p_1 + p_2)^2 = \hat s$. With our shower time definition and the choice of the starting time for the shower from Ref.~\cite{NSThresholdII}, the starting shower time is $t_0 = \log(4\hat s/(9 p_\LT^2))$. Thus $t_0 = \log((8/9)(1 + \cosh(y_1 - y_2))$. We generate hard scatterings with a wide range of transverse momenta, from $P_\LT = 20 \GeV$ to $P_\LT = 5 \TeV$. The smallest possible value for $t_0$ is about 0.6 and, with small values of $P_\LT$, $t_0$ can range up to more than 10. We generate a range of $t_0$ values and plot results in bins of $t_0$, so that $t_0$ has a definite value in Eq.~(\ref{eq:unitaritytest}).

By default, \textsc{Deductor} has an operator $\cU_\cV$ that comes between the hard state $\sket{\rho(t_0)}$ and the probability preserving shower operator $\cU(t,t_0)$. This operator sums threshold logarithms and thereby changes the cross section. For this test, we turn $\cU_\cV$ off.

We wish to test Eq.~(\ref{eq:unitaritytest}) in the presence of approximations with respect to color. However, we caution that exact agreement with Eq.~(\ref{eq:unitaritytest}) is not to be expected, since there are small systematic errors within the \textsc{Deductor} calculation that come from sources other than limits on the color treatment. For instance, there are inevitably approximations in our use of the parton distribution functions. Additionally, there are some functions defined as integrals that are too complicated to be performed analytically; for these, we use gaussian numerical integration. We believe that these systematic errors are smaller than 1\%, but we have not systematically checked their size.

We will test Eq.~(\ref{eq:unitaritytest}) in three ways. First, we turn off $\cV_{\mi \pi}$ and use only $\Delta\cH$ and $\Delta\cV_\textrm{Re}$. Then we turn off $\Delta\cH$ and $\Delta\cV_\textrm{Re}$ and use only $\cV_{\mi \pi}$. Finally, we use $\Delta\cH$ and $\Delta\cV_\textrm{Re}$ and $\cV_{\mi \pi}$ together.

We now try the first test in which we turn off $\cV_{\mi \pi}$ and consider up to four insertions of $\Delta\cH$ and $\Delta\cV_\textrm{Re}$. That is, we set $N_\textrm{Re} = 4$, $N_{\mi \pi} = 0$, and $N_\Delta = 4$.

\begin{figure}
\begin{center}
\ifusefigs 

\begin{tikzpicture}
  \begin{axis}[title = {LC+ contribution},
    xlabel={$t_0$}, ylabel={$\sigma(\text{showered})/\sigma(\text{Born})$},
    legend cell align=left,
    every axis legend/.append style = {
      anchor=north west
    },
    ]
    
    \ratioplot[red,semithick]{fill=red!30!white, opacity=0.5}
    {data/UHV0.dat}{1}{5}{6}{3}{4}

\end{axis}
\end{tikzpicture}

\fi

\end{center}
\caption{
LC+ contribution to the left-hand side of Eq.~(\ref{eq:unitaritytest}).
}
\label{fig:ULCp}
\end{figure}
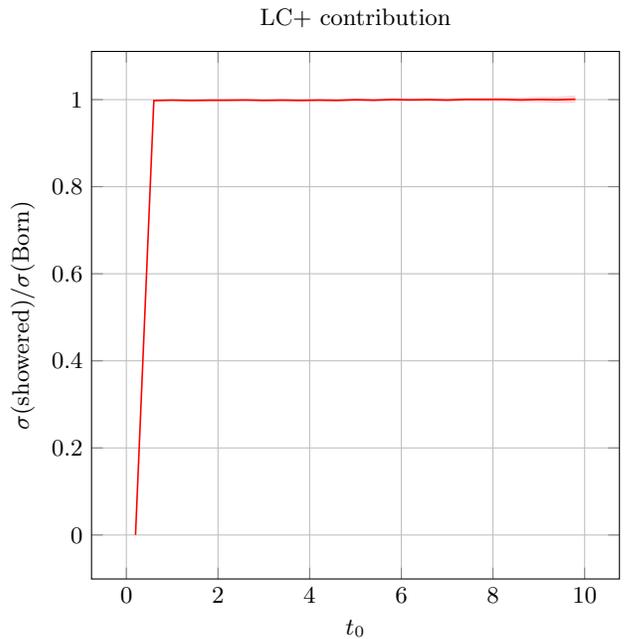

In Fig.~\ref{fig:ULCp}, we show the contributions to the left hand side of Eq.~(\ref{eq:unitaritytest}) from contributions proportional to $[\Delta \cH]^A [\Delta \cV_\textrm{Re}]^B$ with $A = B = 0$ as functions of $t_0$. With $A = B = 0$, we are looking at probability conservation from the LC+ approximation. We see that this contribution is 1 to within small statistical fluctuations. 

Here and in later graphs, we plot error bands that represent the estimated statistical error in the Monte Carlo data. We can also assess the statistical errors by looking at the fluctuations from one bin in the plot to the next. In Fig.~\ref{fig:ULCp}, the statistical errors are hardly visible, but they are more visible in later plots.

\begin{figure}
\begin{center}
\ifusefigs 

\begin{tikzpicture}
  \begin{axis}[title = {$A+B=1$ contributions},
    xlabel={$t_0$}, ylabel={$10^2\cdot\sigma(\text{showered})/\sigma(\text{Born})$},
    legend cell align=left,
    every axis legend/.append style = {at={(axis cs:5,0.039)},
    anchor=north},
    legend columns=3,
    xmin=0, xmax=10,
    ymin=-0.04, ymax=0.04,
    y coord trafo/.code={
      \pgflibraryfpuifactive{
      \pgfmathparse{(#1)*(100)}
      }{
      \pgfkeys{/pgf/fpu=true}
      \pgfmathparse{(#1)*(100)}
      \pgfkeys{/pgf/fpu=false}
      }
    }    
    ]
    
    \ratioplot[red,semithick]{fill=red!30!white, opacity=0.5}
    {data/UHV1.dat}{1}{5}{6}{3}{4}
    \addlegendentry{$\Delta\cH$}

    \ratioplot[blue,semithick]{fill=blue!30!white, opacity=0.5}
    {data/UHV1.dat}{1}{7}{8}{3}{4}
    \addlegendentry{$\Delta\cV_\textrm{Re}$}

    \ratioplot[black,semithick]{fill=gray!30!white, opacity=0.5}
    {data/UHV0.dat}{1}{7}{8}{3}{4}
    \addlegendentry{Total}
        
  \end{axis}
\end{tikzpicture}

\fi

\end{center}
\caption{
$A + B = 1$, $C = 0$ contributions to the left-hand side of Eq.~(\ref{eq:unitaritytest}) times $10^2$.
}
\label{fig:UDelta1}
\end{figure}
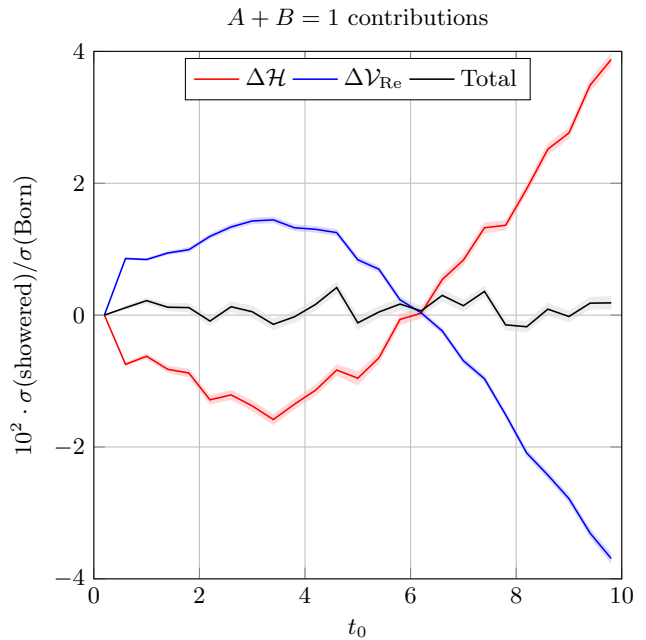

In Fig.~\ref{fig:UDelta1}, we look at the contribution to the left hand side of Eq.~(\ref{eq:unitaritytest}) from contributions with $A + B = 1$. We plot separately the contributions from $\Delta\cH$ and $\Delta\cV_\textrm{Re}$ along with their total. We see that these contributions are typically between $\pm 1\%$ and $\pm 4\%$. However, they cancel, giving a total that is smaller than about $\pm 0.2\%$. It is remarkable to us that these contributions cancel to the extent seen in the figure since the methods of calculation for $\Delta \cH$ and $\Delta \cV_\textrm{Re}$ are very different.

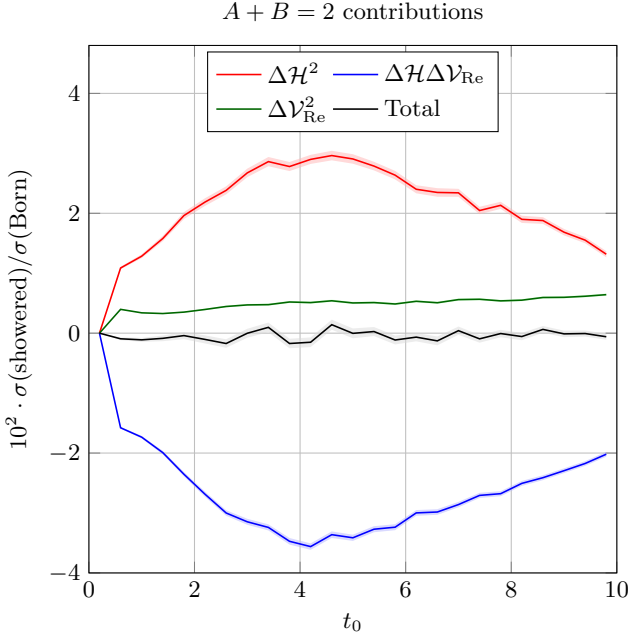
\begin{figure}
\begin{center}
\ifusefigs 

\begin{tikzpicture}
  \begin{axis}[title = {$A+B=2$ contributions},
    xlabel={$t_0$}, 
    ylabel={$10^2\cdot\sigma(\text{showered})/\sigma(\text{Born})$},
    legend cell align=left,
    every axis legend/.append style = {at={(axis cs:5,0.047)},
    anchor=north},
    legend columns=2,
    xmin=0, xmax=10,
    ymin=-0.04, ymax=0.048,
    y coord trafo/.code={
      \pgflibraryfpuifactive{
      \pgfmathparse{(#1)*(100)}
      }{
      \pgfkeys{/pgf/fpu=true}
      \pgfmathparse{(#1)*(100)}
      \pgfkeys{/pgf/fpu=false}
      }
    }    
    ]
    
    \ratioplot[red,semithick]{fill=red!30!white, opacity=0.5}
    {data/UHV1.dat}{1}{9}{10}{3}{4}
    \addlegendentry{$\Delta\cH^2$}
  
    \ratioplot[blue,semithick]{fill=blue!30!white, opacity=0.5}
    {data/UHV1.dat}{1}{11}{12}{3}{4}
    \addlegendentry{$\Delta\cH \Delta\cV_{\textrm{Re}}$}
        
    \ratioplot[darkgreen,semithick]{fill=green!30!white, opacity=0.5}
    {data/UHV1.dat}{1}{13}{14}{3}{4}
    \addlegendentry{$\Delta\cV_{\textrm{Re}}^2$}

    \ratioplot[black,semithick]{fill=gray!30!white, opacity=0.5}
    {data/UHV0.dat}{1}{9}{10}{3}{4}
    \addlegendentry{Total}
        
  \end{axis}
\end{tikzpicture}

\fi

\end{center}
\caption{
$A + B = 2$, $C = 0$ contributions to the left-hand side of Eq.~(\ref{eq:unitaritytest}) times $10^2$.
}
\label{fig:UDelta2}
\end{figure}

In Fig.~\ref{fig:UDelta2}, we look at the contribution to the left hand side of Eq.~(\ref{eq:unitaritytest}) from contributions with $A + B = 2$. We plot separately the contributions from $\Delta\cH^2$, $\Delta\cH\,\Delta\cV_\textrm{Re}$, and $\Delta\cV_\textrm{Re}^2$, along with their total. We see that the individual contributions are typically $\pm 2\%$. However the total is smaller than about $\pm 0.2\%$.

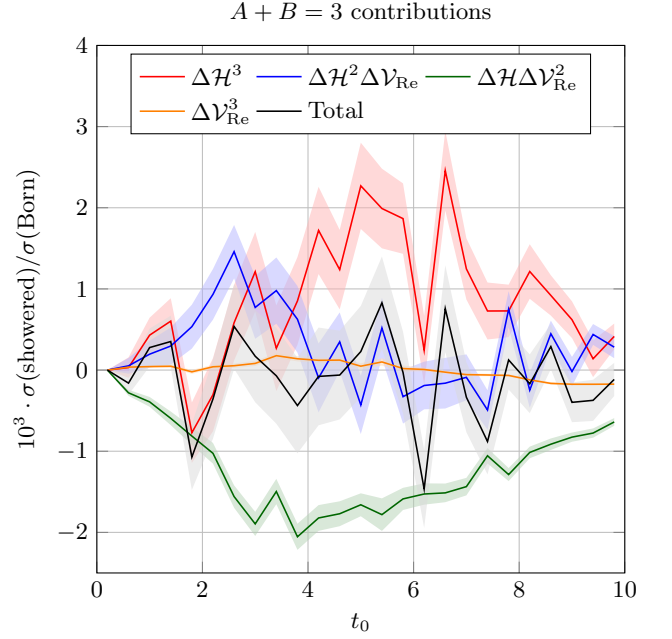
\begin{figure}
\begin{center}
\ifusefigs 

\begin{tikzpicture}
  \begin{axis}[
    title = {$A+B=3$ contributions},
    xlabel={$t_0$}, 
    ylabel={$10^3 \cdot\sigma(\text{showered})/\sigma(\text{Born})$},
    legend cell align=left,
     every axis legend/.append style = {at={(axis cs:5,0.0039)},
     anchor=north},
    legend columns=3,
    xmin=0, xmax=10,
    ymin=-0.0025, ymax=0.004,
    y coord trafo/.code={
      \pgflibraryfpuifactive{
      \pgfmathparse{(#1)*(1000)}
      }{
      \pgfkeys{/pgf/fpu=true}
      \pgfmathparse{(#1)*(1000)}
      \pgfkeys{/pgf/fpu=false}
      }
    }    
    ]
    
    \ratioplot[red,semithick]{fill=red!30!white, opacity=0.5}
    {data/UHV1.dat}{1}{17}{18}{3}{4}
    \addlegendentry{$\Delta\cH^3$}
  
    \ratioplot[blue,semithick]{fill=blue!30!white, opacity=0.5}
    {data/UHV1.dat}{1}{19}{20}{3}{4}
    \addlegendentry{$\Delta\cH^2 \Delta\cV_{\textrm{Re}}$}

    \ratioplot[darkgreen,semithick]{fill=darkgreen!30!white, opacity=0.5}
    {data/UHV1.dat}{1}{21}{22}{3}{4}
    \addlegendentry{$\Delta\cH \Delta\cV_{\textrm{Re}}^2 $}

    \ratioplot[orange,semithick]{fill=orange!30!white, opacity=0.5}
    {data/UHV1.dat}{1}{23}{24}{3}{4}
    \addlegendentry{$\Delta\cV_{\textrm{Re}}^3$}

    \ratioplot[black,semithick]{fill=gray!30!white, opacity=0.5}
    {data/UHV0.dat}{1}{11}{12}{3}{4}
    \addlegendentry{Total}
    
  \end{axis}
\end{tikzpicture}

\fi

\end{center}
\caption{
$A + B = 3$, $C = 0$ contributions to the left-hand side of Eq.~(\ref{eq:unitaritytest}) times $10^3$.
}
\label{fig:UDelta3}
\end{figure}

In Fig.~\ref{fig:UDelta3}, we look at the contributions with $A + B = 3$. We plot separately the contributions from  $\Delta\cH^3$, $\Delta\cH^2\,\Delta\cV_\textrm{Re}$, $\Delta\cH\,\Delta\cV_\textrm{Re}^2$, and $\Delta\cV_\textrm{Re}^3$, along with their total. The individual contributions are small, of order $\pm 10^{-3}$. This small size is understandable because each contribution is proportional to $\as^3$. The statistical fluctuations are almost as large as the individual contributions. The total is smaller than about $\pm 10^{-3}$, certainly smaller than the sum of the absolute values of the individual contributions.

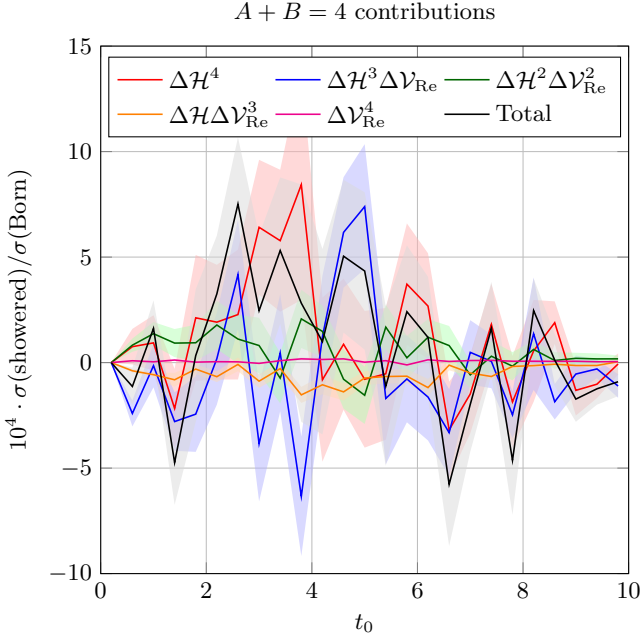
\begin{figure}
\begin{center}
\ifusefigs 

\begin{tikzpicture}
  \begin{axis}[
    title = {$A+B=4$ contributions},
    xlabel={$t_0$}, 
    ylabel={$10^4 \cdot\sigma(\text{showered})/\sigma(\text{Born})$},
    legend cell align=left,
     every axis legend/.append style = {at={(axis cs:5,0.00145)},
     anchor=north},
    legend columns=3,
    xmin=0, xmax=10,
    ymin=-0.001, ymax=0.0015,
    y coord trafo/.code={
      \pgflibraryfpuifactive{
      \pgfmathparse{(#1)*(10000)}
      }{
      \pgfkeys{/pgf/fpu=true}
      \pgfmathparse{(#1)*(10000)}
      \pgfkeys{/pgf/fpu=false}
      }
    }    
    ]
    
    \ratioplot[red,semithick]{fill=red!30!white, opacity=0.5}
    {data/UHV1.dat}{1}{29}{30}{3}{4}
    \addlegendentry{$\Delta\cH^4$}

    \ratioplot[blue,semithick]{fill=blue!30!white, opacity=0.5}
    {data/UHV1.dat}{1}{31}{32}{3}{4}
    \addlegendentry{$\Delta\cH^3 \Delta\cV_{\textrm{Re}}$}

    \ratioplot[darkgreen,semithick]{fill=green!30!white, opacity=0.5}
    {data/UHV1.dat}{1}{33}{34}{3}{4}
    \addlegendentry{$\Delta\cH^2 \Delta\cV_{\textrm{Re}}^2$}

    \ratioplot[orange,semithick]{fill=orange!30!white, opacity=0.5}
    {data/UHV1.dat}{1}{35}{36}{3}{4}
    \addlegendentry{$\Delta\cH \Delta\cV_{\textrm{Re}}^3$}

    \ratioplot[magenta,semithick]{fill=magenta!30!white, opacity=0.5}
    {data/UHV1.dat}{1}{37}{38}{3}{4}
    \addlegendentry{$\Delta\cV_{\textrm{Re}}^4$}
    
    \ratioplot[black,semithick]{fill=gray!30!white, opacity=0.5}
    {data/UHV0.dat}{1}{13}{14}{3}{4}
    \addlegendentry{Total}
    
  \end{axis}
\end{tikzpicture}

\fi

\end{center}
\caption{
$A + B = 4$, $C = 0$ contributions to the left-hand side of Eq.~(\ref{eq:unitaritytest}) times $10^4$.
}
\label{fig:UDelta4}
\end{figure}

Finally, in Fig.~\ref{fig:UDelta4}, we look at the contributions with $A + B = 4$. We plot separately the contributions from $\Delta\cH^4$, $\Delta\cH^3\,\Delta\cV_\textrm{Re}$, $\Delta\cH^2\,\Delta\cV_\textrm{Re}^2$, $\Delta\cH\,\Delta\cV_\textrm{Re}^3$, and $\Delta\cV_\textrm{Re}^4$, along with their total. The individual contributions are again small, of order $\pm 5\times 10^{-4}$. The contributions with the most factors of $\Delta\cH$ have fluctuations roughly as large as the contributions. The total inherits the fluctuations of the individual contributions but is consistent with zero within its fluctuations.

We see that all contributions proportional to $[\Delta \cH]^A [\Delta \cV_\textrm{Re}]^B$ with $A +B$ = 1, 2, 3, or 4 cancel to within their statistical accuracy, confirming Eq.~(\ref{eq:unitaritytest}) for these contributions.

We now try the second test of Eq.~(\ref{eq:unitaritytest}), in which we turn off $\Delta\cH$ and $\Delta\cV_\textrm{Re}$ and consider up to four insertions of $\cV_{\mi \pi}$. That is, we set $N_\textrm{Re} = 0$, $N_{\mi \pi} = 4$, and $N_\Delta = 4$. 

We divide the left hand side of Eq.~(\ref{eq:unitaritytest}) into parts that we can plot separately. The operator $\cV_{\mi \pi}$ in Eq.~(\ref{eq:Vipi}) is the sum of two terms, $\cV_{\mi \pi} = \widetilde{\cV}^\LL_{\mi \pi} + \widetilde{\cV}^\LR_{\mi \pi}$ where in $\widetilde{\cV}^\LL_{\mi \pi}$ the color operator $T_\La\cdot T_\Lb$ operates on ket color states and in $\widetilde{\cV}^\LR_{\mi \pi}$ the color operator $T_\La\cdot T_\Lb$ operates on bra color states. Acting on a color basis state $\ket{\{\hat c\}_m}$, $T_\La\cdot T_\Lb$ produces terms proportional to new basis states $\ket{\{\hat c\}_m}$. When partons ``a'' and ``b'' are color connected in $\ket{\{c\}_m}$, one of these new basis states is a constant times the original basis state $\ket{\{c\}_m}$. Thus we can write $\widetilde{\cV}^\LL_{\mi \pi} = {\cV}^\LL_{\mi \pi} + I^\LL_{\mi \pi}$, where $I^\LL_{\mi \pi}$ is the part of $\widetilde{\cV}^\LL_{\mi \pi}$ that returns a constant times the original color basis state and ${\cV}^\LL_{\mi \pi}$ is the part of $\widetilde{\cV}^\LL_{\mi \pi}$ that changes the color basis state. We apply the same decomposition to $\widetilde{\cV}^\LR_{\mi \pi}$. As a bookkeeping measure, it is convenient to decompose $\cV_{\mi \pi}$ as
\begin{equation}
\cV_{\mi \pi} = {\cV}^\LL_{\mi \pi} + {\cV}^\LR_{\mi \pi}
+ I_{\mi \pi}\;,
\,
\end{equation}
where $I_{\mi \pi} = I_{\mi \pi}^\LL + I_{\mi \pi}^\LR$. That is, we label contributions according to whether the ket color state was changed, ${\cV}^\LL_{\mi \pi}$, or the bra color state was changed, ${\cV}^\LR_{\mi \pi}$, or the color state was unchanged, $I_{\mi \pi}$. In the figure legends, we abbreviate these operators as $L$, $R$, and $I$.

\begin{figure}
\begin{center}
\ifusefigs 

\begin{tikzpicture}
  \begin{axis}[
    title = {$C=1$ contributions},
    xlabel={$t_0$}, 
    ylabel={$10^2 \cdot \text{Im}\sigma(\text{showered})/\sigma(\text{Born})$},
    legend cell align=left,
     every axis legend/.append style = {at={(axis cs:5,0.068)},
     anchor=north},
    legend columns=4,
    xmin=0, xmax=10,
    ymin=-0.06, ymax=0.07,
    y coord trafo/.code={
      \pgflibraryfpuifactive{
      \pgfmathparse{(#1)*(100)}
      }{
      \pgfkeys{/pgf/fpu=true}
      \pgfmathparse{(#1)*(100)}
      \pgfkeys{/pgf/fpu=false}
      }  
      }
    ]
    
    \ratioplot[red,semithick]{fill=red!30!white, opacity=0.5}
    {data/Phase1.dat}{1}{5}{6}{3}{4}
    \addlegendentry{$L$}

    \ratioplot[blue,semithick]{fill=blue!30!white, opacity=0.5}
    {data/Phase1.dat}{1}{7}{8}{3}{4}
    \addlegendentry{$R$}

    \ratioplot[orange,semithick]{fill=orange!30!white, opacity=0.5}
    {data/Phase1.dat}{1}{9}{10}{3}{4}
    \addlegendentry{$I$}

    \ratioplot[black,semithick]{fill=gray!30!white, opacity=0.5}
    {data/Phase0.dat}{1}{7}{8}{3}{4}
    \addlegendentry{Total}    
  \end{axis}
\end{tikzpicture}

\fi

\end{center}
\caption{
$A + B = 0$, $C = 1$ contributions to the imaginary part of the left-hand side of Eq.~(\ref{eq:unitaritytest}) times $10^2$.
}
\label{fig:Phase1}
\end{figure}
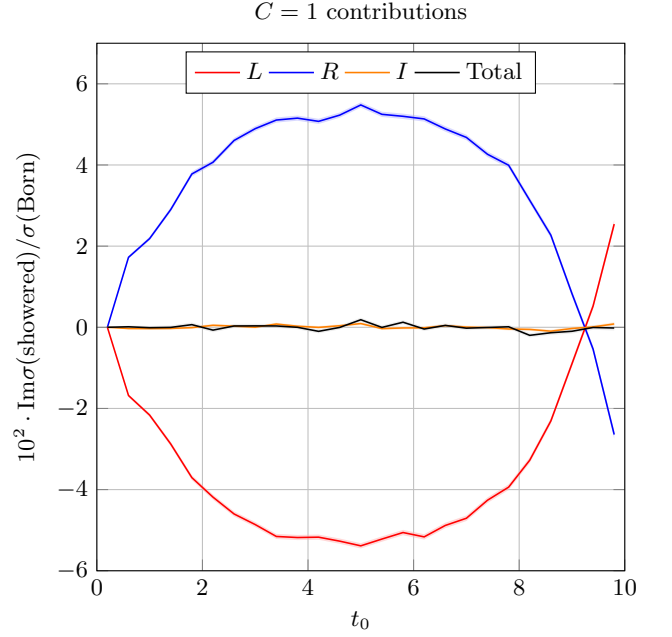

In Fig.~\ref{fig:Phase1}, we look at the contribution to the imaginary part of the left hand side of Eq.~(\ref{eq:unitaritytest}) from contributions proportional to $\cV_{\mi \pi}$. We plot separately the contributions from ${\cV}^\LL_{\mi \pi}$, ${\cV}^\LR_{\mi \pi}$, and $I_{\mi \pi}$ along with their total. We see that the individual contributions are of order $\pm 4\%$ and that these contributions cancel, giving a total that is smaller than about $\pm 0.2 \%$.

\begin{figure}
\begin{center}
\ifusefigs 

\begin{tikzpicture}
  \begin{axis}[
    title = {$C=2$ contributions},
    xlabel={$t_0$}, 
    ylabel={$\sigma(\text{showered})/\sigma(\text{Born})$},
    legend cell align=left,
     every axis legend/.append style = {at={(axis cs:5,1.55)},
     anchor=north},
    legend columns=3,
    xmin=0, xmax=10,
    ymin=-1.1, ymax=1.6,
    ]
    
    \ratioplot[red,semithick]{fill=red!30!white, opacity=0.5}
    {data/Phase1.dat}{1}{11}{12}{3}{4}
    \addlegendentry{$L^2 + R^2$}
  
    \ratioplot[blue,semithick]{fill=blue!30!white, opacity=0.5}
    {data/Phase1.dat}{1}{13}{14}{3}{4}
    \addlegendentry{$R L$}
        
    \ratioplot[darkgreen,semithick, dashed]{fill=green!30!white, opacity=0.5}
    {data/Phase1.dat}{1}{15}{16}{3}{4}
    \addlegendentry{$I^2$}
    
    \ratioplot[magenta,semithick, dashed]{fill=blue!30!white, opacity=0.5}
    {data/Phase1.dat}{1}{17}{18}{3}{4}
    \addlegendentry{$I (L+R)$}

    \ratioplot[black,semithick]{fill=gray!30!white, opacity=0.5}
    {data/Phase0.dat}{1}{9}{10}{3}{4}
    \addlegendentry{Total}
    
\end{axis}
\end{tikzpicture}

\fi

\end{center}
\caption{
$A + B = 0$, $C = 2$ contributions to the left-hand side of Eq.~(\ref{eq:unitaritytest}).
}
\label{fig:Phase2}
\end{figure}
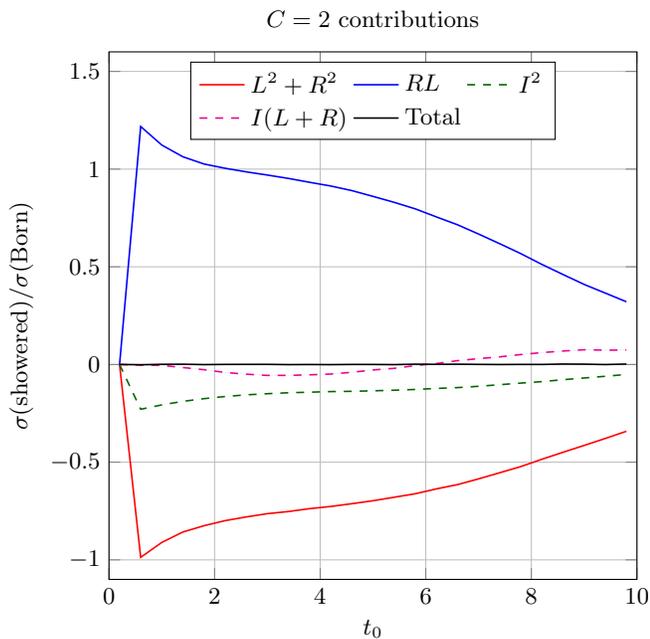

In Fig.~\ref{fig:Phase2}, we look at the contribution to the real part of the left hand side of Eq.~(\ref{eq:unitaritytest}) from contributions proportional to $\cV_{\mi \pi}^2$. We plot separately the contributions from $({\cV}^\LL_{\mi \pi})^2 + ({\cV}^\LR_{\mi \pi})^2$, ${\cV}^\LL_{\mi \pi}\,{\cV}^\LR_{\mi \pi}$, $I_{\mi \pi}^2$ and $I_{\mi \pi} ({\cV}^\LL_{\mi \pi} + {\cV}^\LR_{\mi \pi})$ along with their total. We see that some of the individual contributions are of order $\pm 1$. It is to be expected that the contributions from powers of $\cV_{\mi \pi}$ are larger than those from the same powers of $\Delta \cH$ and $\Delta \cV_\textrm{Re}$ because $\as$ in $\cV_{\mi \pi}$ comes with a factor $4\pi$. The contributions from these terms in $\cV_{\mi \pi}^2$ cancel, giving a total that is smaller than about $\pm 0.2 \%$.

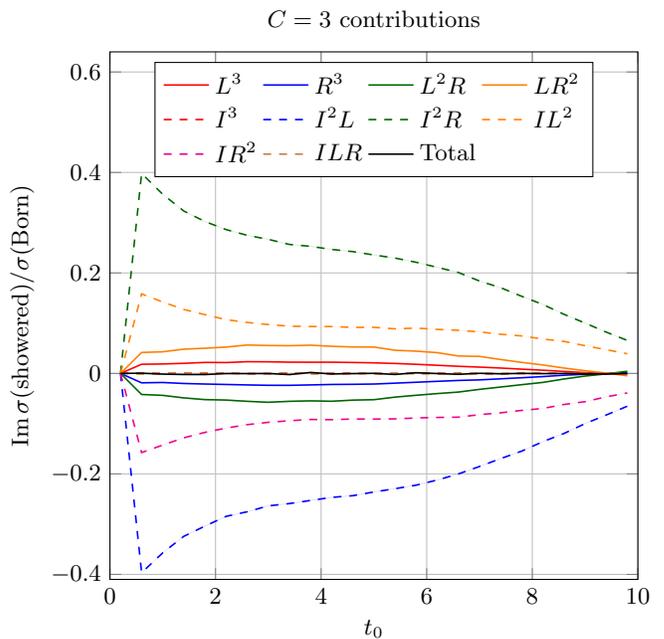
\begin{figure}
\begin{center}
\ifusefigs 

\begin{tikzpicture}
  \begin{axis}[
    title = {$C=3$ contributions},
    xlabel={$t_0$}, 
    ylabel={$\text{Im}\,\sigma(\text{showered})/\sigma(\text{Born})$},
    legend cell align=left,
     every axis legend/.append style = {at={(axis cs:5,0.62)},
     anchor=north},
    legend columns=4,
    xmin=0, xmax=10,
    ymin=-0.41, ymax=0.64,
    ]
    
    \ratioplot[red,semithick]{fill=red!30!white, opacity=0.5}
    {data/Phase1.dat}{1}{19}{20}{3}{4}
    \addlegendentry{$L^3$}
  
    \ratioplot[blue,semithick]{fill=blue!30!white, opacity=0.5}
    {data/Phase1.dat}{1}{21}{22}{3}{4}
    \addlegendentry{$R^3$}

    \ratioplot[darkgreen,semithick]{fill=darkgreen!30!white, opacity=0.5}
    {data/Phase1.dat}{1}{23}{24}{3}{4}
    \addlegendentry{$L^2 R$}

    \ratioplot[orange,semithick]{fill=orange!30!white, opacity=0.5}
    {data/Phase1.dat}{1}{25}{26}{3}{4}
    \addlegendentry{$L  R^2$}
    
    \ratioplot[red,semithick, dashed]{fill=red!30!white, opacity=0.5}
    {data/Phase1.dat}{1}{27}{28}{3}{4}
    \addlegendentry{$I^3$}
    
    \ratioplot[blue,semithick, dashed]{fill=blue!30!white, opacity=0.5}
    {data/Phase1.dat}{1}{29}{30}{3}{4}
    \addlegendentry{$I^2 L$}

    \ratioplot[darkgreen,semithick, dashed]{fill=darkgreen!30!white, opacity=0.5}
    {data/Phase1.dat}{1}{31}{32}{3}{4}
    \addlegendentry{$I^2 R$}

    \ratioplot[orange,semithick, dashed]{fill=orange!30!white, opacity=0.5}
    {data/Phase1.dat}{1}{33}{34}{3}{4}
    \addlegendentry{$I L^2$}
    
    \ratioplot[magenta,semithick, dashed]{fill=magenta!30!white, opacity=0.5}
    {data/Phase1.dat}{1}{35}{36}{3}{4}
    \addlegendentry{$I R^2$}
    
    \ratioplot[brown,semithick, dashed]{fill=brown!30!white, opacity=0.5}
    {data/Phase1.dat}{1}{37}{38}{3}{4}
    \addlegendentry{$I L R$}

    \ratioplot[black,semithick]{fill=gray!30!white, opacity=0.5}
    {data/Phase0.dat}{1}{11}{12}{3}{4}
    \addlegendentry{Total}
    
\end{axis}
\end{tikzpicture}

\fi

\end{center}
\caption{
$A + B = 0$, $C = 3$ contributions to the imaginary part of the left-hand side of Eq.~(\ref{eq:unitaritytest}).
}
\label{fig:Phase3}
\end{figure}

In Fig.~\ref{fig:Phase3}, we look at the contribution to the imaginary part of the left hand side of Eq.~(\ref{eq:unitaritytest}) from contributions proportional to $\cV_{\mi \pi}^3$. We plot separately the contributions from $({\cV}^\LL_{\mi \pi})^3$, $({\cV}^\LL_{\mi \pi})^2\, {\cV}^\LR_{\mi \pi}$, ${\cV}^\LL_{\mi \pi}\,({\cV}^\LR_{\mi \pi})^2$, $({\cV}^\LR_{\mi \pi})^3$, $I_{\mi \pi}^3$, $I_{\mi \pi}^2 {\cV}^\LL_{\mi \pi}$, $I_{\mi \pi}^2 {\cV}^\LR_{\mi \pi}$, $I_{\mi \pi} ({\cV}^\LL_{\mi \pi})^2$, $I_{\mi \pi} ({\cV}^\LR_{\mi \pi})^2$, and $I_{\mi \pi} {\cV}^\LL_{\mi \pi} {\cV}^\LR_{\mi \pi}$, along with their total. We see that some of the individual contributions are of order $\pm 0.3$. Again, the contributions from these terms in $\cV_{\mi \pi}^3$ cancel, giving a total that is smaller than about $\pm 0.2\%$.

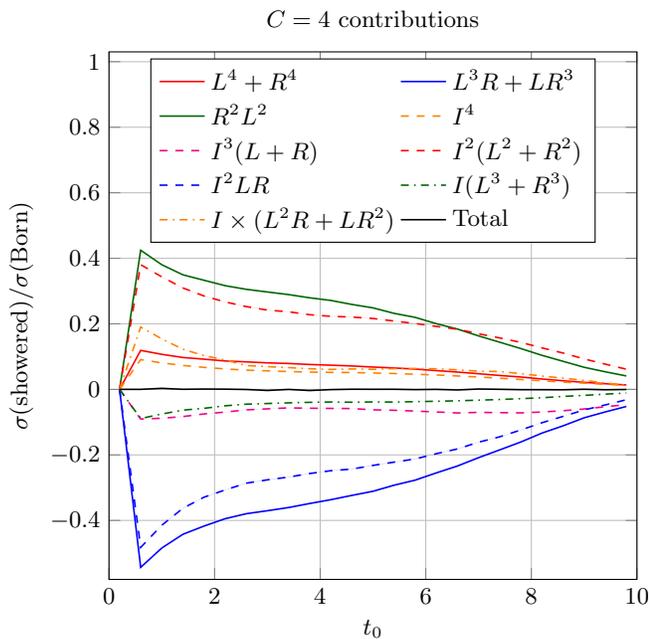
\begin{figure}
\begin{center}
\ifusefigs 

\begin{tikzpicture}
  \begin{axis}[
    title = {$C=4$ contributions},
    xlabel={$t_0$}, 
    ylabel={$\sigma(\text{showered})/\sigma(\text{Born})$},
    legend cell align=left,
     every axis legend/.append style = {at={(axis cs:5,1.01)},
     anchor=north},
    legend columns=2,
    xmin=0, xmax=10,
    ymin=-0.58, ymax=1.03,
    ]
    
    \ratioplot[red,semithick]{fill=red!30!white, opacity=0.5}
    {data/Phase1.dat}{1}{39}{40}{3}{4}
    \addlegendentry{$L^4 + R^4$}

    \ratioplot[blue,semithick]{fill=blue!30!white, opacity=0.5}
    {data/Phase1.dat}{1}{41}{42}{3}{4}
    \addlegendentry{$L^3 R + L R^3$}

    \ratioplot[darkgreen,semithick]{fill=green!30!white, opacity=0.5}
    {data/Phase1.dat}{1}{43}{44}{3}{4}
    \addlegendentry{$R^2  L^2$}

    \ratioplot[orange,semithick,dashed]{fill=orange!30!white, opacity=0.5}
    {data/Phase1.dat}{1}{45}{46}{3}{4}
    \addlegendentry{$I^4$}

    \ratioplot[magenta,semithick,dashed]{fill=magenta!30!white, opacity=0.5}
    {data/Phase1.dat}{1}{47}{48}{3}{4}
    \addlegendentry{$I^3 (L + R)$}

    \ratioplot[red,semithick, dashed]{fill=red!30!white, opacity=0.5}
    {data/Phase1.dat}{1}{49}{50}{3}{4}
    \addlegendentry{$I^2 (L^2 + R^2)$}

    \ratioplot[blue,semithick, dashed]{fill=blue!30!white, opacity=0.5}
    {data/Phase1.dat}{1}{51}{52}{3}{4}
    \addlegendentry{$I^2 L  R$}

    \ratioplot[darkgreen,semithick, dashdotted]{fill=green!30!white, opacity=0.5}
    {data/Phase1.dat}{1}{53}{54}{3}{4}
    \addlegendentry{$I (L^3 + R^3)$}

    \ratioplot[orange,semithick, dashdotted]{fill=orange!30!white, opacity=0.5}
    {data/Phase1.dat}{1}{55}{56}{3}{4}
    \addlegendentry{$I\times (L^2 R + L R^2)$}

    \ratioplot[black,semithick]{fill=gray!30!white, opacity=0.5}
    {data/Phase0.dat}{1}{13}{14}{3}{4}
    \addlegendentry{Total}
    
\end{axis}
\end{tikzpicture}

\fi

\end{center}
\caption{
$A + B = 0$, $C = 4$ contributions to the left-hand side of Eq.~(\ref{eq:unitaritytest}).
}
\label{fig:Phase4}
\end{figure}
In Fig.~\ref{fig:Phase4}, we look at the contribution to the real part of the left hand side of Eq.~(\ref{eq:unitaritytest}) from contributions proportional to $\cV_{\mi \pi}^4$. We plot separately the contributions from 
$({\cV}^\LL_{\mi \pi})^4 + ({\cV}^\LR_{\mi \pi})^4$, 
$({\cV}^\LL_{\mi \pi})^3\, {\cV}^\LR_{\mi \pi} 
  + {\cV}^\LL_{\mi \pi}\, ({\cV}^\LR_{\mi \pi})^3$, 
$({\cV}^\LL_{\mi \pi})^2\,({\cV}^\LR_{\mi \pi})^2$, 
$I_{\mi \pi}^4$,
$I_{\mi \pi}^3 ({\cV}^\LL_{\mi \pi} + {\cV}^\LR_{\mi \pi})$, 
$I_{\mi \pi}^2 (({\cV}^\LL_{\mi \pi})^2 + ({\cV}^\LR_{\mi \pi})^2)$, 
$I_{\mi \pi}^2 {\cV}^\LL_{\mi \pi}{\cV}^\LR_{\mi \pi}$, 
$I_{\mi \pi} (({\cV}^\LL_{\mi \pi})^3 + ({\cV}^\LL_{\mi \pi})^3)$, and
$I_{\mi \pi} (({\cV}^\LL_{\mi \pi})^2 {\cV}^\LR_{\mi \pi} 
  + {\cV}^\LL_{\mi \pi}({\cV}^\LR_{\mi \pi})^2) $,  
along with their total. We see that some of the individual contributions are of order $\pm 0.4$. Yet again, the contributions from these terms in $\cV_{\mi \pi}^4$ cancel, giving a total that is smaller than about $\pm 0.2 \%$.

We see that all contributions proportional to $(\cV_{\mi \pi})^C$ with $C$ = 1, 2, 3, or 4 cancel to within their statistical accuracy, confirming Eq.~(\ref{eq:unitaritytest}) for these contributions.

Finally, we consider up to four insertions of $\Delta\cH$, $\Delta\cV_\textrm{Re}$ and $\cV_{\mi \pi}$ together. That is, we set $N_\textrm{Re} = 4$, $N_{\mi\pi} = 4$, and $N_\Delta = 4$. We examine the real part of the left hand side of Eq.~(\ref{eq:unitaritytest}), considering separately contributions proportional to $[\Delta \cH]^A [\Delta \cV_\textrm{Re}]^B [\cV_{\mi \pi}]^C$ with values of $A+B+C$ up to 4.

When we examine contributions to the real part of the cross section with $A + B + C = 1$, the contribution from $\cV_{\mi \pi}$ is not present because this contribution in imaginary. Thus we obtain the graph shown in Fig.~\ref{fig:UDelta1}, in which the needed cancellations work up to statistical fluctuations, which are smaller than about $\pm 0.2\%$.

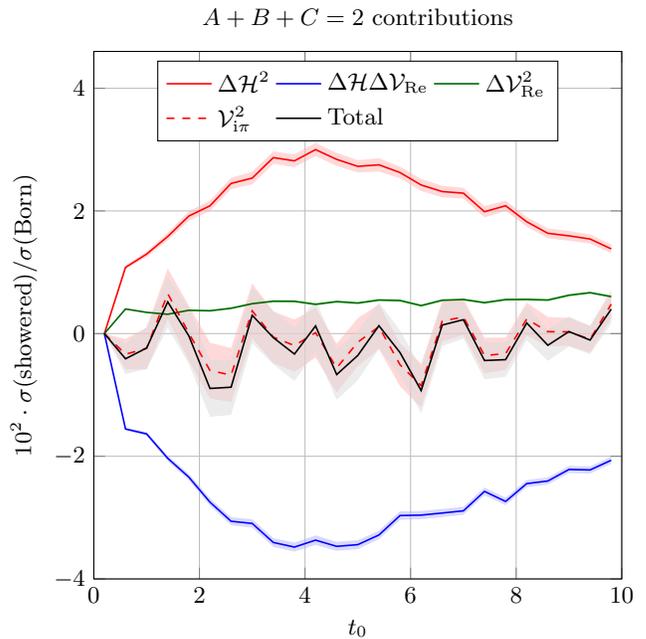
\begin{figure}
\begin{center}
\ifusefigs 

\begin{tikzpicture}
  \begin{axis}[title = {$A+B+C=2$ contributions},
    xlabel={$t_0$}, ylabel={$10^2\cdot\sigma(\text{showered})/\sigma(\text{Born})$},
    legend cell align=left,
    every axis legend/.append style = {at={(axis cs:5,0.0445)},
    anchor=north},
    legend columns=3,
    xmin=0, xmax=10,
    ymin=-0.04, ymax=0.046,
    y coord trafo/.code={
      \pgflibraryfpuifactive{
      \pgfmathparse{(#1)*(100)}
      }{
      \pgfkeys{/pgf/fpu=true}
      \pgfmathparse{(#1)*(100)}
      \pgfkeys{/pgf/fpu=false}
      }
    }    
    ]
    
    \ratioplot[red,semithick]{fill=red!30!white, opacity=0.5}
    {data/Uall1.dat}{1}{9}{10}{3}{4}
    \addlegendentry{$\Delta\cH^2$}
  
    \ratioplot[blue,semithick]{fill=blue!30!white, opacity=0.5}
    {data/Uall1.dat}{1}{11}{12}{3}{4}
    \addlegendentry{$\Delta\cH \Delta\cV_{\textrm{Re}}$}
        
    \ratioplot[darkgreen,semithick]{fill=green!30!white, opacity=0.5}
    {data/Uall1.dat}{1}{13}{14}{3}{4}
    \addlegendentry{$\Delta\cV_{\textrm{Re}}^2$}
    
    \ratioplot[red,semithick, dashed]{fill=red!30!white, opacity=0.5}
    {data/Uall1.dat}{1}{15}{16}{3}{4}
    \addlegendentry{$\cV_{\mi\pi}^2$}

    \ratioplot[black,semithick]{fill=gray!30!white, opacity=0.5}
    {data/Uall0.dat}{1}{9}{10}{3}{4}
    \addlegendentry{Total}
        
  \end{axis}
\end{tikzpicture}

\fi

\end{center}
\caption{
$A + B + C = 2$ contributions to the left-hand side of Eq.~(\ref{eq:unitaritytest}) times $10^2$.
}
\label{fig:Uall2}
\end{figure}

In Fig.~\ref{fig:Uall2}, we look at contributions with $A + B + C = 2$. We plot separately the contributions from $\Delta\cH^2$, $\Delta\cH\,\Delta\cV_\textrm{Re}$, $\Delta\cV_\textrm{Re}^2$, and $\cV_{\mi \pi}^2$ along with their total. We do not break the $\cV_{\mi \pi}^2$ contribution into separate parts as we did in Fig.~\ref{fig:Phase2}. Recall from Fig.~\ref{fig:Phase2} that the separate parts of the $\cV_{\mi \pi}^2$ contribution are of order 1. We see that the total with everything included vanishes up to statistical fluctuations, which are smaller than about $\pm 1\%$ and are dominated by the statistical fluctuations in $\cV_{\mi \pi}^2$.

\begin{figure}
\begin{center}
\ifusefigs 

\begin{tikzpicture}
  \begin{axis}[title = {$A+B+C=3$ contributions},
    xlabel={$t_0$}, ylabel={$10^2\cdot\sigma(\text{showered})/\sigma(\text{Born})$},
    legend cell align=left,
    every axis legend/.append style = {at={(axis cs:5,0.028)},
    anchor=north},
    legend columns=3,
    xmin=0, xmax=10,
    ymin=-0.02, ymax=0.029,
    y coord trafo/.code={
      \pgflibraryfpuifactive{
      \pgfmathparse{(#1)*(100)}
      }{
      \pgfkeys{/pgf/fpu=true}
      \pgfmathparse{(#1)*(100)}
      \pgfkeys{/pgf/fpu=false}
      }
    }    
    ]
    
    \ratioplot[red,semithick]{fill=red!30!white, opacity=0.5}
    {data/Uall1.dat}{1}{17}{18}{3}{4}
    \addlegendentry{$\Delta\cH^3$}
  
    \ratioplot[blue,semithick]{fill=blue!30!white, opacity=0.5}
    {data/Uall1.dat}{1}{19}{20}{3}{4}
    \addlegendentry{$\Delta\cH^2 \Delta\cV_{\textrm{Re}}$}

    \ratioplot[darkgreen,semithick]{fill=darkgreen!30!white, opacity=0.5}
    {data/Uall1.dat}{1}{21}{22}{3}{4}
    \addlegendentry{$\Delta\cH \Delta\cV_{\textrm{Re}}^2 $}

    \ratioplot[orange,semithick]{fill=orange!30!white, opacity=0.5}
    {data/Uall1.dat}{1}{23}{24}{3}{4}
    \addlegendentry{$\Delta\cV_{\textrm{Re}}^3$}
    
    \ratioplot[red,semithick, dashed]{fill=red!30!white, opacity=0.5}
    {data/Uall1.dat}{1}{25}{26}{3}{4}
    \addlegendentry{$\Delta\cH \cV_{\mi\pi}^2$}
    
    \ratioplot[blue,semithick, dashed]{fill=blue!30!white, opacity=0.5}
    {data/Uall1.dat}{1}{27}{28}{3}{4}
    \addlegendentry{$\Delta\cV_{\textrm{Re}} \cV_{\mi\pi}^2$}

    \ratioplot[black,semithick]{fill=gray!30!white, opacity=0.5}
    {data/Uall0.dat}{1}{11}{12}{3}{4}
    \addlegendentry{Total}
            
  \end{axis}
\end{tikzpicture}

\fi

\end{center}
\caption{
$A + B + C = 3$ contributions to the left-hand side of Eq.~(\ref{eq:unitaritytest}) times $10^2$.
}
\label{fig:Uall3}
\end{figure}
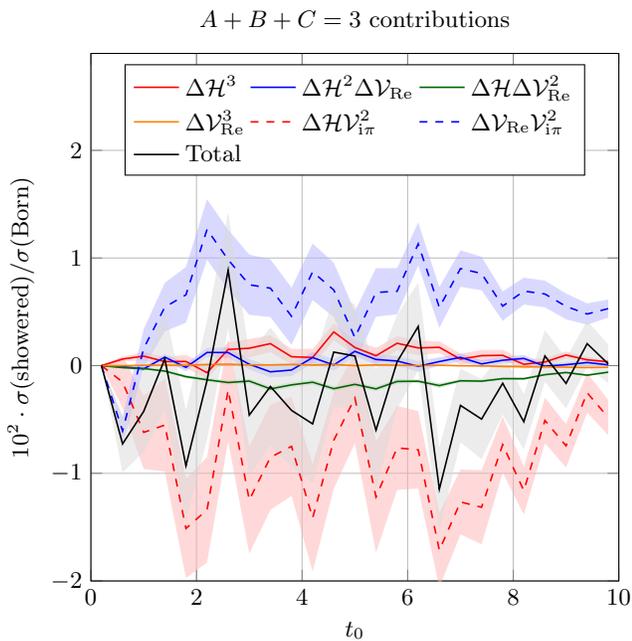

In Fig.~\ref{fig:Uall3}, we look at contributions with $A + B + C = 3$. We plot separately the contributions from $\Delta\cH^3$, $\Delta\cH^2\,\Delta\cV_\textrm{Re}$, $\Delta\cH\,\Delta\cV_\textrm{Re}^2$, $\Delta\cV_\textrm{Re}^3$, $\Delta\cH \cV_{\mi \pi}^2$ and $\Delta\cV_\textrm{Re} \cV_{\mi \pi}^2$ along with their total. We see that the total vanishes up to statistical fluctuations, which are smaller than about $\pm 1\%$ and are dominated by the statistical fluctuations in the terms that contain $\cV_{\mi \pi}^2$.

\begin{figure}
\begin{center}
\ifusefigs 

\begin{tikzpicture}
  \begin{axis}[title = {$A+B+C=4$ contributions},
    xlabel={$t_0$}, ylabel={$10^2\cdot\sigma(\text{showered})/\sigma(\text{Born})$},
    legend cell align=left,
    every axis legend/.append style = {at={(axis cs:5,0.037)},
    anchor=north},
    legend columns=2,
    xmin=0, xmax=10,
    ymin=-0.015, ymax=0.038,
    y coord trafo/.code={
      \pgflibraryfpuifactive{
      \pgfmathparse{(#1)*(100)}
      }{
      \pgfkeys{/pgf/fpu=true}
      \pgfmathparse{(#1)*(100)}
      \pgfkeys{/pgf/fpu=false}
      }
    }    
    ]
    
    \ratioplot[red,semithick]{fill=red!30!white, opacity=0.5}
    {data/Uall1.dat}{1}{29}{30}{3}{4}
    \addlegendentry{$\Delta\cH^4$}

    \ratioplot[blue,semithick]{fill=blue!30!white, opacity=0.5}
    {data/Uall1.dat}{1}{31}{32}{3}{4}
    \addlegendentry{$\Delta\cH^3 \Delta\cV_{\textrm{Re}}$}

    \ratioplot[darkgreen,semithick]{fill=green!30!white, opacity=0.5}
    {data/Uall1.dat}{1}{33}{34}{3}{4}
    \addlegendentry{$\Delta\cH^2 \Delta\cV_{\textrm{Re}}^2$}

    \ratioplot[orange,semithick]{fill=orange!30!white, opacity=0.5}
    {data/Uall1.dat}{1}{35}{36}{3}{4}
    \addlegendentry{$\Delta\cH \Delta\cV_{\textrm{Re}}^3$}

    \ratioplot[magenta,semithick]{fill=magenta!30!white, opacity=0.5}
    {data/Uall1.dat}{1}{37}{38}{3}{4}
    \addlegendentry{$\Delta\cV_{\textrm{Re}}^4$}
    
    \ratioplot[red,semithick, dashed]{fill=red!30!white, opacity=0.5}
    {data/Uall1.dat}{1}{39}{40}{3}{4}
    \addlegendentry{$\Delta\cH^2 \cV_{\mi\pi}^2$}

    \ratioplot[blue,semithick, dashed]{fill=blue!30!white, opacity=0.5}
    {data/Uall1.dat}{1}{41}{42}{3}{4}
    \addlegendentry{$\Delta\cH \Delta\cV_{\textrm{Re}} \cV_{\mi\pi}^2$}

    \ratioplot[darkgreen,semithick, dashed]{fill=green!30!white, opacity=0.5}
    {data/Uall1.dat}{1}{43}{44}{3}{4}
    \addlegendentry{$\Delta\cV_{\textrm{Re}}^2 \cV_{\mi\pi}^2$}

   \ratioplot[magenta,semithick, dashed]{fill=magenta!30!white, opacity=0.5}
    {data/Uall1.dat}{1}{45}{46}{3}{4}
    \addlegendentry{$\cV_{\mi\pi}^4$}

    \ratioplot[black,semithick]{fill=gray!30!white, opacity=0.5}
    {data/Uall0.dat}{1}{13}{14}{3}{4}
    \addlegendentry{Total}
            
  \end{axis}
\end{tikzpicture}

\fi

\end{center}
\caption{
$A + B + C = 4$ contributions to the left-hand side of Eq.~(\ref{eq:unitaritytest}) times $10^2$.
}
\label{fig:Uall4}
\end{figure}
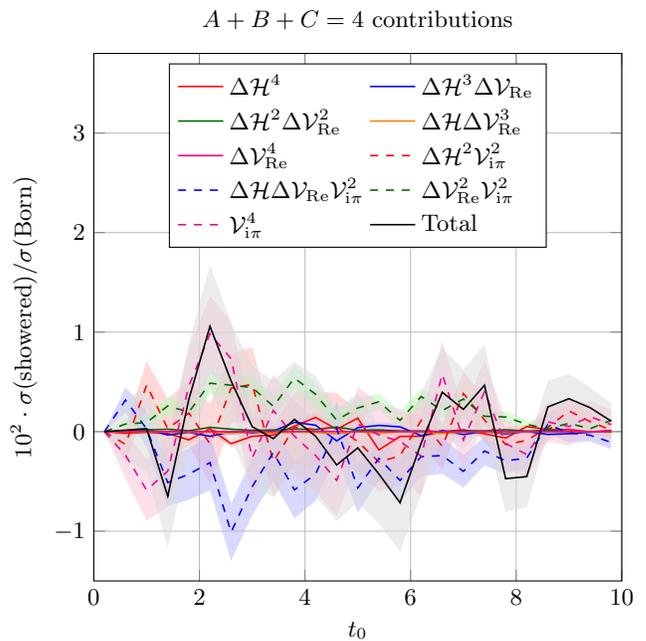

Finally, in Fig.~\ref{fig:Uall4}, we look at contributions with $A + B + C = 4$. We plot separately the contributions from $\Delta\cH^4$, $\Delta\cH^3\,\Delta\cV_\textrm{Re}$, $\Delta\cH^2\,\Delta\cV_\textrm{Re}^2$, $\Delta\cH\,\Delta\cV_\textrm{Re}^3$, $\Delta\cV_\textrm{Re}^4$, $\Delta\cH^2 \cV_{\mi \pi}^2$, $\Delta\cH\,\Delta\cV_\textrm{Re}\,\cV_{\mi \pi}^2$, $\Delta\cV_\textrm{Re}^2\,\cV_{\mi \pi}^2$, and $\cV_{\mi \pi}^4$, along with their total. In this plot, the contributions with no powers of $\cV_{\mi \pi}$ are small. We see that the total vanishes up to statistical fluctuations, which are smaller than about $\pm 1\%$ and are dominated by the statistical fluctuations in the terms that contain powers of $\cV_{\mi \pi}$.

In summary, we have seen in some detail that probability conservation, Eq.~(\ref{eq:unitaritytest}), works.

\section{A sample cross section}
\label{sec:jets}

Version 3.0.0 of \textsc{Deductor}, as described above, can be used to calculate cross sections with color treated beyond the LC+ approximation. We leave an examination of the phenomenology of color corrections to a future work. However, in this section we demonstrate at least that \textsc{Deductor} with non-leading color effects can be used to calculate a physical cross section. We choose a cross section for which we expect that color effects beyond the LC+ approximation will be small, namely the one jet inclusive cross section $d\sigma/dP_\LT$ for jets with rapidities in the range $-2 < y_J < 2$ as a function of the jet transverse momentum $P_\LT$ for proton-proton collisions at $\sqrt s = 13 \TeV$. For the renormalization and factorization scales in the Born cross section with which the shower begins, we choose $\mu_\LR = \mu_\LF = P_\LT^{\textrm{Born}}/\sqrt{2}$. The shower cross section depends on the $\Lambda$-ordering starting scale for the shower, which we choose to be $\mu_\Ls = (3/2) P_\LT^{\textrm{Born}}$, which corresponds to a shower time $t_0 = \log(4\hat s/(9 [P_\LT^{\textrm{Born}}]^2))$. We have examined this cross section in some detail in Ref.~\cite{NSThresholdII} in the LC+ approximation. Here we limit our investigation to the effects of color beyond the LC+ approximation. 

The shower stops when the transverse momentum in a splitting reaches a cutoff $k_\LT^\textrm{min}$. In this section, we choose $k_\LT^\textrm{min} = 5 \GeV$.  In Ref.~\cite{NSThresholdII}, we used $k_\LT^\textrm{min} = 1 \GeV$ for final state splittings and $k_\LT^\textrm{min} = 1.295 \GeV$ for initial state splittings. The 5 GeV choice has the advantage of making the code run faster. If we wanted to correct to the choice in Ref.~\cite{NSThresholdII}, we would apply a correction factor that ranges from about 0.93 at $P_\LT = 300 \GeV$ to 1.0 at $P_\LT = 3.5 \TeV$. We could also apply a correction factor of about 0.98 to account for non-perturbative effects  \cite{NSThresholdII}.

The formula that represents what \textsc{Deductor} does is the following, taken from Eq.~(134) of Ref.~\cite{NSAllOrder} with some simplification of the notation:
\begin{equation}
\begin{split}
\label{eq:sigmaU8}
\sigma[J] ={}& 
\sbra{1} \cO_J\, 
\cU(t_\Lf,t_0)\,
\cU_\cV(t_\Lf,t_0)\,
\sket{\rho_\textrm{hard}}
\;.
\end{split}
\end{equation}
The statistical state $\sket{\rho_\textrm{hard}}$ includes the scattering matrix elements and a factor containing the parton distribution functions for the two incoming partons for a range of transverse momenta and rapidities in the hard scattering. In principle, $\sket{\rho_\textrm{hard}}$ should include next-to-leading order (NLO) corrections with their accompanying subtractions, as described in Ref.~\cite{NSAllOrder}. However, we have not implemented the NLO corrections in \textsc{Deductor} 3.0.0. This limits the accuracy in the calculation. The initial shower time $t_0$ is determined from the kinematics of the initial hard scattering: $t_0 = \log(4\hat s/(9 [P_\LT^\textrm{Born}]^2))$. Thus $t_0$ in Eq.~(\ref{eq:sigmaU8}) is really an operator that gives $\log(4\hat s/(9 [P_\LT^\textrm{Born}]^2))$ as its eigenvalue for a statistical basis state in the expansion of $\sket{\rho_\textrm{hard}}$.

The operator $\cU_\cV(t_\Lf,t_0)$ implements the approximate summation of threshold logarithms, as described in Refs.~\cite{NSThresholdII} and \cite{NSAllOrder}. We have, however, dropped some numerically unimportant terms from $\cU_\cV(t_\Lf,t_0)$ compared to Ref.~\cite{NSThresholdII}. The next operator, $\cU(t_\Lf,t_0)$, generates the shower. The shower stops when the transverse momentum in a splitting is smaller than a cutoff value $k_\LT^\textrm{min} = 5 \GeV$. Then the operator $\cO_J$ specifies the jet measurement that we want to make. Finally, the bra state $\sbra{1}$ is an instruction to integrate and sum over all of the parton variables, including taking the trace over the color variables.

Since $d\sigma/dP_\LT$ is a steeply falling cross section, we display results for the ratio to the next-to-leading order cross section \cite{EKS},
\begin{equation}
K(P_\LT) = \frac{d\sigma(\textrm{shower})/dP_\LT}{d\sigma(\textrm{NLO})/dP_\LT}
\;.
\end{equation}
In the plots that follow, we choose the maximum color suppression index to be $I_\textrm{max} = 4$. This applies both within the LC+ approximation and when operators $\Delta \cH$ and $\Delta \cV$ are allowed.

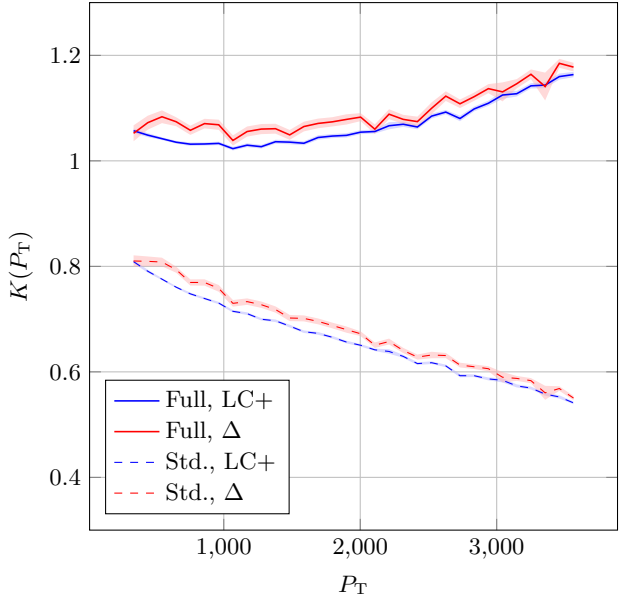
\begin{figure}
\begin{center}
\ifusefigs 

\begin{tikzpicture}
\begin{axis}[title = {One jet cross section, ratios to NLO},
   xlabel={$P_\LT$}, ylabel={$K(P_\LT)$},
   legend cell align=left,
   every axis legend/.append style = {at={(axis cs:130,0.33)},anchor=south west},
   ymin= 0.30, ymax= 1.3
]

\errorband[blue,semithick]{fill=blue!30!white, opacity=0.5}
{data/JETS21.dat}{1}{5}{6}
\addlegendentry{Full, LC+} 

\errorband[red,semithick]{fill=red!30!white, opacity=0.5}
{data/JETS21.dat}{1}{13}{14}
\addlegendentry{Full, $\Delta$} 

\errorband[blue,dashed]{fill=blue!30!white, opacity=0.5}
{data/JETS21.dat}{1}{3}{4}
\addlegendentry{Std., LC+} 

\errorband[red,dashed]{fill=red!30!white, opacity=0.5}
{data/JETS21.dat}{1}{11}{12}
\addlegendentry{Std., $\Delta$} 

\end{axis}
\end{tikzpicture}

\fi

\end{center}
\caption{
Jet cross section ratios with $N_\Delta^\textrm{thr.}= 1$ and with $N_\Delta = N_\textrm{Re} = N_{\mi\pi} = 2$ for $t_0 < t < t_0 + 5$.
}
\label{fig:jets21}
\end{figure}

In the upper red curve in Fig.~\ref{fig:jets21}, we plot $K(P_\LT)$ with two units of color beyond the LC+ approximation in the shower and one unit of extra color in the threshold factor: $N_\Delta = N_\textrm{Re} = N_{\mi\pi} = 2$ and $N_\Delta^\textrm{thr.}= 1$. In the shower, the inclusion of extra color applies in the first 5 units of shower time, $t_0 < t < t_0 + 5$. After that, we use an LC+ shower. The upper blue curve is $K(P_\LT)$ in the LC+ approximation, $N_\Delta = N_\textrm{Re} = N_{\mi\pi} = 0$ and $N_\Delta^\textrm{thr.}= 0$. We label these as ``Full'' calculations, indicating that they include the threshold factor $\cU_\cV$. 

The threshold factor is quite important: in the lower, dashed red and blue curves, we plot $K(P_\LT)$ with $\cU_\cV$ turned off, so that we have only the standard (``Std.''), probability preserving shower generated by $\cU(t_\Lf,t_0)$. The dashed red curve is calculated with $N_\Delta = N_\textrm{Re} = N_{\mi\pi} = 2$ in the first 5 units of shower time while the dashed blue curve is calculated in the LC+ approximation, $N_\Delta = N_\textrm{Re} = N_{\mi\pi} = 0$. 

We see that there is a small effect from going beyond the LC+ approximation both with and without the threshold factor. This approximately 3\% effect appears both in the Full calculation and the Std. calculation.

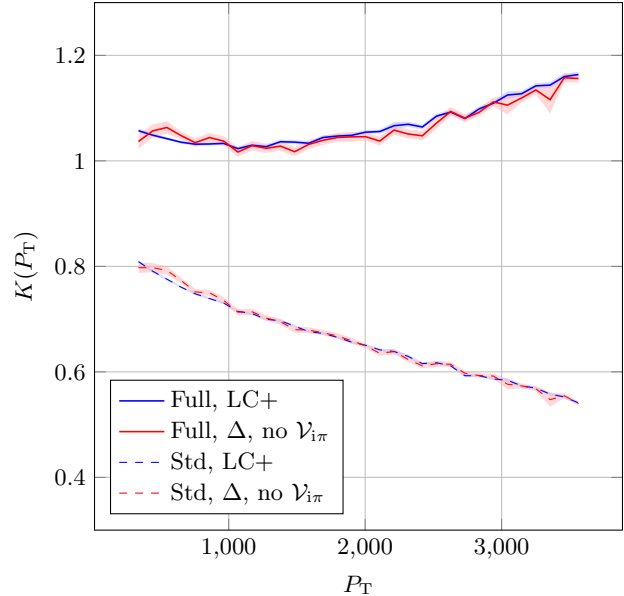
\begin{figure}
\begin{center}
\ifusefigs 

\begin{tikzpicture}
\begin{axis}[title = {One jet cross section, ratios to NLO},
   xlabel={$P_\LT$}, ylabel={$K(P_\LT)$},
   legend cell align=left,
   every axis legend/.append style = {at={(axis cs:130,0.33)},anchor=south west},
   ymin= 0.30, ymax= 1.3
]

\errorband[blue,semithick]{fill=blue!30!white, opacity=0.5}
{data/JETS21.dat}{1}{5}{6}
\addlegendentry{Full, LC+} 

\errorband[red,semithick]{fill=red!30!white, opacity=0.5}
{data/JETS21.dat}{1}{9}{10}
\addlegendentry{Full, $\Delta$, no $\cV_{\mi\pi}$} 

\errorband[blue,dashed]{fill=blue!30!white, opacity=0.5}
{data/JETS21.dat}{1}{3}{4}
\addlegendentry{Std, LC+} 

\errorband[red,dashed]{fill=red!30!white, opacity=0.5}
{data/JETS21.dat}{1}{7}{8}
\addlegendentry{Std, $\Delta$, no $\cV_{\mi\pi}$} 

\end{axis}
\end{tikzpicture}

\fi

\end{center}
\caption{
Jet cross section ratios with $N_\Delta^\textrm{thr.}= 1$ and with $N_\Delta = N_\textrm{Re} = 2$ for $t_0 < t < t_0 + 5$, as in Fig.~\ref{fig:jets21}, but with $N_{\mi\pi} = 0$ so that factors of $\cV_{\mi\pi}$ do not appear.
}
\label{fig:jets21noipi}
\end{figure}

We find that the small effect from extra color comes from the $\cV_{\mi\pi}$ operators. To see this, we plot in Fig.~\ref{fig:jets21noipi} the same curves as in Fig.~\ref{fig:jets21} but without $\cV_{\mi\pi}$. Now there is no difference, within statistical fluctuations, between the red and blue curves.

It is understandable that $\cV_{\mi\pi}$ affects the jet cross section. We have seen that $\cV_{\mi\pi}$, with its factor of $4\pi\as$, is not a small operator. Thus contributions from $\cV_{\mi\pi}^2$ in Fig.~\ref{fig:Phase2} are of order 0.2. These contributions cancel when we use a completely inclusive measurement operator, as we saw in Fig.~\ref{fig:Phase2}. However the different contributions have different color states, so that we can radiate more or fewer gluons out of the jet cone when we use a jet measurement operator. This can change the jet cross section.

What happens if we add more powers of $\Delta \cH$ and $\Delta \cV$?  We can double the amount of extra color by using $N_\Delta^\textrm{thr.}= 2$ and by using $N_\Delta = N_\textrm{Re} = N_{\mi\pi} = 4$ in the first 5 units of shower time. We have seen that having two powers of $\Delta \cH$ and $\Delta \cV$ in $\cU(t_\Lf,t_0)$ changes the cross section by a factor of only about 1.03. Thus we might expect that adding two more powers of $\Delta \cH$ and $\Delta \cV$ will further change the cross section by a factor of only $1 + (0.03)^2 \approx 1.001$. However, this expectation could be wrong. Perhaps an unanticipated effect will change the cross section by a much larger factor, say 1.1. We can check by simply doing the calculation.

\begin{figure}
\begin{center}
\ifusefigs 

\begin{tikzpicture}
\begin{axis}[title = {One jet cross section, ratios to NLO},
   xlabel={$P_\LT$}, ylabel={$K(P_\LT)$},
   legend cell align=left,
   every axis legend/.append style = {at={(axis cs:130,0.33)},anchor=south west},
   ymin= 0.30, ymax= 1.3
]

\errorband[blue,semithick]{fill=blue!30!white, opacity=0.5}
{data/JETS42.dat}{1}{5}{6}
\addlegendentry{Full, LC+} 

\errorband[red,semithick]{fill=red!30!white, opacity=0.5}
{data/JETS42.dat}{1}{13}{14}
\addlegendentry{Full, $\Delta$} 

\errorband[blue,dashed]{fill=blue!30!white, opacity=0.5}
{data/JETS42.dat}{1}{3}{4}
\addlegendentry{Std., LC+} 

\errorband[red,dashed]{fill=red!30!white, opacity=0.5}
{data/JETS42.dat}{1}{11}{12}
\addlegendentry{Std., $\Delta$} 

\end{axis}
\end{tikzpicture}

\fi

\end{center}
\caption{
Jet cross section ratios with $N_\Delta^\textrm{thr.}= 2$ and with $N_\Delta = N_\textrm{Re} = N_{\mi\pi} = 4$ for $t_0 < t < t_0 + 5$.
}
\label{fig:jets42}
\end{figure}
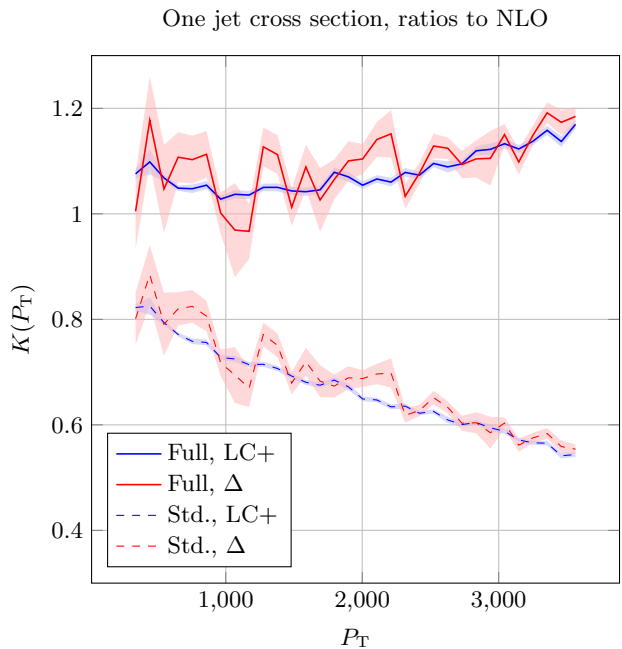

In Fig.~\ref{fig:jets42}, we plot results as in Fig.~\ref{fig:jets21} but with $N_\Delta = N_\textrm{Re} = N_{\mi\pi} = 4$ and $N_\Delta^\textrm{thr.}= 2$. The red curves show results with the extra color insertions while the blue curves show the results with just the LC+ approximation. The upper, solid curves are for the Full calculation including the threshold factor while the lower, dashed curves are for the Std. calculation with only the probability preserving parton shower. We see that the statistical fluctuations are much larger than they were with just two units of extra color.\footnote{With twice as much added color, it takes about ten times more computer time to generate the same number of Monte Carlo events. To produce Fig.~\ref{fig:jets42}, we used four times as much computer time as Fig.~\ref{fig:jets42}, so we generated only about 1/3 as many points.} We now cannot see the approximately 3\% change in the cross section that resulted from adding two units of extra color. However, if four units of extra color added 10\% to the cross section, we could see that even with the greater statistical fluctuations.

\section{Conclusions}
\label{sec:outlook}

We can offer three arguments for pursuing corrections to a leading order parton shower beyond the leading color approximation.

When one uses an NLO calculation for the hard scattering that initiates a shower, the hard scattering calculation should be matched to the parton shower. (\textsc{Deductor} does not currently include this matching.) The NLO calculation naturally includes contributions beyond the leading order in $1/N_\Lc^2$. It is not impossible to perform the matching when the shower lacks these contributions. However, it is clearly best if the shower includes the same color accuracy as the NLO hard scattering calculation for the color density matrix.

One can also hope to eventually have a parton shower algorithm in which the splitting kernels are correct to order $\as^2$ instead of just $\as$. However, $1/N_\Lc^2$ is of roughly the same size as $\as$. Thus it would seem as important to include $1/N_\Lc^2$ corrections to the order $\as$ splitting functions as it is to include $\as^2$ contributions to the shower splitting functions.

Perhaps most importantly, there may be processes in which corrections beyond the leading order in $1/N_\Lc^2$ are numerically important because these corrections multiply large logarithms \cite{Manchester2009, EarlyGap1, EarlyGap2, Manchester2005, NonGlobal1, NonGlobal2, NonGlobal3, NonGlobal4, SuperLeading1, SuperLeading2, Manchester2011, Seymour2018}. A parton shower is a promising way to investigate these effects, but evidently the shower must incorporate $(1/N_\Lc^2)^k$ corrections.

Now, we briefly review the argument of this paper. 

We view a parton shower as an application of the renormalization group, proceeding from harder interactions to softer interactions. We take the proper framework for the shower to be quantum statistical mechanics. Simple classical statistical mechanics is not sufficient for two reasons. First, we need to account for quantum interference between emission of a soft gluon from one parton and emission from another gluon. Second, parton color is a quantum degree of freedom. Using quantum statistical mechanics for color requires one to consider the $\ket{\{c\}_m}\bra{\{c'\}_m}$ density matrix in color. (The density matrix in parton spin space is required also, but in this paper we ignore spin.)

For a lowest order parton shower, we need two operators, which we call $\cH_I(t)$ and $\cV(t)$. $\cH_I(t)$ describes parton splittings, which increase the number of partons by one. It has a certain color structure, which one simply reads off from the Feynman rules for QCD. $\cV(t)$ leaves the number of partons unchanged.  It represents an approximation to one loop virtual graphs. It has a certain color structure, which one simply reads off from the Feynman rules for QCD.

The operators $\cH_I(t)$ and $\cV(t)$ are related by the equation, in the notation of this paper, $\sbra{1}\cH_I(t) = \sbra{1}\cV(t)$. This means that the ideal shower based on $\cH_I(t)$ and $\cV(t)$ preserves probabilities. Fundamentally, this property arises from the fact the infrared divergences of QCD cancel between real and virtual graphs, which in turn follows from the fact that the quantum evolution operator $U(t_2,t_1)$ is unitary \cite{LibbySterman}.

We know the color structure of $\cH_I(t)$, so it is straightforward to incorporate $\cH_I(t)$ into computer code, as in Refs.~\cite{PlatzerSjodahl, Isaacson:2018zdi, PlatzerSjodahlThoren}. However, we need also to incorporate $\cV(t)$ in order to be consistent with quantum mechanics.

It is, unfortunately, not easy to incorporate $\cV(t)$ into computer code for a parton shower. The reason is that, in the traditional formulation of a parton shower, one needs a Sudakov factor consisting of the exponential of an integral of $\cV(t)$, but this exponential is an operator on the color space. As the number of partons increases, the dimensionality of the color space becomes enormous, so the exponential of an integral of $\cV(t)$ becomes difficult to calculate.

One commonly applies the leading color (LC) approximation in a practical parton shower. Here $\cH_I(t)$ is approximated by an operator $\cH^\textsc{lc}(t)$ and $\cV(t)$ is approximated by an operator $\cV^\textsc{lc}(t)$. This approximation gets cross sections right to leading power in $1/N_\Lc^2$, where $N_\Lc = 3$ is the number of colors. It has the property that $\cV^\textsc{lc}(t)$ is diagonal in color, so that is easy to exponentiate. 

The LC approximation also has the crucial property $\sbra{1}\cH^\textsc{lc}(t) = \sbra{1}\cV^\textsc{lc}(t)$. For this reason the LC shower preserves probabilities.

\textsc{Deductor} uses an improved approximation, the LC+ approximation, as its starting point for treating color. This approximation retains some contributions that the LC approximation drops.  In this approximation $\cH_I(t)$ is approximated by an operator $\cH^\LCP(t)$ and $\cV(t)$ is approximated by an operator $\cV^\LCP(t)$.  The approximation has the property that $\cV^\LCP(t)$ is diagonal in color, so that is easy to exponentiate. It also has the property that $\sbra{1}\cH^\LCP(t) = \sbra{1}\cV^\LCP(t)$, so that the LC+ shower preserves probabilities.

The LC+ approximation becomes exact in the limit of collinear emissions or soft$\times$collinear emissions. That is, the only singular limit for emissions in which the LC+ approximation is not exact is the limit of fixed angle soft emissions. This feature, which is not shared by the LC approximation, is important for the working of the algorithm described in this paper.

\textsc{Deductor} is organized as a dipole shower, in which there is quantum interference between emission of a gluon from a parton with label $l$ and another parton with label $k$. The symmetry between $l$ and $k$ is removed in such a way that there is a singularity when the emitted gluon becomes collinear with parton $l$ but not with parton $k$. The momentum mapping used in \textsc{Deductor} depends on the choice of $l$ but not the choice of $k$. This feature is important in the algorithm used in this paper.

In this paper, we define $\Delta \cH(t)$ and $\Delta \cV(t)$ by $\cH_I(t) = \cH^\LCP(t) + \Delta\cH(t)$ and $\cV(t) = \cV^\LCP(t) + \Delta\cV(t)$, with $\Delta\cV(t) = \Delta\cV_\textrm{Re}(t)+ \cV_{\mi \pi}(t)$. Then we work order by order in $\Delta \cH(t)$ and $\Delta \cV(t)$ as advocated in Ref.~\cite{NScolor}.

We can retain terms with up to $N_\Delta$ powers of $\Delta \cH$ and $\Delta \cV$ in the first $t_\Delta$ units of shower time and we can retain $N_\Delta^\textrm{thr.}$ powers of $\Delta\cV$ in the threshold operator $\cU_\cV$. We also impose a limit $I_\textrm{max}$ on how large the color suppression index $I$, Eq.~(\ref{eq:Idef}), can grow. This limits the number of powers of $1/N_\Lc$ that we keep. 

The resulting algorithm for a parton shower is approximate with respect to color but the approximation is systematically improvable by making $N_\Delta$, $N_\Delta^\textrm{thr.}$, $t_\Delta$, and $I_\textrm{max}$ larger, at the cost of requiring more computer memory or more computer time to reach the same level of Monte Carlo statistical accuracy. This algorithm is implemented in public computer code.\footnote{Version 3.0.0 of the code, 
  used in this paper, is available at 
  \href{http://www.desy.de/~znagy/deductor/}
  {http://www.desy.de/$\sim$znagy/deductor/}
  and
  \href{http://pages.uoregon.edu/soper/deductor/}
  {http://pages.uoregon.edu/soper/deductor/}.} 

We have tested whether the cancellations between $\Delta \cH$ and $\Delta \cV$ actually work so as to produce a probability preserving shower. Within the accuracy of the calculations, these cancellations do work.

We have not used the new version of \textsc{Deductor} to investigate cross sections in which color beyond the LC+ approximation might play an important role. We expect to carry out such investigations in future work. However, we have calculated the one jet inclusive cross section beyond the LC+ approximation just to check how well the program works in calculating a physical cross section.

We now turn to the outlook for future work.

We expect that the algorithm presented here will not be the last word in algorithms for this purpose. Surely it is possible to do better. Indeed, \'Angeles Mart\'inez, De Angelis, Forshaw, Pl\"atzer, and Seymour \cite{Seymour2018} have provided a formalism for the description of soft gluon emissions that is similar in some ways to the general formalism \cite{NSI, NScolor} on which this paper is based. If the approach of Ref.~\cite{Seymour2018} can be extended to include the collinear singularities of QCD, then it will be of great interest to see if there can be a practical implementation of the resulting formalism. Perhaps such an implementation will be able to outperform what this paper provides.

Our numerical investigations suggest that $\cV_{\mi\pi}$ is effectively a larger operator than $\cV_\textrm{Re}$. For this reason, it may be better to include $\cV_{\mi\pi}$ in $\cV^\LCP$ instead of $\Delta \cV$. This means that one would need to numerically exponentiate $\cV_{\mi\pi}$. This will cost computer resources, so it remains to be seen if this is a better option.

\acknowledgments{ 
This work was supported in part by the United States Department of Energy under grant DE-SC0011640. This project has benefited from the participation of the authors at the Munich Institute for Astro- and Particle Physics program ``Automated, resummed and effective: precision computations for the LHC and beyond'' as well as the Kavli Institute for Theoretical Physics at the University of California, Santa Barbara, program ``LHC Run II and the Precision Frontier'' which was supported by the U.\ S.\ National Science Foundation under Grant No.\ NSF PHY11-25915.  We thank the MIAPP and the KITP for providing stimulating research environments. DS thanks the Erwin Schr\"odinger Institute of the University of Vienna for its hospitality at its program ``Challenges and Concepts for Field Theory and Applications in the Era of the LHC Run-2.'' This work benefited from access to the University of Oregon high performance computer cluster, Talapas.
}

\vfil
\appendix

\section{The action of $\cH_I$ and $\cV$}
\label{sec:HandVresult}

We need a convenient formula for $\cH_I(t)$ and for its LC+ version. We start with Eq.~(5.7) of Ref.~\cite{NScolor}, and write this as
\begin{widetext}
\begin{equation}
\begin{split}
\label{eq:HIdef2}
\big(\{\hat p,\hat f,\hat c',\hat c\}_{m+1}{}&\big|
\cH_I(t)\sket{\{p,f,c',c\}_{m}}
\\={}&
\sum_{l,k}
\delta(t - T_l(\zeta_\Lp,\{p\}_{m}))\,
(m+1)
\sbra{\{\hat p,\hat f\}_{m+1}}\cP_l\sket{\{p,f\}_m}\,
\lambda_{lk}(\{p,f\}_{m},\zeta)/N(k,l,\zeta_\Lf)
\\& \times
\big[\theta(k = l) - \theta(k \ne l)\big]
\\& \times
\bigg[
\sbra{\{\hat c',\hat c\}_{m+1}}
t^\dagger_l(f_l \to \hat f_l + \hat f_{m+1})\otimes 
t_k(f_k \to \hat f_k + \hat f_{m+1})
\sket{\{c',c\}_m}
\\&\qquad
+ 
\sbra{\{\hat c',\hat c\}_{m+1}}
t^\dagger_k(f_k \to \hat f_k + \hat f_{m+1}) \otimes 
t_l(f_l \to \hat f_l + \hat f_{m+1})
\sket{\{c',c\}_m}
\bigg]\;.
\end{split}
\end{equation}
\end{widetext}
Acting on a basis state $\sket{\{p,f,c',c\}_{m}}$ for $m$ partons, $\cH_I(t)$ produces a linear combination of basis states $\sket{\{\hat p,\hat f,\hat c',\hat c\}_{m+1}}$ for $m+1$ partons. There is a sum over the index $l$ of the emitting parton and the index $k$ of a parton that participates in quantum interference in the splitting. There is a new parton, labeled $m+1$. The function $\sbra{\{\hat p,\hat f\}_{m+1}}\cP_l\sket{\{p,f\}_m}$ specifies the momentum mapping for the splitting. It contains delta functions that, for given initial state momenta $\{p\}_m$, restrict the momenta $\{\hat p\}_{m+1}$ after the splitting to a three dimensional surface in momentum space. The points in this space can be labeled by splitting variables $\zeta_\Lp$ which can be chosen to be a virtuality variable $y$ (Eqs.~(\ref{eq:ydefFS}) and (\ref{eq:ydefIS})), a momentum fraction $z$ (Eqs.~(\ref{eq:zdefFS}) and (\ref{eq:zdefIS})), and an azimuthal angle $\phi$. Additionally, this function restricts the final state flavors $\{\hat f\}_{m+1}$. The splitting can then be characterized by flavor splitting variables $\zeta_\Lf$ (for instance $\Lu \to \Lu + \Lg$). We denote the momentum and flavor splitting variables collectively by $\zeta$. The function $T_l(\zeta_\Lp,\{p\}_{m})$ specifies our choice of the shower time $t$ (Eq.~(\ref{eq:showertime})).

The last factor in Eq.~(\ref{eq:HIdef2}) contains color factors. The notation is from Ref.~\cite{NSI}. For instance, the operator $t^\dagger_l(f_l \to \hat f_l + \hat f_{m+1})$ in $t^\dagger_l(f_l \to \hat f_l + \hat f_{m+1})\otimes 
t_k(f_k \to \hat f_k + \hat f_{m+1})$ acts on the ket color state and supplies the color matrices to split parton $l$ into a new parton $l$ and parton $m+1$, with color representations according to the specified flavors. For a $\Lg \to \Lg + \Lg$ splitting, we have (from Eq.~(7.24) of Ref.~\cite{NSI}),
\begin{equation}
t^\dagger_l(\Lg \to \Lg + \Lg) =
\sqrt{C_\LF}\, a_+^\dagger(l) - \sqrt{C_\LF}\, a_-^\dagger(l)
\;, 
\end{equation}
where $a_+^\dagger(l)$ inserts the new gluon just to the right of parton $l$ on whatever string in the color basis state contains parton $l$ and $a_-^\dagger(l)$ inserts the new gluon just to the left of parton $l$. Here we define ``left'' to be the direction of the quark line in the color string. For the emission of a gluon from a quark line, we define (from Eq.~(7.25) of Ref.~\cite{NSI}),
\begin{equation}
t^\dagger_l(q \to q + \Lg) =
\sqrt{C_\LF}\, a_+^\dagger(l)
\;, 
\end{equation}
where $ a_+^\dagger(l)$ inserts the gluon at the quark end of the string to which the quark $l$ belongs.

The functions $\lambda_{lk}(\{p,f\}_{m},\zeta)$ are
\begin{widetext}
\begin{equation}
\label{eq:partiallambda}
\begin{split}
    \lambda_{lk}(\{p,f\}_{m},\zeta)
    ={}&
    \frac{n_\Lc(a) n_\Lc(b)\,\eta_{\La}\eta_{\Lb}}
    {n_\Lc(\hat a) n_\Lc(\hat b)\,\hat \eta_{\La}\hat \eta_{\Lb}}\,
    \frac{f_{\hat a/A}(\hat\eta_\La,\mu^2_F)
      f_{\hat b/B}(\hat\eta_\Lb,\mu^2_F)}
    {f_{a/A}(\eta_\La,\mu^2_F)f_{b/B}(\eta_\Lb,\mu^2_F)}\,
    \frac{1}{2} N(k,l,\zeta_\Lf)
    \\
    &\times
    \Big\{
    \theta(k=l)\theta(\hat f_{m+1} \ne g)\,
    \overline w_{ll}(\{\hat p,\hat f\}_{m+1})
    \\
    &\qquad 
    + \theta(k=l) \theta(\hat f_{m+1} = g)\,
    [\overline w_{ll}(\{\hat p,\hat f\}_{m+1}) 
    - \overline w_{ll}^{\rm eikonal}
    (\{\hat p,\hat f\}_{m+1})]
    \\
    &\qquad 
    + \theta(k\ne l) \theta(\hat f_{m+1} = g)
    A'_{lk}(\{\hat p\}_{m+1})
    \overline w_{lk}^{\rm dipole}(\{\hat p,\hat f\}_{m+1})
    \Big\}
\;.
\end{split}
\end{equation}
\end{widetext}
The function $\overline w_{lk}^{\rm dipole}$ is the familiar eikonal splitting function,
\begin{equation}
\label{eq:wlkdipole}
\overline w_{lk}^{\rm dipole}(\{\hat p,\hat f\}_{m+1}) =
4\pi\as\ \frac{2\hat p_k\cdot\hat p_l}{\hat p_{m+1}\cdot \hat p_k\
\hat p_{m+1}\cdot \hat p_l}
\;.
\end{equation}
(See Eq.~(5.3) of Ref.~\cite{NScolor}. Here and throughout this paper, we assume massless partons.) The function $A'_{lk}$ partitions the dipole splitting function, which is symmetric under $k \leftrightarrow l$, into a part considered to be a splitting of parton $l$ and a part considered to be associated with a splitting of parton $k$. Our preferred choice is given in Eq.~(7.12) of Ref.~\cite{NSspin},
\begin{equation}
\label{eq:Aprimelk}
A'_{lk}(\{\hat p\}_{m+1}) =
\frac{\hat p_{m+1}\cdot \hat p_k\ \hat p_l\cdot \hat Q}
{\hat p_{m+1}\cdot \hat p_k\ \hat p_l\cdot \hat Q
+ \hat p_{m+1}\cdot \hat p_l\ \hat p_k\cdot \hat Q}
\;.
\end{equation}
The function $\overline w_{ll}^{\rm eikonal}$, from Eq.~(2.10) of Ref.~\cite{NSII}, is
\begin{equation}
\label{eq:wlleikonal}
\overline w_{ll}^{\rm eikonal}(\{\hat p,\hat f\}_{m+1}) = 
4\pi\as\ \frac{\hat p_l\cdot D(\hat p_{m+1};\hat Q) \cdot \hat p_l}
{(\hat p_{m+1}\cdot \hat p_l)^2}
\;.
\end{equation}
Here the total momentum of all the partons before the splitting is $Q$ and after the splitting is $\hat Q$. The tensor $D(\hat p_{m+1};\hat Q)^{\mu\nu}$ is
\begin{equation}
\begin{split}
\label{eq:Dmunu}
D^{\mu\nu}(\hat p_{m+1};\hat Q) ={}&
-g^{\mu\nu} 
+ \frac{\hat p_{m+1}^\mu \hat Q^\nu + \hat Q^\mu \hat p_{m+1}^\nu}
{\hat p_{m+1} \cdot \hat Q}
\\
&- \frac{\hat Q^2\ \hat p_{m+1}^\mu \hat p_{m+1}^\nu}{(\hat p_{m+1} \cdot \hat Q)^2}
\;,
\end{split}
\end{equation}
which is the polarization sum for gluon $m+1$ in timelike axial gauge, $\hat Q \cdot A(x) = 0$. 

The function $\overline w_{ll}$ is given in Refs.~\cite{NSI} and \cite{NSII}.\footnote{The definitions of $y$ and $z$ in Ref.~\cite{NSII} are in some cases different from the definition that we use in this paper.} It is rather complicated, but it is simple in the limit $y \to 0$ with fixed $z$. For example, for a final state $q \to q + \Lg$ splitting, we find (See Ref.~\cite{NSII}, Eq.~(2.24))
\begin{equation}
\overline w_{ll} \sim \frac{4\pi \as}{y\, p_l\cdot Q}\
\frac{1 + z^2}{1-z}
\;.
\end{equation}
Here we recognize that $(1+z^2)/(1-z)$ is the DGLAP splitting function for this splitting. For a final state $\Lg \to \Lg + \Lg$ splitting, we find (See Ref.~\cite{NSII}, Eq.~(2.52))
\begin{equation}
\overline w_{ll} \sim \frac{4\pi \as}{y\, p_l\cdot Q}\
\left[\frac{2 z}{1-z} + z(1-z)\right]
\;.
\end{equation}
If we add this quantity and the same quantity with $z \leftrightarrow (1-z)$, corresponding to interchanging the two identical final state gluons with labels $l$ and $m+1$, the sum of the quantities in square brackets is the DGLAP splitting function for finding a gluon in a gluon. For an initial state  $q \to q + \Lg$ splitting, we find (See Ref.~\cite{NSII}, Eq.~(2.38))
\begin{equation}
\overline w_{ll} \sim \frac{4\pi \as}{z y\, p_l\cdot Q}\
\frac{1 + z^2}{1-z}
\;.
\end{equation}
Again, $\overline w_{ll}$ is proportional to the DGLAP kernel $({1 + z^2})/({1-z})$ for finding a quark in a quark. For an initial state  $\Lg \to \Lg + \Lg$ splitting, we find (See Ref.~\cite{NSII}, Eq.~(2.59))
\begin{equation}
\overline w_{ll} \sim \frac{4\pi \as}{z y\, p_l\cdot Q}\
\left[
\frac{2z}{1-z}
+ \frac{2(1-z)}{z}
+ 2 z(1-z)
\right]
\;.
\end{equation}
Thus, the limiting form of $\overline w_{ll}$ is proportional to the DGLAP kernel for finding a gluon in a gluon. Although $\overline w_{ll}$ is proportional to the relevant DGLAP kernel in the limit $y \to 0$ at fixed $z$, the full functions $\overline w_{ll}$ are obtained directly from the relevant Feynman diagrams and are markedly different from the DGLAP kernels when $y$ is not small, and particularly when $y$ is comparable to $z$ or $1-z$.

We have multiplied and divided by a factor $N(k,l,\zeta_\Lf)$ that depends on whether $k = l$ and on the flavors $\zeta_\Lf$ in the splitting:
\begin{equation}
\begin{split}
\label{eq:Nlkdef}
N(k,l,\zeta_\Lf)
={}&
\begin{cases}
T_\LR & k=l,\      \Lg \to q + \bar q \\
C_\LF & k=l,\      q \to q + \Lg \textrm{ or } \bar q \to \bar q + \Lg \\
C_\LA & k=l,\      \Lg \to \Lg + \Lg   \\
C_\LF & k\ne l,\   q \to q + \Lg \textrm{ or } \bar q \to \bar q + \Lg   \\
C_\LA/2 & k\ne l,\ \Lg \to \Lg + \Lg  
\end{cases}
\;.
\end{split}
\end{equation}
This factor, from Eq.~(6.12) of Ref.~\cite{NScolor}, plays a role when we apply the LC+ approximation to the splitting.

We can rewrite Eq.~(\ref{eq:HIdef2}) by using the completeness relation from Eq.~(3.28) of Ref.~\cite{NSI} (with spin omitted),
\begin{equation}
  \label{eq:completeness}
  \begin{split}
    1 = \sum_m &
    \sum_{\{\hat c',\hat c\}_{m+1}}
    \frac{1}{(m+1)!}
    \int [d\{\hat p,\hat f\}_{m+1}]
    \\
    &\times
    \sket{\{\hat p,\hat f,\hat c',\hat c\}_{m+1}}
    \sbra{\{\hat p,\hat f,\hat c',\hat c\}_{m+1}}
    \;.
  \end{split}
\end{equation}
Here we have an integration over the momenta of the partons after the splitting along with sums over their flavors and colors. The factor $\sbra{\{\hat p,\hat f\}_{m+1}}\cP_l\sket{\{p,f\}_m}$ in Eq.~(\ref{eq:HIdef2}) contains delta functions, so that, using Eq.(12.2) of Ref.~\cite{NSI}, we are left with an integration over splitting variables $\zeta$, including a sum over the flavors in the splitting,
\begin{equation}
  \begin{split}
    &\frac{m+1}{(m+1)!}\,
    [d\{\hat p,\hat f\}_{m+1}]\
    \sbra{\{\hat p,\hat f\}_{m+1}}\cP_l\sket{\{p,f\}_m}
    = d\zeta
    \;.
  \end{split}
\end{equation}
For a final state splitting, we have (from Ref.~\cite{NSThreshold}, Eq.~(B.11))
\begin{equation}
  \label{eq:dzetaF}
  \begin{split}
    \int\! d\zeta\, \cdots
    ={}& 
    \frac{p_l\cdot Q}{8\pi^2}
    \int\!dy\, \sqrt{(1+y)^2 -  \frac{4 Q^2 y}{2 p_l\cdot Q}}\
    \\
    &\times
    \int\!dz \int\!\frac{d\phi}{2\pi}
    \sum_{f_{m+1}}
    \cdots
    \,.
  \end{split}
\end{equation}
For an initial state splitting, we have (from Ref.~\cite{NSThreshold}, Eq.~(B.41))
\begin{equation}
\label{eq:dzetaI}
\int\! d\zeta\, \cdots = 
\frac{Q^2}{16\pi^2}
\int\!dy\, 
\int\!\frac{dz}{z} \int\!\frac{d\phi}{2\pi}
\sum_{f_{m+1}}
\cdots
\,.
\end{equation}
A sum over the new colors remains from Eq.~(\ref{eq:completeness}). 

\vskip 1cm
\begin{widetext}
This gives us a more compact version of Eq.~(\ref{eq:HIdef2}),
\begin{equation}
\begin{split}
\label{eq:HIdef4}
\cH_I(t)\sket{\{p,f,c',c\}_{m}}
={}&
\sum_{l,k} \sum_{\{\hat c',\hat c\}_{m+1}}
\int\!d\zeta\
\delta(t - T_l(\zeta_\Lp,\{p\}_{m}))\,
\sket{\{\hat p,\hat f,\hat c',\hat c\}_{m+1}}\,
\\&\times
\lambda_{lk}(\{p,f\}_{m},\zeta)\,
G(k,l,\zeta_\Lf, \{\hat c',\hat c\}_{m+1},\{c',c\}_m)
\;,
\end{split}
\end{equation}
where $G$ is a color factor,
\begin{equation}
\begin{split}
\label{eq:Gdef}
G(k,l,\zeta_\Lf, \{\hat c',\hat c\}_{m+1},\{c',c\}_m) 
={}& 
\frac{\theta(k = l) - \theta(k \ne l)}{N(k,l,\zeta_\Lf)}
\\& \times
\bigg[
\sbra{\{\hat c',\hat c\}_{m+1}}
t^\dagger_l(f_l \to \hat f_l + \hat f_{m+1})\otimes 
t_k(f_k \to \hat f_k + \hat f_{m+1}) 
\sket{\{c',c\}_m}
\\&\quad
+ 
\sbra{\{\hat c',\hat c\}_{m+1}}
t^\dagger_k(f_k \to \hat f_k + \hat f_{m+1}) \otimes 
t_l(f_l \to \hat f_l + \hat f_{m+1})
\sket{\{c',c\}_m}
\bigg]\;.
\end{split}
\end{equation}

This was for the full splitting operator $\cH_I(t)$. With the LC+ approximation, the color factor here is modified \cite{NScolor}:
\begin{equation}
\begin{split}
\label{eq:HILCPdef2}
\cH^\LCP(t)\sket{\{p,f,c',c\}_{m}}
={}&
\sum_{l,k} \sum_{\{\hat c',\hat c\}_{m+1}}
\int\!d\zeta\
\delta(t - T_l(\zeta_\Lp,\{p\}_{m}))\,
\sket{\{\hat p,\hat f,\hat c',\hat c\}_{m+1}}\,
\\&\times
\lambda_{lk}(\{p,f\}_{m},\zeta)\,
G^\LCP(k,l,\zeta_\Lf, \{\hat c',\hat c\}_{m+1},\{c',c\}_m)
\;,
\end{split}
\end{equation}
where $G^\LCP$ is a color factor,
\begin{equation}
\begin{split}
G^\LCP(k,l,\zeta_\Lf,{}& \{\hat c',\hat c\}_{m+1},\{c',c\}_m) 
\\={}& 
\frac{\theta(k = l) - \theta(k \ne l)}{N(k,l,\zeta_\Lf)}
\\& \times
\bigg[
\sbra{\{\hat c',\hat c\}_{m+1}}
t^\dagger_l(f_l \to \hat f_l + \hat f_{m+1})\otimes 
t_k(f_k \to \hat f_k + \hat f_{m+1}) 
C^\dagger(l,m+1)
\sket{\{c',c\}_m}
\\&\quad
+ 
\sbra{\{\hat c',\hat c\}_{m+1}}
C(l,m+1)  t^\dagger_k(f_k \to \hat f_k + \hat f_{m+1}) \otimes 
t_l(f_l \to \hat f_l + \hat f_{m+1})
\sket{\{c',c\}_m}
\bigg]\;.
\end{split}
\end{equation}
The added color operator $C(l,m+1)$ restricts the allowed states after the splitting to those in which partons $l$ and $m+1$ are color connected or in which they are a quark and the corresponding antiquark in a $\Lg \to q + \bar q$ splitting. See Eq.~(6.4) of Ref.~\cite{NScolor}.

In the case of gluon emission, $\hat f_{m+1} = \Lg$, the color factor $G^\LCP$ is non-zero only if parton $k$ is color connected to parton $l$ in the ket state or the bra state or both. That is, the result is zero unless the function $\chi(k,l,\{c',c\}_m)$ defined in Eq.~(\ref{eq:chidef1}) is nonzero.

The factor $N(k,l,\zeta_\Lf)$ in Eq.~(\ref{eq:Nlkdef}) is defined so that the total probability associated with the color factor $G^\LCP$ is just $\chi(k,l,\{c',c\}_m)$ times the probability associated with the starting color state:
\begin{equation}
\sum_{\{\hat c',\hat c\}_{m+1}} \brax{\{\hat c'\}_{m+1}}\ket{\{\hat c\}_{m+1}}\, 
G^\LCP(k,l,\zeta_\Lf, \{\hat c',\hat c\}_{m+1},\{c',c\}_m) 
= \brax{\{c'\}_{m}}\ket{\{c\}_{m}}\,\chi(k,l,\{c',c\}_m)
\;.
\end{equation}

We define $\Delta \cH(t)$ as $\Delta \cH(t) = \cH_I(t) - \cH^\LCP(t)$. Then
\begin{equation}
\begin{split}
\label{eq:DeltaHI}
\Delta\cH(t)\sket{\{p,f,c',c\}_{m}}
={}&
\sum_{l} \sum_{k \ne l}\sum_{\{\hat c',\hat c\}_{m+1}}
\int\!d\zeta\
\delta(t - T_l(\zeta_\Lp,\{p\}_{m}))\,
\sket{\{\hat p,\hat f,\hat c',\hat c\}_{m+1}}\,
\\&\times
\lambda_{lk}(\{p,f\}_{m},\zeta)\, 
\Delta G(k,l,\zeta_\Lf, \{\hat c',\hat c\}_{m+1},\{c',c\}_m)
\;,
\end{split}
\end{equation}
where
\begin{equation}
  \begin{split}
    \Delta G(k,l,\zeta_\Lf, \{\hat c',\hat c\}_{m+1},\{c',c\}_m)
    ={}&
    G(k,l,\zeta_\Lf, \{\hat c',\hat c\}_{m+1},\{c',c\}_m)
    -
    G^\LCP(k,l,\zeta_\Lf, \{\hat c',\hat c\}_{m+1},\{c',c\}_m)
    \;.
  \end{split}
\end{equation}
We have noted that $\Delta G(k,l,\zeta_\Lf, \{\hat c',\hat c\}_{m+1},\{c',c\}_m)$ is non-zero only for $k \ne l$.

We need also the operator $\cV(t)$, which does not change the number of partons, their momenta, or their flavors and is related to $\cH_I(t)$ by $\sbra{1}\cH_I(t) =  \sbra{1}\cV(t)$. Following Ref.~\cite{NScolor} we define
\begin{equation}
  \begin{split}
    \label{eq:Vdef}
    \cV(t)\sket{\{p,f,c',c\}_{m}}
    ={}&
    \sum_{l,k} \sum_{\{\hat c',\hat c\}_m}
    \sket{\{p, f,\hat c',\hat c\}_{m}}\,
    \int\!d\zeta\
    \delta(t - T_l(\zeta_\Lp,\{p\}_{m}))\,
    \frac{\lambda_{lk}(\{p,f\}_{m},\zeta)}{N(k,l,\zeta_\Lf)}
    \\&\times
    [\theta(k = l) - \theta(k \ne l)]
    \\& \times
    \bigg[
    \sbra{\{\hat c',\hat c\}_{m}}
    1\otimes 
    t_k(f_k \to \hat f_k + \hat f_{m+1})
    t^\dagger_l(f_l \to \hat f_l + \hat f_{m+1})
    \sket{\{c',c\}_m}
    \\&\qquad
    + 
    \sbra{\{\hat c',\hat c\}_{m}}
    t_l(f_l \to \hat f_l + \hat f_{m+1})
    t^\dagger_k(f_k \to \hat f_k + \hat f_{m+1}) \otimes 
    1\sket{\{c',c\}_m}
    \bigg]\;.
  \end{split}
\end{equation}
This creates a practical problem because the color operators are not diagonal in the basis that we use, so that it is not practical to calculate matrix elements of an exponential of $\cV(t)$.

With the LC+ approximation, we define $\cV^\LCP(t)$, using $\sbra{1}\cH^\LCP(t) =  \sbra{1}\cV^\LCP(t)$. This gives us
\begin{equation}
  \begin{split}
    \label{eq:VLCPdef}
    \cV^\LCP(t)\sket{\{p,f,c',c\}_{m}}
    ={}&
    \sum_{l,k} \sum_{\{\hat c',\hat c\}_m}
    \int\!d\zeta\
    \delta(t - T_l(\zeta_\Lp,\{p\}_{m}))\,
    \sket{\{p, f, c', c\}_{m}}\,
    \frac{\lambda_{lk}(\{p,f\}_{m},\zeta)}{N(k,l,\zeta_\Lf)}
    \\&\times
    [\theta(k = l) - \theta(k \ne l)]
    \\& \times
    \bigg[
    \sbra{\{\hat c',\hat c\}_{m}}
    1\otimes 
    t_k(f_k \to \hat f_k + \hat f_{m+1})
    C^\dagger(l,m+1)
    t^\dagger_l(f_l \to \hat f_l + \hat f_{m+1})
    \sket{\{c',c\}_m}
    \\&\qquad
    +
    \sbra{\{\hat c',\hat c\}_{m}}
    t_l(f_l \to \hat f_l + \hat f_{m+1})
    C(l,m+1)
    t^\dagger_k(f_k \to \hat f_k + \hat f_{m+1}) \otimes 
    1\sket{\{c',c\}_m}
    \bigg]\;.
  \end{split}
\end{equation}
Now the color factors are much simpler. For instance for $k \ne l$, for the case in which all of the partons $k$, $l$ and $m+1$ are gluons, in the term
\begin{equation*}
\sbra{\{\hat c',\hat c\}_{m}}
t_l(f_l \to \hat f_l + \hat f_{m+1})
C(l,m+1)
t^\dagger_k(f_k \to \hat f_k + \hat f_{m+1}) \otimes 
1\sket{\{c',c\}_m}
\;,
\end{equation*}
parton $k$ splits into two gluons in the ket state $\ket{\{c\}_m}$. When we expand the result into basis states (now with $m+1$ partons), the new gluon can go in two places. The operator $C(l,m+1)$ selects one of these choices as long as $l$ was color connected to $k$ in $\ket{\{c\}_m}$. Otherwise no choice survives. Thus the result is proportional to $\chi(k,l,\{c\}_m)$. The added gluon then also attaches to parton $l$. The color algebra then gives us a factor $C_\LA/2$ times $\ket{\{c\}_m}$. Continuing with this analysis as in Ref.~\cite{NScolor}, we find that the entire color factor in the last three lines of Eq.~(\ref{eq:VLCPdef}) is an eigenvalue $\chi(k,l,\{c', c\}_{m})\,N(k,l,\zeta_\Lf)$. This gives us a very simple result,
\begin{equation}
  \begin{split}
    \label{eq:VLCP}
    \cV^\LCP(t)\sket{\{p,f,c',c\}_{m}}
    ={}&
    \sket{\{p, f, c', c\}_{m}}\,
    \sum_{l,k}
    \int\!d\zeta\
    \delta(t - T_l(\zeta_\Lp,\{p\}_{m}))\,
    \lambda_{lk}(\{p,f\}_{m},\zeta) 
    \chi(k,l,\{c', c\}_{m})
    \;.
  \end{split}
\end{equation}
%

\section{The soft splitting functions}
\label{sec:softsplitting}

In this appendix, we work out just what the soft splitting functions are. We can start with Eq.~(5.7) of Ref.~\cite{NScolor}:
\begin{equation}
  \begin{split}
    \label{eq:HIdef}
    \big(\{\hat p,\hat f,\hat c',\hat c\}_{m+1}{}&\big|
    \cH_I(t)\sket{\{p,f,c',c\}_{m}}
    \\={}&
    \sum_{l,k}
    \delta(t - T_l(\zeta_\Lp,\{p\}_{m})))\,
    (m+1)
    \sbra{\{\hat p,\hat f\}_{m+1}}\cP_l\sket{\{p,f\}_m}
    \\&\times
    \frac
    {n_\Lc(a) n_\Lc(b)\,\eta_{\La}\eta_{\Lb}}
    {n_\Lc(\hat a) n_\Lc(\hat b)\,
      \hat \eta_{\La}\hat \eta_{\Lb}}\,
    \frac{
      f_{\hat a/A}(\hat \eta_{\La},\mu^{2}_{F})
      f_{\hat b/B}(\hat \eta_{\Lb},\mu^{2}_{F})}
    {f_{a/A}(\eta_{\La},\mu^{2}_{F})
      f_{b/B}(\eta_{\Lb},\mu^{2}_{F})}
    \\&\times \frac{1}{2}\bigg[
    \theta(k = l)\,
    \theta(\hat f_{m+1} \ne g)\,
    \overline w_{ll}(\{\hat p,\hat f\}_{m+1})
    \\&\qquad +
    \theta(k = l)\,
    \theta(\hat f_{m+1} = g)\,
    [\overline w_{ll}(\{\hat p,\hat f\}_{m+1})
    - \overline w_{ll}^{\rm eikonal}(\{\hat p,\hat f\}_{m+1})]
    \\&\qquad -
    \theta(k\ne l)\,
    \theta(\hat f_{m+1} = g)
    A'_{lk}(\{\hat p\}_{m+1})\,\overline w_{lk}^{\rm dipole}(\{\hat p,\hat f\}_{m+1})
    \bigg]
    \\&\hskip 0.2 cm \times
    \bigg[
    \sbra{\{\hat c',\hat c\}_{m+1}}
    t^\dagger_l(f_l \to \hat f_l + \hat f_{m+1})\otimes 
    t_k(f_k \to \hat f_k + \hat f_{m+1})
    \sket{\{c',c\}_m}
    \\&\qquad
    +
    \sbra{\{\hat c',\hat c\}_{m+1}}
    t^\dagger_k(f_k \to \hat f_k + \hat f_{m+1}) \otimes 
    t_l(f_l \to \hat f_l + \hat f_{m+1})
    \sket{\{c',c\}_m}
    \bigg]\;.
  \end{split}
\end{equation}
This corresponds to the way that we actually compute, but it is not suited for our present investigations. Instead, we replace
\begin{equation}
\begin{split}
- \overline w_{ll}^{\rm eikonal}&(\{\hat p,\hat f\}_{m+1})\,
2\,[t^\dagger_l(f_l \to \hat f_l + \Lg)\otimes 
t_l(f_l \to \hat f_l + \Lg)]
=
\sum_{k\ne l}
\overline w_{ll}^{\rm eikonal}(\{\hat p,\hat f\}_{m+1})
\\&\times
[t^\dagger_l(f_l \to \hat f_l + \Lg)\otimes 
t_k(f_k \to \hat f_k + \Lg)
+
t^\dagger_k(f_l \to \hat f_l + \Lg)\otimes 
t_l(f_l \to \hat f_k + \Lg)]
\;.
\end{split}
\end{equation}
This gives us
\begin{equation}
\begin{split}
\label{eq:HImod1}
\big(\{\hat p,\hat f,\hat c',\hat c\}_{m+1}{}&\big|
\cH_I(t)\sket{\{p,f,c',c\}_{m}}
\\={}&
\sum_{l,k}
\delta(t - T_l(\zeta_\Lp,\{p\}_{m})))\,
(m+1)
\sbra{\{\hat p,\hat f\}_{m+1}}\cP_l\sket{\{p,f\}_m}
\\&\times
\frac
{n_\Lc(a) n_\Lc(b)\,\eta_{\La}\eta_{\Lb}}
{n_\Lc(\hat a) n_\Lc(\hat b)\,
 \hat \eta_{\La}\hat \eta_{\Lb}}\,
\frac{
f_{\hat a/A}(\hat \eta_{\La},\mu^{2}_{F})
f_{\hat b/B}(\hat \eta_{\Lb},\mu^{2}_{F})}
{f_{a/A}(\eta_{\La},\mu^{2}_{F})
f_{b/B}(\eta_{\Lb},\mu^{2}_{F})}
\\&\times \frac{1}{2}\bigg[
\theta(k = l)\,
\theta(\hat f_{m+1} \ne g)\,
\overline w_{ll}(\{\hat p,\hat f\}_{m+1})
\\&\qquad +
\theta(k = l)\,
\theta(\hat f_{m+1} = g)\,
\overline w_{ll}(\{\hat p,\hat f\}_{m+1})
\\&\qquad -
\theta(k\ne l)\,
\theta(\hat f_{m+1} = g)
\\ &\quad\quad \times
\left\{
A'_{lk}(\{\hat p\}_{m+1})\,\overline w_{lk}^{\rm dipole}(\{\hat p,\hat f\}_{m+1})
- \overline w_{ll}^{\rm eikonal}(\{\hat p,\hat f\}_{m+1})
\right\}
\bigg]
\\&\hskip 0.2 cm \times
\bigg[
\sbra{\{\hat c',\hat c\}_{m+1}}
t^\dagger_l(f_l \to \hat f_l + \hat f_{m+1})\otimes 
t_k(f_k \to \hat f_k + \hat f_{m+1})
\sket{\{c',c\}_m}
\\&\qquad
+
\sbra{\{\hat c',\hat c\}_{m+1}}
t^\dagger_k(f_k \to \hat f_k + \hat f_{m+1}) \otimes 
t_l(f_l \to \hat f_l + \hat f_{m+1})
\sket{\{c',c\}_m}
\bigg]\;.
\end{split}
\end{equation}

The combination
\begin{equation}
\begin{split}
\overline w_{lk}^{\rm soft}(\{\hat p,\hat f\}_{m+1}) ={}&
A'_{lk}(\{\hat p\}_{m+1})\,
\overline w_{lk}^{\rm dipole}(\{\hat p,\hat f\}_{m+1})
- \overline w_{ll}^{\rm eikonal}(\{\hat p,\hat f\}_{m+1})
\end{split}
\end{equation}
is the soft splitting function describing the emission of a soft gluon from parton $l$ with interference from emitting the gluon from parton $k$. The function $\overline w_{lk}^{\rm dipole}$ was given in Eq.~(\ref{eq:wlkdipole}). The function $A'_{lk}$ was given in Eq.~(\ref{eq:Aprimelk}). The function $w_{ll}^{\rm eikonal}$ was given in Eqs.~(\ref{eq:wlleikonal}) and (\ref{eq:Dmunu}). Hatted vectors represent momenta of the partons after the splitting and $\hat Q$ is the total momentum of the final state partons after the splitting.

We can assemble this expression:
\begin{equation}
\begin{split}
\overline w_{lk}^{\rm soft}(\{\hat p,\hat f\}_{m+1}) ={}& 
\frac{4\pi\as}{\hat p_{m+1}\cdot \hat p_l}\bigg\{
\frac{2\hat p_k\cdot\hat p_l\,\hat p_l\cdot \hat Q}
{\hat p_{m+1}\cdot \hat p_k\ \hat p_l\cdot \hat Q
+ \hat p_{m+1}\cdot \hat p_l\ \hat p_k\cdot \hat Q}
- \frac{2\hat p_l\cdot \hat Q}{\hat p_{m+1} \cdot \hat Q}
+ \frac{\hat Q^2\,\hat p_{m+1}\cdot \hat p_l}{(\hat p_{m+1} \cdot \hat Q)^2}
\bigg\}
\;.
\end{split}
\end{equation}
An instructive way to examine the collinear limit is to write
\begin{equation}
\begin{split}
\overline w_{lk}^{\rm soft}(\{\hat p,\hat f\}_{m+1}) ={}& 
\frac{4\pi\as}{2\hat p_{m+1}\cdot \hat p_l\,\hat Q^2}
\frac{(\hat p_l\cdot \hat Q)^2\,\hat p_k\cdot \hat Q}
{\hat p_{m+1}\cdot \hat p_k\ \hat p_l\cdot \hat Q
+ \hat p_{m+1}\cdot \hat p_l\ \hat p_k\cdot \hat Q}\
h_{lk}
\;,
\end{split}
\end{equation}
where
\begin{equation}
\begin{split}
h_{lk} ={}& 
\frac{4\hat p_k\cdot\hat p_l\,\hat Q^2}
{\hat p_l\cdot \hat Q\,\hat p_k\cdot \hat Q}
- \frac{4 \hat Q^2}
{\hat p_{m+1} \cdot \hat Q\,\hat p_l\cdot \hat Q\,\hat p_k\cdot \hat Q}
\left(\hat p_{m+1}\cdot \hat p_k\ \hat p_l\cdot \hat Q
+ \hat p_{m+1}\cdot \hat p_l\ \hat p_k\cdot \hat Q\right)
\\&
+ \frac{2(\hat Q^2)^2\,\hat p_{m+1}\cdot \hat p_l}
{(\hat p_l\cdot \hat Q)^2\,(\hat p_{m+1} \cdot \hat Q)^2\,\hat p_k\cdot \hat Q}
\left(\hat p_{m+1}\cdot \hat p_k\ \hat p_l\cdot \hat Q
+ \hat p_{m+1}\cdot \hat p_l\ \hat p_k\cdot \hat Q\right)
\;.
\end{split}
\end{equation}
\end{widetext}

We look at this in the rest frame of $\hat Q$. Writing $\hat p_l = (E_l, E_l\, \vec u_l)$, where $\vec u_l^2 = 1$, and similarly for the other vectors, this is
\begin{equation}
\begin{split}
\overline w_{lk}^{\rm soft}(\{\hat p,&\hat f\}_{m+1})
\\
 ={}& 
\frac{4\pi\as}{2E_{m+1}^2 (1- \vec u_{m+1} \cdot \vec u_l)}
\\&\times
\frac{h_{lk}}
{(1- \vec u_{m+1} \cdot \vec u_k) + (1- \vec u_{m+1} \cdot \vec u_l)}
\;,
\end{split}
\end{equation}
with
\begin{equation}
\begin{split}
h_{lk} ={}& 
4(1- \vec u_k \cdot \vec u_l) 
\\&
- 4(1- \vec u_{m+1} \cdot \vec u_k) -4 (1- \vec u_{m+1} \cdot \vec u_l)
\\&
+ 2(1- \vec u_{m+1} \cdot \vec u_l)
[(1- \vec u_{m+1} \cdot \vec u_k) 
\\&
+ (1- \vec u_{m+1} \cdot \vec u_l)]
\;.
\end{split}
\end{equation}
With a little manipulation, this becomes
\begin{align}
&{\!\!\!\!\!}\overline w_{lk}^{\rm soft}(\{\hat p,\hat f\}_{m+1}) 
\\={}& 
\frac{4\pi\as}{E_{m+1}^2 (\vec u_{m+1} - \vec u_l)^2}
\frac{h_{lk}}
{(1- \vec u_{m+1} \!\cdot \!\vec u_k) + (1- \vec u_{m+1}\!\cdot\!\vec u_l)}
\;,
\notag
\end{align}
with
\begin{align}
h_{lk} ={}& 
4\,(\vec u_{m+1} - \vec u_l) \cdot  (\vec u_{k} - \vec u_l)
\\&
- (\vec u_{m+1} - \vec u_l)^2
[(1 + \vec u_{m+1} \!\cdot\!\vec u_k) + (1 + \vec u_{m+1}\!\cdot\! \vec u_l)]
\;.
\notag
\end{align}
There is an overall factor $1/E_{m+1}^2$, so that there is a singularity in the soft limit. There is also an overall factor $1/(\vec u_{m+1} - \vec u_l)^2$, so that there would be a singularity in the collinear limit $\vec u_{m+1} \to \vec u_l$ if this factor were not cancelled. The first term in $h_{lk}$ has a linear zero as $\vec u_{m+1} \to \vec u_l$, while the second term has a quadratic zero. This leaves an integrable collinear singularity, which disappears if one averages over the direction of $(\vec u_{m+1} - \vec u_l)$. Also, note that $h_{lk}$ can have either sign.

We could, if we liked, take the soft limit in $\overline w_{lk}^{\rm soft}(\{\hat p,\hat f\}_{m+1})$ by setting $\hat p_l \to p_l$ and $\hat p_k \to p_k$. For an initial state emission, we can also set $\hat Q \to Q$ and we can replace $\hat \eta \to \eta$ in the parton distribution functions that multiply $\overline w_{lk}^{\rm soft}(\{\hat p,\hat f\}_{m+1})$ in Eq.~(\ref{eq:HImod1}). This doesn't help in the \textsc{Deductor} code, but it may be useful for analyses.


\newpage 

\end{document}